\pdfoutput=1
\documentclass[journal,twoside,web]{ieeecolor}  % Comment this line out if you need a4paper
\usepackage{generic}
\usepackage{cite}
\usepackage{amsmath,amssymb,amsfonts}
\usepackage{graphicx}
\usepackage{textcomp}
\usepackage{longtable}% for long tables
\usepackage{mathtools}
\usepackage{amsfonts,amsmath,amssymb}

\usepackage{amsthm}
\makeatletter
\renewcommand*\env@matrix[1][*\c@MaxMatrixCols c]{%
	\hskip -\arraycolsep
	\let\@ifnextchar\new@ifnextchar
	\array{#1}}
\makeatother
\usepackage{booktabs}
\usepackage{mathrsfs}  
\usepackage{tikz}
\usepackage{pgfplots}
\usetikzlibrary{positioning,calc,fit,arrows}
\usepackage{tkz-fct}
\usepackage{calc}
\usetikzlibrary{intersections}
\usetikzlibrary{decorations.shapes,shapes.geometric,shadows,arrows,automata,positioning,calendar,mindmap,backgrounds,scopes,chains,er,patterns,pgfplots.groupplots}
\tikzset{initial text={}, % remove text for initial state
	double distance=2pt, % adjust appearance of accepting state
	every state/.style = {draw = black, fill = grayfilling} % sets properties for each state
}
\usetikzlibrary{calc}
\usepackage[load-configurations=version-1]{siunitx} % newer version
\usepackage{tikz-3dplot}
\usepackage{subfig}
\usepackage{url}
\pgfmathdeclarefunction{normal}{2}{%
	\pgfmathparse{1/(#2*sqrt(2*pi))*exp(-((x-#1)^2)/(2*#2^2))}%
}
\pgfmathdeclarefunction{dnorm}{2}{%
	\pgfmathparse{1/(#2*sqrt(2*pi))*exp(-((x-#1)^2)/(2*#2^2))}%
}
\usepackage[author={Oliver}]{pdfcomment}
\pdfcommentsetup{color=red}
\usepackage[framemethod=TikZ]{mdframed}
\usepackage{algcompatible}
\usepackage{algorithm}

\makeatletter
\renewcommand{\fnum@algorithm}{\fname@algorithm{} \thealgorithm:}
\makeatother

\usepackage{graphicx}
\newcommand{\Refi}{\mathbf{s}}

\newtheoremstyle{theoremdd}% name of the style to be used
{\topsep}% measure of space to leave above the theorem. E.g.: 3pt
{\topsep}% measure of space to leave below the theorem. E.g.: 3pt
{\itshape}% name of font to use in the body of the theorem
{0pt}% measure of space to indent
{\bfseries}% name of head font
{:}% punctuation between head and body
{ }% space after theorem head; " " = normal interword space
{\thmname{#1}\thmnumber{ #2}\boldmath\textbf{\thmnote{ (#3)}}} %{\thmname{#1}\thmnumber{ #2}\textnormal{\thmnote{ (#3)}}}
\theoremstyle{theoremdd}
\makeatletter % Use colon for proof env
\renewenvironment{proof}[1][\proofname]{\par
	\pushQED{\qed}%
	\normalfont \topsep6\p@\@plus6\p@\relax
	\trivlist
	\item\relax
	{\itshape
		#1\@addpunct{:}}\hspace\labelsep\ignorespaces
}{%
	\popQED\endtrivlist\@endpefalse
}
\makeatother

\usepackage{xcolor}

\definecolor{lightblue}{RGB}{0,141,225} %TACcyan
\definecolor{lightgreen}{rgb}{0.67, 0.88, 0.69}
\definecolor{lightpink}{rgb}{1.0, 0.72, 0.77}
\definecolor{lightpurple}{rgb}{0.96, 0.73, 1.0}
\definecolor{lightyellow}{rgb}{0.98, 0.93, 0.37}
\definecolor{grayfilling}{gray}{0.95} % Used for tikz graphics as a filling
\definecolor{grayshadow}{gray}{0.5} % Used for tikz graphics as a filling
\colorlet{cherryred}{red!80!black}

\newtheorem{problem}{Problem}
\newtheorem{definition}{Definition}
\newtheorem{theorem}{Theorem}
\newtheorem{proposition}{Proposition}
\newenvironment{prob}
{\begin{mdframed}[backgroundcolor=grayfilling, shadow=true, shadowsize=5.5pt, shadowcolor=grayshadow]
		\begin{problem}}
		{\end{problem}\vspace{.4em}\end{mdframed}}
\newcount\savedtheorem
\newenvironment{subproblem}{%
	\savedtheorem=\value{problem}%
	\edef\prevthetheorem{\theproblem}%
	\setcounter{problem}{0}%
	\renewcommand\theproblem{\prevthetheorem.\arabic{problem}}%
}
{%
	\setcounter{problem}{\savedtheorem}%
}
       
\newtheorem{remark}{Remark}
\newtheorem{example}{Example}
\newcommand{\1}{\mathbf 1} % Indicator function
\newcommand{\R}{\mathcal R} % Simulation relation

\newcommand{\norm}[1]{\left\lVert#1\right\rVert} % Norm
\newcommand{\normalkernel}[2]{\frac{1}{(2\pi)^{\frac{n}{2}}} e^{-\frac{\norm{#1}^2}{2}}#2} % Multivariate normal kernel
\newcommand{\cdf}[1]{\mathrm{cdf}\left( #1 \right)}
\newcommand{\offset}{\gamma}

\newcommand{\T}{^{\top\!}}
\newcommand{\dxp}{d\xp}
\newcommand{\dxhp}{d\xhp}
\newcommand{\xp}{x^{+\!}}
\newcommand{\xhp}{\hat{x}^{+\!}}
\newcommand{\xtp}{\tilde{x}^{+\!}}
\ifdefined\U
\renewcommand{\U}{\mathbb{U}}
\else
\newcommand{\U}{\mathbb{U}}
\fi

\newcommand{\Uh}{\hat{\mathbb{U}}}
\newcommand{\given}{\;|\;}

\newcommand{\fdel}[1]{{f}_\delta(#1)}
\newcommand{\fhdel}[1]{\hat{{f}}_\delta (#1)}

\newcommand{\absstwo}{{\Mh\rightarrow\Mt}}
\newcommand{\out}[1]{\mathbf d_\Y(#1)}

\newcommand{\Z}{\mathbb{Z}}
\newcommand{\satisfies}{\vDash}

\newcommand{\Tr}{\mathbf{t}}
\newcommand{\Trh}{\hat{\mathbf{t}}}

\newcommand{\X}{\mathbb{X}}
\newcommand{\Xh}{\hat{\mathbb{X}}}
\newcommand{\x}[1]{{x}_{#1}}
\newcommand{\xs}{\mathbf{x}}
\newcommand{\xin}{ {x}_0}
\newcommand{\M}{\mathbf M}
\newcommand{\Mh}{\widehat{\mathbf M}}
\newcommand{\Mt}{\widetilde{\mathbf M}}
\newcommand{\A}{\mathbb{U}}
\newcommand{\Ah}{\hat{\mathbb{U}}}

\newcommand{\ac}[1]{u_{#1}}
\newcommand{\acs}{\mathbf{u}}
\newcommand{\rel}{\mathcal{R}}

\newcommand{\Hist}{\mathbb{H}}

\newcommand{\ind}{\mathbf 1}
\newcommand{\AP}{\mathsf{AP}}

\newcommand{\notltl}{\neg}
\newcommand{\andltl}{\wedge}
\newcommand{\orltl}{\vee}

\newcommand{\Next}{\ensuremath{\bigcirc}}

\newcommand{\Event}{\ensuremath{\ \diamondsuit\ }}
\newcommand{\Until}{\ \mathcal{U}\ }

\newcommand{\cdotx}{\,\cdot\,}
\newcommand{\alphabeth}{\Sigma}
\newcommand{\word}{\boldsymbol{\omega}}

\newcommand{\letter}{l}
\newcommand{\True}{\operatorname{\mathsf{true}}}

{} % normal distribution

\newcommand{\meas}{\nu}     % Probability of an event
\newcommand{\po}{p}     % Probability of an event
\newcommand{\pok}{\mathbf{p}}     % Probability kernel of an event
\newcommand{\pk}[1]{\pok\left(#1\right)}     % Probability kernel of an event
\newcommand{\borel}[1]{\mathcal{B}\left(#1\right)}
\newcommand{\eps}{\varepsilon}
\newcommand{\InF}{\mathbf{i}} %\mathcal{F}_{u}

\newcommand{\Ca}{{\mathbf{C}}}
\newcommand{\Cah}{\widehat{\mathbf{C}}}
\newcommand{\Y}{\mathbb{Y}}%{The set of output values}     % Observation space
\newcommand{\Yh}{\hat{\mathbb{Y}}}%{The set of output values}     % Observation space
\newcommand{\support}{\operatorname{supp}}

\newcommand{\W}{v}
\newcommand{\Wt}{\boldsymbol{v}}
\newcommand{\fW}{\bar{\W}} % full probability coupling
\newcommand{\sW}{{\W}} % sub-probability coupling
\newcommand{\fWt}{\bar{\Wt}} % full probability coupling
\newcommand{\sWt}{{\Wt}} % sub-probability coupling

\newcommand{\Vb}{V} % Value function derivate

\newcommand{\trans}{\tau}
\newcommand{\Lim}{\mathbf L}

\newcommand{\xh}[1]{\hat{x}_{#1}}
\newcommand{\xt}[1]{\tilde{x}_{#1}}
\newcommand{\uh}[1]{\hat{u}_{#1}}
\newcommand{\ut}[1]{\tilde{u}_{#1}}
\newcommand{\ach}[1]{\hat{u}_{#1}}

\renewcommand{\P}{\mathbb{P}}

\usepackage{tcolorbox}
\usepackage{mathtools}

\newtcbox{\blueb}{nobeforeafter,tcbox raise base,boxrule=0.4pt,top=0mm,bottom=0mm,
	right=0mm,left=0mm,arc=1pt,boxsep=2pt,before upper={\vphantom{dlg}},
	colframe=blue!50!black,coltext=black!25!black,colback=blue!10!white}
\newtcbox{\redb}{nobeforeafter,tcbox raise base,boxrule=0.4pt,top=0mm,bottom=0mm,
	right=0mm,left=0mm,arc=1pt,boxsep=2pt,before upper={\vphantom{dlg}},
	colframe=red!50!black,coltext=black!25!black,colback=red!10!white}

\newtcbox{\bluebs}{nobeforeafter,tcbox raise base,boxrule=0.4pt,top=0mm,bottom=0mm,
	right=0mm,left=0mm,arc=1pt,boxsep=.5pt,before upper={\vphantom{dlg}},
	colframe=blue!50!black,coltext=black!25!black,colback=blue!10!white}
\newtcbox{\redbs}{nobeforeafter,tcbox raise base,boxrule=0.4pt,top=0mm,bottom=0mm,
	right=0mm,left=0mm,arc=1pt,boxsep=.5pt,before upper={\vphantom{dlg}},
	colframe=red!50!black,coltext=black!25!black,colback=red!10!white}

\robustify{\bluebs}
\robustify{\redbs}
\robustify{\blueb}
\robustify{\redb}

\newcommand{\red}[1]{{\color{red} #1}}

\usepackage{hyperref}%Added for clickable links
\hypersetup{
	colorlinks   = true,    % Colours links instead of ugly boxes
	urlcolor     = lightblue,    % Colour for external hyperlinks
	linkcolor    = lightblue,    % Colour of internal links
	citecolor    = lightblue      % Colour of citations
}

\newcommand{\new}[1]{{\color{blue} #1}} % Regex replace all instances of new: \\new\{([^{}]+)\}      $1
\newcommand{\Oliver}[1]{{\color{red!80!black} [Oliver]: #1}}
\newcommand{\Sadegh}[1]{{\color{magenta} [Sadegh]: #1}}
\newcommand{\Birgit}[1]{{\color{green!70!black} [Birgit]: #1}}
\usepackage{tikz}

\tikzstyle{epibox} = [
fill=grayfilling,
drop shadow=grayshadow,
rectangle,
rounded corners,
inner sep=10pt,
inner ysep=10pt
]

\def\titlestring{
Bayesian Formal Synthesis of Unknown Systems via Robust Simulation Relations
}
\def\BibTeX{{\rm B\kern-.05em{\sc i\kern-.025em b}\kern-.08em
		T\kern-.1667em\lower.7ex\hbox{E}\kern-.125emX}}
\begin{document}
\title{\titlestring}
%\author{First A. Author, \IEEEmembership{Fellow, IEEE}, Second B. Author, and Third C. Author, Jr., \IEEEmembership{Member, IEEE}
\author{Oliver Sch\"{o}n,
%\IEEEmembership{Student Member, IEEE},
Birgit van Huijgevoort,
Sofie Haesaert,
%\IEEEmembership{Member, IEEE},
and Sadegh Soudjani
%\IEEEmembership{Member, IEEE}
%\red{(Add IEEE membership info)}%
%\author{Oliver Sch\"{o}n, \IEEEmembership{Student Member, IEEE}, Birgit van Huijgevoort, \IEEEmembership{Student Member, IEEE}, Sofie Haesaert, \IEEEmembership{Member, IEEE}, and Sadegh Soudjani, \IEEEmembership{Member, IEEE}
\thanks{
% Paper submitted for review on XX/XX/XXXX. 
This work is supported by
the UK EPSRC New Investigator Award CodeCPS (EP/V043676/1),
the NWO Veni project CODEC (18244), EIC project SymAware (101070802), and ERC project Auto-CyPheR (101089047).
}% <-this % stops a space
\thanks{Oliver Sch\"{o}n is with the School of Computing, Newcastle University, Newcastle, NE4 5TG, United Kingdom
        (o.schoen2@ncl.ac.uk)}%
\thanks{Birgit van Huijgevoort and Sofie Haesaert are with the Electrical Engineering Department, TU Eindhoven, The Netherlands
        (bhuijgevoort@mpi-sws.org, s.haesaert@tue.nl)}%
\thanks{Sadegh Soudjani is with the Max Planck Institute for Software Systems, Kaiserslautern, D-67663, Germany,
        (sadegh@mpi-sws.org)}%
}

%\pagestyle{plain}
%\onecolumn

\maketitle
%%%%%%%%%%%%%%%%%%%%%%%%%%%%%%%%%%%%%%%%%%%%%%%%%%%%%%%%%%%%%%%%%%%%%%%%%%%%%%%%
\begin{abstract}
%When designing a controller for a system, we often have limited information about the system dynamics.
This paper addresses the problem of data-driven computation of controllers that are correct by design for safety-critical systems and can provably satisfy (complex) functional requirements.
With a focus on continuous-space stochastic systems with parametric uncertainty, 
%\Birgit{Alternative: continuous-state stochastic systems with parameteric uncertainty}
%\Sofie{Make it clear that this is for stochastic systems with continuous states. Done.}
we propose a two-stage approach that decomposes the problem into a learning stage and a robust formal controller synthesis stage. The first stage utilizes available Bayesian regression results to compute robust credible sets for the true parameters of the system. For the second stage, we introduce methods for systems subject to both stochastic and parametric uncertainties.
We provide simulation relations for enabling correct-by-design control refinement that are founded on coupling uncertainties of stochastic systems via sub-probability measures. The presented relations are essential for constructing abstract models that are related to not only one model but to a set of parameterized models.
%\Sofie{Don't overdo the novel, new etc. Done.}
%\red{We present theoretical results for specifying the required data size with respect to a given confidence, and for establishing this new class of relations and the associated closeness guarantees for nonlinear systems with additive Gaussian uncertainty.} \Sofie{I would skip this part in the introduction. Okay, it is to detailed. Done.}
The results are demonstrated on three case studies, including a nonlinear and a high-dimensional system.
% The results are demonstrated on a linear model and the nonlinear model of the Van der Pol Oscillator.
% \Sadegh{more case studies?}
%\Sofie{I think the abstract is already quite detailed, I would not make it too precise here. Done.}
%In this paper, we develop a new method that enables symbolic correct-by-control design via abstraction and refinement for models with \emph{epistemic uncertainty}.
%In this work, we aim to address the lack of full knowledge of the probability distributions and state transitions, which is called \emph{epistemic uncertainty}.
%\Sadegh{revise the abstract.}
%\Oliver{Currently, the abstract doesn't mention data at all.}
\end{abstract}

%%%%%%%%%%%%%%%%%%%%%%%%%%%%%%%%%%%%%%%%%%%%%%%%%%%%%%%%%%%%%%%%%%%%%%%%%%%%%%%%
\section{Introduction}
\IEEEPARstart{T}{he} rapid adaptation of AI and learning-based methods has changed the face of modern technology.
%\IEEEPARstart{A}{utonomous}
Autonomous cars, smart grids, robotic systems, and medical devices are just a few
%fragment of a plethora of
examples of
engineered systems powered by this technology. %that leverage those techniques.
Most of these systems operate in safety-critical environments, with operational scenarios being uncertain.
Despite the undisputed impact of data-driven methods, their premature adaptation can lead to %has led to a number of
severe incidents \cite{Axelrod2013CPSRisks}.
Ensuring safe operation, or more generally, designing safety-critical systems that behave in some desired manner even if the environment is uncertain, entails synthesizing robust controllers
such that the controlled system exhibits the desired behavior
with the satisfaction being formally verifiable.
%\Sofie{[Check grammar] Done.}
%and robust w.r.t. all possible operational scenarios.
Therefore, there is an ever-growing demand for so-called \emph{correct-by-design} approaches, giving formal guarantees on the absence of any undesired behavior of the controlled system.

However, designing control software with robust satisfaction guarantees proves to be very challenging.
Most safety-critical systems are large in scale, operate in an uncertain environment (i.e., their state evolution is subject to uncertainty), and comprise both continuous and discrete state variables.
%In general, the synthesis of controllers for systems on continuous and hybrid spaces does not grant analytical or closed-form solutions even when an exact model of the system is known.
Designing controllers for such systems does not grant analytical or closed-form solutions even when an exact model of the system is known.
%\pdfmargincomment{For this specification type specifically? There exists robust MPC.}
The survey paper \cite{lavaei2021automated} provides an overview of the current state of the art in formal controller synthesis for stochastic systems.
%this line of research.
%In the past two decades, formal controller synthesis for stochastic systems has witnessed a growing interest.
%
Abstraction methods represent a promising solution and enable formal control synthesis w.r.t. high-level requirements \cite{belta2017formal,tabuada09}.
Existing approaches, however, as well as most available results, are limited to small-scale systems and require prior knowledge of the exact stochastic model of the system. These two major shortcomings still prevent the implementation of correct-by-design controllers into real-world systems.

%The abstract models are built using model order reduction and space discretizations and are better suited for formal verification and control synthesis due to their finite space being amenable to exact, efficient, symbolic computational methods \cite{BK08,majumdar2020symbolic,SA13}.
%Controllers designed on these finite-state abstractions can be refined to the respective original models by leveraging (approximate) similarity relations and control refinements \cite{haesaert2020robust}.
In the pursuit of improving the scalability of formal synthesis approaches, abstraction-based techniques such as model order reduction, adaptive, and compositional methods are exploited \cite{lavaei2021automated}.
%\Sofie{have been exploited...? By whom? Otherwise: Have been developed. }Added survey reference
%Furthermore, latter have proven to be useful for learning a model
All of these necessitate formally quantifying the similarity between an \emph{abstract} model and its latent true counterpart.
%\Sofie{Is the word latent use correctly in this context? }\Oliver{The meaning is "existing in hidden or dormant form", so yes.}
%\Sofie{These two sentences dont connect well.}Done
One means of relating the systems is using \emph{simulation relations} \cite{BK08}. Whilst there exist different definitions \cite{lavaei2021automated}, in essence, simulation relations allow for quantifying the behavioral similarity of two systems and refining controllers designed on the (finite-state) abstractions to the respective original model whilst transferring any guarantees obtained on the abstract model to the original system.
%\red{With the introduction of SySCoRe, synthesizing simulation-relation-based robust controllers for stochastic systems has been made accessible \cite{syscore}.}

Requiring knowledge of the exact stochastic model of the system implies that any guarantees on the correctness of the closed-loop system only hold for that specific model.
Unfortunately, obtaining an exact model of the system of interest is either not possible or expensive and time-consuming. The type of uncertainty arising from incomplete knowledge of the system is called \emph{epistemic uncertainty}.
Data-driven identification methods for learning stochastic systems are well studied \cite{keesman2011system,van1995identification}. These system identification approaches try to find the best parameters that minimize an appropriate distance (either in time or frequency domain) between the output data and the output trajectories of the identified model. For control synthesis of stochastic systems, there are limited results available to relate the satisfaction of temporal requirements by the identified model to that of the original unknown model w.r.t. the size of the dataset (cf. the related work section).
%In comparison to our approach, most of the available work is restricted to linear or affine systems, finite-horizon specifications, or bounded noise, as exposed in the related work in Sec.~\ref{sec:relatedwork}.
%\Oliver{We give results that are not directly based on number samples (which can be good) but on the model that is obtained using the data (which will naturally be a better estimated model the more data is available).}

The problem considered in this paper is as follows.
%\pdfmargincomment[color=blue]{Reorganized s.t. first problem formulation and then revealing our solution.}
\begin{prob}\label{prob:prob1}
	Design a controller using data from the unknown true system
	such that the controlled system satisfies a given temporal logic specification with at least probability $p$ and confidence $(1-\alpha)$.% if the unknown true model belongs to a given set of models?
\end{prob}
%\pdfmargincomment{Can we combine the two levels of probability?}
The main contribution of this paper is to provide an abstraction-based scheme for answering this question for the class of parameterized discrete-time stochastic systems and the class of syntactically co-safe linear temporal logic (scLTL) specifications \cite{he2015towards}.
The first stage of our scheme is to utilize available system identification techniques to
learn the parameter set that contains the true parameters with a given confidence level from a finite amount of data.
%find a set of parameters such that this set contains the true parameters with the given confidence. \Sofie{This is repetitive.}
In the second stage, we
provide closeness guarantees over the whole set of associated models
by defining a simulation relation between a set of models and an abstract model, which is founded on coupling uncertainties in stochastic systems via sub-probability measures. 
As one of the main contributions of this paper, this type of simulation relation allows us to use a control refinement that is independent of the uncertain parameters.
%\Birgit{This is a little bit dangerous, since the control refinement in this paper depends on the uncertain parameters.}
We provide theoretical results for establishing this new relation and the associated closeness guarantees for nonlinear parametric systems with additive Gaussian uncertainty.
This allows us to design a controller alongside quantified guarantees such that the controlled system satisfies a given probabilistic temporal specification uniformly on this parameter set.
Whilst being generally compatible with any robust estimation method providing credible parameter sets \cite{Jaynes1976ConfvsCredible}, we demonstrate our approach using Bayesian linear regression \cite{Bishop2006ML}.

This paper extends further upon the results presented in the conference paper \cite{Schoen2022ParamUncert}
%A limited subset of the results has been presented in \cite{Schoen2022ParamUncert}.
%\Sofie{Rephrase. Remove the "CDC'22 conference" This paper extends further upon the preliminary results presented in ..[] Done.}
%\pdfmargincomment{Birgit: At the moment it is not possible to find this paper. Make sure that this paper is published before submitting the journal version or change it to the arXiv version for now! Oliver: Good point, thanks! Done.}
in the following main directions:
(a) We give a complete theoretical basis for stochastic sub-simulation relations together with the proofs of statements;
(b) We integrate parameter learning techniques with robust abstraction-based methods;
(c) We demonstrate the trade-off between the size of the dataset and the conservatism introduced by the robust treatment of the parametric uncertainty; and
(d) The implementations are extended to include data-driven estimation of parameter sets w.r.t. a given confidence.
%\Sadegh{Additional case studies?}
%Proof of statements are provided.  \Sofie{What do you mean with proof of statements? Furthermore, the paper contains previously omitted proofs and details. Done.}
%\Sadegh{Can we integrate the implementation as part of SySCoRe?} \Oliver{Personally, I don't think this is worth the effort right now but there are several changes I made to the code that will make it into SySCoRe v2.}
%\begin{enumerate}
%	\item We integrate for the first time system identification techniques with robust abstraction-based methods;
%	\item Implementation results are extended to include
%	% \item Combining our proposed notion of sub-simulation relations with robust data-driven parameter estimation;
%%	 to improve scalability and mitigate conservatism;
%%	\item using recent compositionality results to allow exploiting knowledge of the structure of the targeted system; and
%	\item an illustrative examination of sub-probability couplings and information on possible extensions;
%	\item providing the proofs of the theorems.
%\end{enumerate}
%The extensions improve the scalability and introduced conservatism of the proposed approach. Indeed,
%
%To showcase its performance, an \red{extended} collection of case studies is given.

%%%%%%%%%%%%%%%%%%%%%%%%%%%%%%%%%%%%
\subsection{Paper structure}%\medskip %%%%%%%%%%%%%%%%%%%%%%%%%%%%%%%
%%%%%%%%%%%%%%%%%%%%%%%%%%%%%%%%%%%%

% Build up of the paper
The rest of the paper is organized as follows.
After reviewing related work, we introduce in Sec.~\ref{sec:prelem} the necessary notions to deal with stochasticity and parametric uncertainty. We also give the class of models, the class of specifications, and the problem statement.
The system identification 
%\Birgit{parameter estimation is more in line with the section title}\Oliver{Changed section title}
framework is presented in Sec.~\ref{sec:BLR}.
In Sec.~\ref{sec:framework}, we introduce our new notion of sub-simulation relations and control refinement that is based on partial coupling. We also show how to design a controller and use this new relation to give lower bounds on the satisfaction probability of the specification.
In Sec.~\ref{sec:establish_relation}, we utilize the introduced system identification method
%parameter estimation methods 
to obtain credible sets and establish the relation between parametric nonlinear models and their simplified abstract models. 
%\Birgit{Consider revising description of Sec. V. I believe the main focus in this section is on nonlinear models, which is not what I conclude from the description here.}
Finally, we demonstrate the application of the proposed approach on a linear, high-dimensional, and nonlinear system in Sec.~\ref{sec:case_study}. 
%\Birgit{Shorter: .... the proposed approach on a linear and nonlinear system..... Does it matter that it is a Van der Pol oscillator?}
%before concluding the paper in Sec.~\ref{sec:concl}.
We conclude the paper with a discussion in Sec.~\ref{sec:discussAndExtend}.

%%%%%%%%%%%%%%%%%%%%%%%%%%%%%%%%%%%%
\subsection{Related work}\label{sec:relatedwork}
%\medskip %%%%%%%%%%%%%%%%%%%%%%%%%%%%%%%
%%%%%%%%%%%%%%%%%%%%%%%%%%%%%%%%%%%%
%\noindent\textbf{Related work:}
%{\emph{Abstraction-based approaches:}}
%\textcolor{gray}{\emph{Parametric deterministic systems:}}
% \Birgit{Ask Sofie or Sadegh to carefully check this section. I think it is too long and detailed at the moment. Perhaps we can make the reviewers happy with less (complicated) text.}
% \noindent\emph{Deterministic systems:}
Data-driven formal approaches
%There are several approaches
for systems with no stochastic state transitions are studied recently in \cite{haesaert2017data,makdesi2021efficient,kazemi2022datadriven,fan2020statistical,pedrielli2023part}.
%
% \new{Notably, \cite{fan2020statistical} presents a verification approach for deterministic systems subject to epistemic uncertainty satisfying signal temporal logic specifications that is based on conformal inference and Gaussian processes (GPs). 
%Conformal methods allow us to quantify prediction intervals of regression models without placing assumptions on the underlying parameter distributions \cite{romano2019conformalized}.
%Extending our approach using conformal inference and GP regression is an interesting direction for further research. Furthermore, state-dependent confidence parameters can be obtained using quantile regression \cite{romano2019conformalized} or locally-weighted conformal inference \cite{lei2018distribution}.
% One major issue with conformal methods is that available results can generally only provide rigorous guarantees for infinitely many samples \cite{romano2019conformalized}. In contrast, we obtain finite-sample guarantees using Bayesian linear regression.
% \Birgit{Shorten this new part. In my opinion the part Conforming methods ....  conformal inference [], can be removed.}
% }
%
% \noindent\emph{Abstraction-based approaches:}
%\textcolor{gray}{\emph{Uncertain stochastic systems:}}
%\textcolor{gray}{\emph{Two-player games finite state:}}
For stochastic systems, one approach for dealing with epistemic uncertainty is to model it as a stochastic two-player game, where the objective of the first player is to create the best performance considering the worst-case epistemic uncertainty.
%\pdfmargincomment{How is that different to what we are doing?}
%The literature on solving stochastic two-player games is relatively mature for finite state systems \cite{chatterjee2016perfect,chatterjee2012survey}.
%
%\textcolor{gray}{\emph{Two-player games continuous state:}}
%There is a limited body of papers addressing this problem for continuous-state systems. The papers \cite{majumdar2021symbolic,majumdar2020symbolic} look at satisfying temporal logic specifications on nonlinear systems utilizing mu-calculus and space discretization. The results rely on direct access to the full state and are hence incompatible with model-order reduction techniques.
%
%\new{Conversely, the authors in \cite{kazemi2022datadriven} compute a growth bound of the unknown system and design controllers founded on estimating a bound on the Lipschitz constant of the system.}
%
%\textcolor{gray}{\emph{Interval MDPs:}}
The work \cite{Badings2022epistemicUncert} addresses epistemic uncertainty by abstracting the system to an interval Markov decision process (MDP) and is limited to finite horizons, reach-avoid specifications, and linear systems.
%\textcolor{gray}{\emph{Scenario optimization:}}
%\Oliver{Many approaches limited to bounded stochastic noise!}
%Similarly, \cite{Lavaei2022DDMDP} extends upon this by constructing an interval MDP for uncertain nonlinear systems by solving a scenario optimization problem. %This is helpful when only data of the system is available or the dynamics are very complex.
The paper \cite{jagtap2020control} uses Gaussian processes (GPs) to learn a model based on data, which is then used for designing barrier certificates while being restricted to safety specifications. The authors of the papers \cite{Jackson2020safety,jiang2022safe} use Gaussian processes (GPs) to learn a model of the system based on data and then use finite abstractions to satisfy logical properties. This is extended with deep kernel learning in \cite{reed2023promises}, where the GP kernel is augmented with a neural network preprocessing its inherent feature map.
All these approaches require appropriate information on the function class and complexity of the dynamics.
Controllers for systems captured as stochastic neural network dynamic models (NNDMs) are designed in \cite{adams2022nndm} by translating the NNDMs to interval MDPs.
The starting point, however, is the trained NNDM and no guarantees on the correctness of the learned model w.r.t. to the data-generating system are considered.

Capturing epistemic uncertainty using interval MDPs relies on the assumption that the epistemic uncertainty is state-wise independent, leading to inconsistent and overly conservative results.
In contrast, we relate epistemic uncertainty explicitly using parametric MDPs.
%Furthermore, if the noise is unbounded, implementations based on interval MDPs generate trivial results for infinite horizon specifications such as safety. 
%\Birgit{This is a crucial point to make! Most approaches do not work for stochastic systems where the noise is additive and unbounded, \emph{especially} when the specifications have a finite horizon. Perhaps we can try to categorize the different techniques and shorten by saying: we have deterministic approaches [],[],[]. Approaches that work based on ...., but they are limited to ....}
%
%\textcolor{gray}{\emph{Model-free:}}
%\textcolor{gray}{\emph{Reinforcement learning:}}
In an effort of reducing the number of required assumptions, the paper \cite{thorpe2022cmechance} uses conditional mean embeddings (CMEs) to embed the conditional distribution of a non-Markovian random trajectory into an reproducing kernel Hilbert space. 
%	Whilst this direction seems generally promising, 
%	the authors do not provide any rigorous end-to-end guarantees for their results and obtain their approximate chance constrained optimization problem by using an empirical CME and disregarding its statistical correctness. 
However, the authors obtain their approximate chance-constrained optimization by using an empirical CME without providing guarantees of its statistical correctness.
%	\Birgit{Too harsh? Reformulate}
The paper \cite{romao2023distributionally} provides a similar distributionally-robust scheme, leaving the necessary theoretical results for quantifying the size of the required ambiguity set for future work.

% \noindent\emph{Abstraction-free approaches:} %{\emph{Abstraction-free approaches:}}
There exists a large body of abstraction-free approaches via control barrier certificates (CBCs) \cite{prajna2007BC}.
Latest work on learning CBCs from data is limited to control-affine systems \cite{Lopez2022UnmatchCBC,Cohen2022RCBF} or requires exponentially large data sets and knowledge of bounds on the regularity of the system \cite{salamati2024data}.
% Latest work in the area of abstraction-free approaches develop control barrier certificates (CBCs) that are based on constructing a set of feasible models \cite{Cohen2022RCBF}.
% Similarly, adaptive control for continuous-time control-affine parametric systems using so-called unmatched control barrier functions is studied in \cite{Lopez2022UnmatchCBC}.
% \new{In \cite{salamati2024data}, the authors find CBCs using robust scenario optimization. Whilst this necessitates them to sample transitions between every pair of partitions multiple times, incurring a high sample burden, they also require knowledge of the Lipschitz constants of several intermediate functions.}
%Analogous to our approach, the authors generate a family of control barrier functions and use online parameter adaptation to select a safety controller from the obtained set.
Whilst CBCs are potentially more scalable than abstraction-based solutions, finding a valid CBC for systems with non-control-affine and non-polynomial dynamics is generally hard.
%Even more, CBCs for specifications beyond simple reachability yields, e.g., sequential reachability problems \cite{anand2021CBComegaregular}, greatly impeding their applicability.
%\Oliver{Add \cite{Ameneh2023DataDriven}?}
%Further work on model-free reinforcement learning studies synthesizing robust controllers for temporal logic specifications without constructing a model of the system \cite{kazemi2020fullLTL,lavaei2020RL}.
%Moreover, the work \cite{Cohen2022RCBF} is limited to polyhedral over-approximations of the credible set that introduces another degree of conservatism.
%Similarly, adaptive control for continuous-time control-affine parametric systems using the so-called unmatched control barrier functions is studied in \cite{Lopez2022UnmatchCBC}. Analogous to our approach, the authors generate a family of control barrier functions and use online parameter adaptation to select a safety controller from the obtained set. However, the work is limited to forward invariance specifications.
%\color{black}
An approach based on computing reachable sets leverages random set theory to obtain infinite-sample guarantees \cite{lew2021sampling}. 
%	Analyzing convergence rates is left for future work. \Birgit{Same for us, right? If yes, remove this sentence. }
Similarly, further works construct reachable sets efficiently using knowledge of a growth bound \cite{ajeleye2023data}, and GPs under the assumption of known bounds on the unknown dynamics components \cite{cao2022efficient}, respectively.
% There exist a body of work on verifying models based on neural networks \cite{harapanahalli2023forward, jafarpour2023interval}. However, similar as mentioned before, no guarantees of the correctness of the data-driven model w.r.t. the data-generating system are provided.
% \noindent\emph{Other related work:} %\Birgit{alt: Other related work:}
% \new{There is a large body on model predictive control (MPC) based approaches trying to address the challenges arising from uncertain systems \cite{dyro2021particleMPC, sinha2022adaptive}. For example, in \cite{dyro2021particleMPC}, the authors propose a particle MPC solves using sequential convex programming. Note that these approaches can generally not give any guarantees for the robustness of the resulting controller.
% As one example of a wealth of work on planning, probabilistically safe planning under uncertain human interactions via a real-time Bayesian framework is studied in \cite{fisac2018probabilistically}.}
There is also a growing body of work on utilizing polynomial chaos expansion to propagate stochastic distributions through dynamics \cite{pan2023polychaos, kim2013polychaos, seshadribayesian}, which is based on representing random variables as a function of a finite-dimensional random vector using an orthogonal polynomial basis.
%We use a similar parametrization of the models without requiring the employed basis functions to be orthogonal.}
% \red{I don't get this. Are you sure they are limited to linear systems? They have orthogonal basis functions.}

Our approach utilizes the techniques from the system identification literature to compute parameter sets w.r.t. a given confidence for stochastic systems. We then build a theoretical foundation on the concepts presented in \cite{haesaert2017verification,haesaert2020robust} to design robust controllers using abstraction methods that are compatible with both model-order reduction and space discretization.

\section{Preliminaries and Problem Statement}
\label{sec:prelem}
\subsection{Preliminaries}
%\red{reorder the stuff here to make the best flow based on what we have in the paper}

The following notions are used.
%We indicate the spectral norm of a matrix $A$ as $||A||$.
The transpose of a matrix $A$ is indicated by $A\T$.
We write $I_n$ to denote the identity matrix of dimension $n\times n$. 
%\Birgit{Why not write $I_n$ instead?}
%
A measurable space is a pair $(\X,\mathcal{ F})$ with sample space $\X$ and $\sigma$-algebra $\mathcal{F}$ defined over $\X$,
 which is equipped with a topology.
 In this work, we restrict our attention to Polish sample spaces~\cite{bogachev2007measure}.
% \Sofie{Which space will be Polish?}\Oliver{$\X$ (and therefore later-on also $\U$)?!}\Oliver{I put this here to say that all sample spaces should be Polish}
 As a specific instance of $\mathcal F$, consider
%In this work, we only consider
 Borel measurable spaces, i.e., $(\X,\mathcal{B}(\X))$, where $\mathcal{B}(\X)$ is the Borel $\sigma$-algebra on $\X$, that is the smallest $\sigma$-algebra containing open subsets of $\X$.
% to have well-defined measurable events over unbounded trajectories of systems.
A positive \emph{measure} $\meas$ on $(\X,\mathcal{B}(\X))$ is a non-negative map
$\nu:\mathcal{B}(\X)\rightarrow \mathbb R_{\ge 0}$ such that for all countable collections $\{A_i\}_{i=1}^\infty$ of pairwise disjoint sets in $\mathcal{B}(\X)$
it holds that
%every two disjoint events $A,B\in \mathcal{F}$: $\p{A\cup B}=\p{A}+\p{B}$.
$\nu({\bigcup_i A_i })=\sum _i \nu({A_i})$.
A positive measure $\nu$ is called a \emph{probability measure} if $\nu(\X)=1$, and is called a \emph{sub-probability measure} if $\nu(\X)\leq 1$.
%A \emph{{(sub-)}probability measure} $\p{\cdot}$ is a measure $\po:\mathcal{B}(\X)\rightarrow [0,1]$ for which $\p{\X}=1$ (respectively, $\p{\X}\leq 1$).
%\red{The \emph{support} of a sub-probability measure $\po$ is the set of all points $x\in\X$ for which every open neighborhood of $x$ has a positive measure. This is denoted as $\support(\po)\subset \X$.}\pdfmargincomment[color=red]{Not used anymore in the paper. Consider removing}
%\Oliver{@Sadegh: Please specify that "$\subset$" is is not meant as a strict subset.}

A probability measure $\po$ together with the measurable space $(\X,\mathcal{B}(\X))$ define a \emph{probability space} denoted by $(\X,\mathcal{B}(\X),\po)$ and has realizations  $x\sim \po$.
%\pdfmargincomment{To stay uniform with the rest of the paper, we could use $p_x(\cdot)$ instead of $p$.}
We denote the set of all probability measures for a given measurable space $(\X,\mathcal{B}(\X))$ as $\mathcal P (\X)$.
For two measurable spaces $(\X,\mathcal{B}(\X))$ and $(\Y,\mathcal{B}(\Y))$, a \emph{kernel} is a mapping $\pok: \X \times \mathcal B(\Y)\rightarrow \mathbb R_{\geq 0}$
%\pdfmargincomment[color=red]{Why whole R?}
such that $\pk{x,\cdotx}:\mathcal B(\Y)\rightarrow\mathbb{R}_{\geq 0}$ is a measure for all $x\in\X$, and $\pk{\cdotx, B}: \X\rightarrow \mathbb R_{\geq 0}$ is measurable for all  $B\in\mathcal B(\Y)$.
%\pdfmargincomment{Notation of kernel once with and once without comma?!}
A kernel associates to each point $x\in\X$ a measure denoted by $\pk{\cdotx|x}$.
We refer to $\pok$ as a (sub-)probability kernel if in addition $\pk{\cdotx|x}:\mathcal B(\Y)\rightarrow [0,1]$ is a (sub-)probability measure.
The \emph{Dirac delta} measure $\delta_a:\mathcal{B}(\X)\rightarrow [0,1]$ concentrated at a point $a\in\X$ is defined as $\delta_a(A)=1$ if $a\in A$ and $\delta_a(A)=0$ otherwise, for any measurable $A\in \borel{\X}$.
%\begin{equation*}
%	\delta_a(A) := \begin{cases}
%		1 \text{, if $a\in A$}\\
%		0  \text{, otherwise,}
%	\end{cases}
%\end{equation*}
%for all measurable $A$.
The normal stochastic kernel with mean $\mu\in\mathbb{R}^n$ and covariance matrix $\Sigma\in\mathbb{R}^{n\times n}$ is defined by
\begin{equation*}
	\mathcal N(dx|\mu, \Sigma) \!:=\!\frac{dx}{ \sqrt{(2\pi)^n \!\left|\Sigma\right| }}\exp\!\left[\!-\frac{1}{2}(x\!-\!\mu)\T\Sigma^{-1}\!(x\!-\!\mu)\!\right],
\end{equation*}
with $|\Sigma|$ being the determinant of $\Sigma$.

For given sets $A$ and $B$, a relation $\rel\subset A\times B$ is a subset of the Cartesian product $A\times B$. The relation $\rel$ relates $x\in A$ with $y\in B$ if $(x,y)\in\rel$, written equivalently as $x\rel y$.
%\new{If $x\rel y$, we denote $x\in\R(y)$ and $y\in\R^{-1}(x)$.} \Sadegh{It should be the other way around. This is not good at all. We may have $x\rel y_1$ and $x\rel y_2$ at the same time. How come that you define $y=\R(x)$? This is a relation not a function.}
%We use the following notation for the mappings $\rel(\tilde A):=\{y: x\rel y,\  x\in \tilde A\}$ and $\rel^{-1}(\tilde B):=\{x: x\rel y,\  y\in \tilde B\}$  for $\tilde A\subset A$ and $\tilde B\subset B$.
For a given set $\Y$, a metric or distance function $\mathbf d_\Y$ is a function $\mathbf{d}_\Y: \Y\times \Y\rightarrow \mathbb R_{\ge 0}$
satisfying the following conditions for all $y_1,y_2,y_3\in\Y$:
$\mathbf d_\Y(y_1,y_2)=0$ iff $y_1=y_2$;
$\mathbf d_\Y(y_1,y_2)=\mathbf d_{\Y}(y_2,y_1)$;  and
$\mathbf d_\Y(y_1,y_3)\leq \mathbf d_\Y(y_1,y_2) +\mathbf d_\Y(y_2,y_3)$.
The indicator function $\1_A(x)$ for a set $A$ is defined as  $\1_A(x)=1$ if $x\in A$ and $\1_A(x)=0$ otherwise.
%We denote the determinant of a matrix $A$ as $\left|A\right|$ and the identity matrix as $I$.

%\section{Problem statement: control synthesis for uncertain stochastic systems}
%\label{sec:prob}
%\red{This section seems to be fine.}

\subsection{Discrete-time uncertain stochastic systems and control policies}
%\Oliver{Do we want to remove this section or push it before the data-driven sysId part? And if we push it, do we want to remove the parametrization?}
%\Oliver{Do we want to start with the data-driven part here and also put a sketch of the credible set here? Ten we have the setting and can unravel all the theory as before whilst knowing that the first step is system identification.}

%\Sadegh{re-order this subsection properly.}
%\Oliver{@Sadegh: Connect and elaborate on the connection/notation of the models in this section and the beginning of section 3. Maybe use $\M_{\mathfrak n}$.}
We consider discrete-time nonlinear systems perturbed by additive stochastic noise under model-parametric uncertainty.
% This modeling formalism is essential if we can only access an uncertain model of this stochastic system.\Sofie{I dont follow this reasoning.}\Oliver{Rephrasing the problem for an uncertain model to a model with parametric uncertainty allows us to tackle it using existing techniques.}Removed.
%Consider the set of models $\{\M(\theta)$ with $\theta\in\Theta\}$, parametrized with $\theta$:
Consider the model $\M(\theta)$ parametrized with $\theta$:
\begin{equation}
	\label{eq:model}
	\M(\theta): \left\{ \begin{array}{ll}
		x_{k+1}&= f(x_k,u_k;\theta) + w_k,\\
		y_k& = h(x_k\red{}),%\pdfmargincomment[color=red]{We do not consider such systems!}
	\end{array} \right.
\end{equation}
where the system state, input, and observation at the $k^{\text{th}}$ time step are denoted by $x_k\in\X, \,u_k\in\U$, and $y_k\in\Y$, respectively. The functions $f$ and $h$ specify, respectively, the parameterized state evolution of the system and the observation map. The additive noise is denoted by $w_k\in\mathbb{W}$, which is an independent, identically distributed (i.i.d.) noise sequence with a (potentially unbounded) distribution $w_k\sim p_w(\cdotx)$.
%\Oliver{Mention assumptions such as Markovian dynamics, full-state observability, Gaussian noise here but extension (also utilizing structure) \cite{Schon2023GMM}, etc?}
%We assume that the parameter set $\Theta$ is a bounded polytope.
%\new{We impose the following assumption:
	%\begin{assumption}\label{asm:uncertparam}
	%	The uncertain parameter $\theta$ belongs to a known, bounded polytope $\Theta
	%	\pdfmargincomment{should we add that Theta is a subset of R\^{}p for example? That emphasizes that it is more than one scalar parameter}
	%	\subset\mathbb{R}^{\mathfrak{p}}
	%	$.
	%\end{assumption}
	%}
%We assume throughout this paper that the uncertain parameter $\theta$ belongs to a known, bounded set\pdfmargincomment[color=blue]{Before it was polytope.} $\Theta$%\subset\mathbb{R}^{\mathfrak{p}}$ for some $\mathfrak{p}$.
%\pdfmargincomment{Birgit: I think it is better to remove this statement or rephrase it. We are going to compute a bounded set, such that with a certain confidence we have $\theta \in \Theta$. So this assumption does not make sense anymore.}

%\red{It doesnt seem necessary to cite [19] for this. Requiring $\Theta$ to be a bounded set is quite common. Also the requirement that it is closed it quite unnecessary, as you could always just work with the closure of the set. I think the original version was quite fine:"We assume that the parameter set $\Theta$ is a known, bounded polytope."}\Oliver{Done.}

%We denote the class of all MDPs with the same metric output space $\Y$ as $\mathcal{M}_\Y$.
We indicate the input sequence of a model $\M$ by $\acs:= \ac{0},\ac{1},\ac{2},\ldots$
%$u_{\bullet}:\mathbb N\rightarrow\A$ is given, where $\mathbb N:=\{0,1,2,\ldots\}$ is the set of natural numbers.
and we define its (finite) \emph{executions} as sequences of states $\xs=\x{0},\x{1},\x{2},\ldots$ (respectively, $\xs_N=\x{0},\x{1},\x{2},\ldots, \x{N}$) initialized with the initial state  $\xin$ of $\M$ at $k=0$.
In each execution, the consecutive state $x_{k+1}\in\X$
is obtained as a realization $x_{k+1}= f(x_k,u_k;\theta) + w_k$, $w_k\sim p_w(\cdotx)$.
% of the controlled Borel-measurable stochastic kernel. % $\Tr\left(\cdot\mid x_t, u_t \right)$.
%Thus, an execution of $\M$ is a time sequence of these states $x_{\bullet}:\mathbb N\rightarrow\X$. % is referred to as a path or execution.
%Note that for a parametrized MDP $\M(\theta)$ its transition kernel also depends on $\theta$ denoted as $\Tr\left(\cdot| x_k, u_k; \theta \right)$.

The execution history $(x_0, u_0, x_1,\ldots, u_{N-1}, x_N)$ grows with the number of observations $N$ and takes values in the \emph{history space} $\Hist_N := (\X \times \A )^{N} \times \X$.
A control policy or controller for $\M(\theta)$ is a sequence of policies mapping the current execution history to a control input.
\begin{definition}[Control policy]
	\label{def:markovpolicy}
	A control policy $\boldsymbol{\eta}$ is a sequence $\boldsymbol{\eta}=(\eta_0,\eta_1,\eta_2,\ldots)$ of universally measurable maps $\eta_k:\Hist_k\rightarrow \mathcal P(\A)$, $k\in\mathbb N:=\{0,1,2,\ldots\}$, from the execution history to a distribution on the input space.
\end{definition}
As special types of control policies, we differentiate Markov policies and finite memory policies.
%\begin{definition}[Markov policy]
%	\label{def:markovpolicy}
A \emph{Markov policy} $\boldsymbol{\pi}$ is a sequence $\boldsymbol{\pi}=(\pi_0,\pi_1,\pi_2,\ldots)$ of universally measurable maps $\pi_k:\X\rightarrow  \mathcal P(\A)$, $k\in\mathbb N$, from the state space $\X$ to a distribution on the input space.
%\end{definition}
We say that a Markov policy is \emph{stationary}, if $\boldsymbol{\pi}=(\pi,\pi,\pi,\ldots)$ for some $\pi$.
\emph{Finite memory policies} first map the finite state execution of the system to a finite set (memory). The input is then chosen similar to the Markov policy as a function of the system state and the memory state. This class of policies is needed for satisfying temporal specifications on the system executions.

In the following, a control policy for the model \eqref{eq:model} is denoted by $\Ca$ and is implemented as a state-feedback controller, i.e., $u_k=\Ca(x_k)$.
%depicted in Fig.~\ref{fig:control}. 
We denote the composition of $\Ca$ with the model $\M(\theta)$ as $ \Ca\times \M(\theta)$.

%\begin{figure}[htp]
%	\centering
%	\scalebox{1}{
%		\includegraphics{Control_fig.pdf}
%	}
%	\caption{Control design for a parameterized stochastic model. \Birgit{Figure can be removed to save space.}}
%	\label{fig:control}
%\end{figure}

%\Sofie{Should you be more specific about what $\mathbf C$ is?}\Oliver{What do you want to add? It's basically just a lookup table.}Done.

%In the next subsection, we formally define the class of specifications studied in this paper. \Birgit{Sentence can be removed to save space.}
\subsection{Temporal logic specifications}
Consider a set of atomic propositions $AP := \{ p_1, \ldots, p_L \}$ that defines an \emph{alphabet} $\alphabeth := 2^{AP}$, where any \emph{letter} $\letter\in\alphabeth$ is composed of a set of atomic propositions.  An infinite string of letters forms a \emph{word} $\word=\letter_0\letter_1\letter_2\ldots\in\alphabeth^{\mathbb{N}}$. We denote the suffix of $\word$ by $\word_i = \letter_i\letter_{i+1}\letter_{i+2}\ldots$ for any $i\in\mathbb N$.
Specifications imposed on the behavior of the system are defined as formulas composed of atomic propositions and operators. We consider the co-safe subset of linear-time temporal logic properties \cite{kupferman2001model} abbreviated as scLTL.
%
%\begin{definition}[LTL]
%  \label{def:LTL}
%  Formulas given as linear time temporal logic properties are constructed according to the syntax
%  \begin{equation*}
	%    \label{eq:LTL}
	%    \psi ::=  p \ |\ \notltl \psi \ |\ \psi_1 \vee\psi_2  \ |\ \psi_1 \andltl \psi_2 \ |\ \psi_1 \Until \psi_2 \ |\ \Next \psi,
	%  \end{equation*}
%\end{definition}
This subset of interest consists of temporal logic formulas 
for which violation can be determined from a finite bad prefix of a word. scLTL formulas can be
constructed according to the following syntax
%\red{I don't understand why you changed the definition of scLTL? why do you put negation beside any specification??????}
\begin{equation*}
	%	\label{eq:scLTL}
	\psi ::=  \True\ |\  p \ |\ \notltl p\ |\ \psi_1 \vee\psi_2  \ |\ \psi_1 \andltl \psi_2 \ |\ \psi_1 \Until \psi_2 \ |\ \Next \psi,
\end{equation*}   where $p\in \AP$ is an atomic proposition.
%\begin{definition}
The \emph{semantics} of scLTL are defined recursively over $\word_i$ as
% \begin{itemize}
	$\word_i \satisfies p$ iff $p \in \letter_i$;
	$\word_i \satisfies \psi_1 \andltl  \psi_2  $ iff $ ( \word_i \satisfies \psi_1 ) \andltl ( \word_i \satisfies \psi_2 ) $;
	$\word_i \satisfies \psi_1 \orltl  \psi_2  $ iff $ ( \word_i \satisfies \psi_1 ) \orltl ( \word_i \satisfies \psi_2 ) $;
	$\word_i \satisfies  \psi_1 \Until \psi_2 $ iff $\exists j \geq i \text{ subject to } (\word_j \satisfies \psi_2 ) $ and $\word_k \satisfies \psi_1, \forall k \in \{i, \ldots j-1\}$; and
	$\word_i \satisfies \Next \psi$ iff $\word_{i+1} \satisfies \psi$.
	The eventually operator  $\Event \psi$ is used in the sequel as a shorthand for $\True\Until  \psi $.
	We say that $\word\satisfies\psi$ iff $\word_0\satisfies\psi$.
%\Birgit{I would not at this here but in the Discussion section (before or after you mention $LTL_f$). You can say that in discrete-time STL is an extension of LTL and due to its time-valued things it is often finite horizon. Hence STL specifications can be written as DFAs and can therefore be handled by this approach. However, the number of states and transitions in the DFA can get quite large.}
%	\new{Specifications of the scLTL subset can be represented as deterministic finite automatons (DFAs), introduced in Definition~\ref{def:dfa}. 
%		This allows us to reformulate the satisfaction of an scLTL specification as a reachability problem on the product system composed of the DFA and the original system.
%		\red{Note that scLTL subsumes quantitative linear temporal logic languages such as discrete-time signal temporal logic (STL).}}
	We give more information on possible extensions in Sec.~\ref{sec:discussAndExtend}. 
	
	Consider a labeling function $\mathcal L: \Y\rightarrow \Sigma$ that assigns a letter to each output. Using this labeling map, we can define temporal logic specifications over the output of the system.
	Each output trace of the system $\mathbf y \!=\! y_0,y_1,y_2,\ldots$ can be translated to a word as $\word \!=\! \mathcal L(\mathbf y)$.
	We say that a system satisfies the specification $\psi$ with the probability of at least $p_\psi$ if
	$\mathbb P(\word\satisfies \psi) \ge p_\psi$, where $\word$ is any %a random 
		word produced by the system.
	When the labeling function $\mathcal L$ is known from the context, we write $\mathbb P(\Ca\times\M\satisfies \psi)$ to emphasize that the output traces of $\Ca\times\M$ are used for checking the satisfaction.
%	 \Oliver{This should be defined over $\Ca\times\M$!! Error in previous papers.}
	Similarly, we denote by $\P\left(\Ca\times \M(\theta) \satisfies \psi \right)$ the probability that the controlled system $\Ca\times \M(\theta)$ satisfies $\psi$.

%\color{blue!70!black}
\subsection{Problem statement}\label{sec:prob_statem}
	%Consider a parameterized model in \eqref{eq:model} with $\theta\in\Theta$.
	Consider an unknown system $\M^\ast$ for which we impose a parametrized form \eqref{eq:model}, i.e., $\M^\ast:=\M(\theta^\ast)$, where $f,g$ are known but the underlying fixed true parameter $\theta^\ast$ is unknown. 
%	Birgit{for some readers it is unclear that $f$ and $h$ are know, so you could add this.}
	We are interested in designing a controller $\Ca$ to satisfy temporal specifications $\psi$ on the output of the model, i.e, $\Ca\times \M(\theta^\ast) \satisfies \psi$.
	To infer the unknown parameter, we collect data $\mathcal{D}$ from $\M^\ast$
	by sampling a finite number of data points:
	\begin{align}
		\begin{split}
			&\mathcal{D} := \left\lbrace (x^{(i)},u^{(i)},x^{+(i)}),\quad i\in\{1,\ldots,N\}\right\rbrace ,\text{ where } \\
			&x^{+(i)} = f(x^{(i)},u^{(i)}; \theta^\ast)  + w_k^{(i)},\quad w_k^{(i)} \sim p_w(\cdotx),
		\end{split}
		\label{eq:data}
	\end{align}
	for arbitrary state-input pairs $(x^{(i)},u^{(i)})$.
	Note that we assume access to data of the full state. 
%		Relaxing this assumption is an open question.
	 
%\Birgit{Remove last sentence or put in Discussion section.}
	%
	We are interested in designing a controller that ensures the satisfaction of $\psi$ with at least probability $p_\psi$ using only data $\mathcal{D}$ from the true system.
	In particular, we require the controller not to depend on the unknown $\theta^\ast$ directly.
	This is the objective of Problem~\ref{prob:prob1}, which we decompose into the following two sub-problems.
%	Can we design a controller that does not depend on $\theta^\ast$ and that ensures the satisfaction of $\psi$ with at least probability $p_\psi$ using only data $\mathcal{D}$ from the true system? This is equivalent to Problem~\ref{prob:prob1}.
%	%\pdfmargincomment{Birgit: How is this different from Problem 1? I was surprised that the reference to problem 1 does not occur here. Can we move the reference to Problem 1 up? Can we state that we split up Problem 1 in sub-problems Problem 2 and 3? Oliver: Done.}
%	%We formalize this problem as follows:
%	To address this problem, we decompose it into the following two sub-problems:
	%and provide solutions to the following :
%	\color{blue!70!black}
	\begin{subproblem}
		\begin{prob}\label{prob:confset}
			\setlength{\belowdisplayskip}{0pt}
			Construct a set $\Theta$ from $\mathcal{D}$ that contains the true parameters $\theta^\ast$ with a given probability $1-\alpha\in(0,1)$, i.e.,
			\begin{equation*}
				\P(\theta^\ast\in\Theta)\geq 1-\alpha.
			\end{equation*}
		\end{prob}
		Data-driven parameter identification provides a solution to Problem~\ref{prob:confset}.
		\begin{remark}[Assumptions] 
			For the scope of this paper, we restrict our attention to systems linear in the unknown parameters $\theta$ and additive noise (see Eq.~\ref{eq:sysDataGenNonlin}), acknowledging that this can be relaxed by choosing any suitable alternative parameter identification method that provides finite-sample guarantees as the proposed coupling approach is generally applicable.
			Moreover, whilst we restrict ourselves to parametric uncertainty in the deterministic part of the dynamics only, and Gaussian noise, it is shown in \cite{Schon2023GMM} how these assumptions can be lifted.
		\end{remark}
		%\pdfmargincomment{Birgit: The original caption is Problem, so it is nicer to not abbreviate when referring to it. If you agree, change this in the remainder of the paper. Oliver: Leaving it for now.}
		Based on the set $\Theta$ we construct a parametrized set of models $\{\M(\theta)$ with $\theta\in\Theta\}$.
		We solve Problem~\ref{prob:prob1} by designing a controller valid for this parametrized set of models, formulated as follows.
		\begin{prob}\label{prob:ssr}
			\setlength{\belowdisplayskip}{0pt}
			For a given specification $\psi$ and a threshold $p_\psi\in [0,1]$,
%			\Birgit{Why not $[0,1]$?}
			design a controller $\mathbf C$ for $\M(\theta)$ that does not depend on the parameter $\theta$ explicitly and that
			\[ \P\left(\Ca\times \M(\theta) \satisfies \psi \right)\geq   p_\psi,\quad  \forall \theta\in\Theta. \]
		\end{prob}
	\end{subproblem}

	%\fbox{\begin{minipage}{7cm}
			%	Can we design a controller $\mathbf C$ such that $ \Ca\times \M(\theta) \satisfies \psi$ for all $\theta\in\Theta$ with probability at least $p_\psi$ ?
			%	\end{minipage}}
	%\begin{prob}\label{prob:ssr}
	%	\setlength{\belowdisplayskip}{0pt}
	%	 For a given specification $\psi$ and a threshold $p_\psi\in (0,1)$, design a controller $\mathbf C$ for $\M(\theta)$ that does not depend on the parameter $\theta$ and that
	%	 \[ \P\left(\Ca\times \M(\theta) \satisfies \psi \right)\geq   p_\psi,\quad  \forall \theta\in\Theta. \]
	%%	 test
	%\end{prob}
	%\red{[Sofie:] I removed the notion of parameterized stochastic models for [10], since we do not work with that  in [10].}
	%\begin{remark}
	The controller synthesis for stochastic models is studied in \cite{haesaert2020robust} through coupled simulation relations. Although these simulation relations can relate one abstract model to a set of parameterized models $\M(\theta)$, these relations would lead to a control refinement that is still dependent on the true model or true parameter $\theta^\ast$. Therefore, this approach is unfit to solve Problem~\ref{prob:ssr}, since $\theta^\ast$ is unknown.
    \begin{remark}[Key idea]
        As one of the main contributions of this paper, we consider a parameter-independent control refinement and compute a novel simulation relation based on a sub-probability coupling (see Sec.~\ref{sec:framework}) to synthesize a single controller for all $\theta\in\Theta$.
    \end{remark}

\section{Data-Driven System Identification via Bayesian Linear Regression}\label{sec:BLR}
In this section, we present a solution to Problem~\ref{prob:confset} for the class of nonlinear systems that are linear in the unknown parameters
$\theta\T=[\theta_1,\ldots,\theta_n]\T\in\mathbb{R}^{n\times m}$, i.e.,
%\pdfmargincomment{[Birgit:] You switched notation here. In the preliminaries you use mathbb R. I believe this R is used for simulation relation only. Besides that it is a bit surprising that you start using spaces here, instead of when you first introduce the model. But I guess that is fine, if you only need them at this point.}
\begin{equation}
	\label{eq:sysDataGenNonlin}
	\M(\theta): \left\{ \begin{array}{ll}
		x_{k+1} &= \theta\T f(x_k, u_k) + w_k,\\
		y_k &= h(x_k),%+D_vv_k,
	\end{array} \right.
\end{equation}
%\Oliver{Assume additive zero-mean Gaussian noise! $w_k\sim\mathcal{N}(0, \sigma_w^2)$.}
where $x\in\mathbb{R}^n$, the noise is i.i.d. Gaussian, $w_k\sim\mathcal{N}(\cdotx|0, \Sigma)$, with zero mean and a given covariance matrix $\Sigma\in\mathbb{R}^{n\times n}$, %$w_k\sim p_w(\cdot)$,
%\Oliver{Write normal distribution argument?!}
and the functions $h$ and $f:\X\times\U\rightarrow\mathbb{R}^{m}$ are known.
%Without loss of generality, we assume that the noise $w_k\sim \mathcal N(0,\sigma_w^2 I)$ has Gaussian distribution with a full rank covariance matrix.
%We denote by $\sigma_w^2$ and $I$, respectively, the noise variance and identity matrix.
Note that $f:=[f_1,\ldots,f_m]\T$ can be a comprehensive library of nonlinear functions, hence, many systems can be expressed in the form of \eqref{eq:sysDataGenNonlin}. %\new{This representation is also used in the works that utilize polynomial chaos expansion while assuming orthogonality of the basis functions~\cite{pan2023polychaos, kim2013polychaos,seshadribayesian}.}

\begin{figure}
	\centering
	\begin{tikzpicture}[scale=.5]
		\draw[draw=black] plot[smooth, tension=.8] coordinates {(-3.5,0.5) (-1,3) (4.5,3.5) (2,1) (1,-2) (-3,-1) (-3.5,0.5)};

		\node at (-2.5,-1) [black,right] {$\Theta$};

		\node at (0,1+.5) [gray,right] {$\hat\theta$};
		\fill[gray] (0,1) circle [radius=.1];

		\node at (1.8,2.5+.5) [cherryred,right] {$\theta^\ast$};
		\fill[cherryred] (1.8,2.5) circle [radius=.1];
	\end{tikzpicture}
	\caption{The unknown true parameters $\theta^\ast$ are approximated with an estimate $\hat\theta$ and are contained within a credible set $\Theta$ with a confidence of $(1-\alpha)$.}
	\label{fig:confSetSketch}
\end{figure}
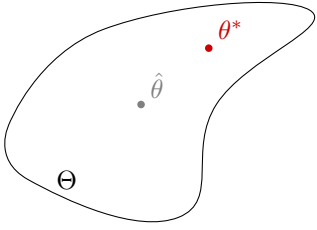

Based on state-input data $\mathcal{D}$ as in \eqref{eq:data}, \emph{Bayesian linear regression} (e.g., \cite{Bishop2006ML}, Sec.~3.3) allows us to find an \emph{estimate} $\hat\theta$ that approximates the true unknown parameters $\theta^\ast$ and to construct a \emph{credible set} $\Theta$ for $\theta^\ast$ with a given confidence $(1-\alpha)\in(0,1)$ (cf. Fig.~\ref{fig:confSetSketch}), defined as follows.
%that contains the true parameters $\theta^\ast$ with some given confidence $(1-\alpha)\in(0,1)$, i.e., $\P(\theta^\ast\in\Theta)\geq 1-\alpha$ (cf. Fig.~\ref{fig:confSetSketch}).
%Let $\hat\theta\in\Theta$.

\begin{definition}[Credible set]\label{def:credible_set}
		A set $\Theta$ is called a \emph{credible set} of $\theta^\ast$ with confidence level $(1-\alpha)$ if for the given confidence $(1-\alpha)$ we have $\P(\theta^\ast\in\Theta)\geq 1-\alpha$.
\end{definition}

\begin{remark}[Credible vs confidence sets]
	Credible sets and \emph{confidence sets} are two related notions stemming from two fundamentally different statistical paradigms.
	As a product of Bayesian statistics, a credible set
	of a random variable
	represents a set of plausible values characterizing the uncertainty in the estimation of the random variable up to a desired confidence, i.e.,
 	a fixed quantile of the posterior distribution that contains a prescribed fraction of the posterior mass.
%	represents the
%	fixed super-level set of the likelihood function of a random variable at a given confidence level. \red{modify this set.}
	A confidence set, on the other hand, is based on frequentist theory and is computed based on the frequency of random observations of a fixed true parameter.
	Bayesian credible sets treat their bounds as fixed and the estimated parameter as a random variable, whereas frequentist confidence sets treat their bounds as random variables and include the true fixed parameter with certain probability. 
	In the following, with an abuse of terminology, we may talk about the \emph{true} value of a random variable in the context of credible sets.
 %\Birgit{This statement is unclear to me.}
	We refer the reader to \cite{Jaynes1976ConfvsCredible} for a thorough exposition.
	%Indeed, the paper \cite{Jaynes1976ConfvsCredible} shows that the best possible confidence set attainable is the corresponding Bayesian credible set.
	%
\end{remark}

\begin{remark}
	The sub-simulation relations presented in subsequent sections are not restricted to the class of systems in $\eqref{eq:sysDataGenNonlin}$ and
	can be used in conjunction with
	any other estimation method that gives a credible set $\Theta$ with $\hat\theta\in\Theta$.
\end{remark}

%\color{blue!70!black}
\subsection{Parameter estimate}
To obtain an estimate of the parameters $\theta=[\theta_1,\ldots,\theta_n]\in\mathbb{R}^{m\times n}$,
%we define prior probability distributions
%\[ p(\theta_j)=\mathcal{N}(\theta_j|m_{0,j},\Sigma_{0,j}),\quad j\in\{1,\ldots,n\}, \]
we define a prior probability distribution of the weight vector $\bar\theta:=[\theta_1;\ldots;\theta_n]$ as
\begin{equation}
	p(\bar\theta)=\mathcal{N}(\bar\theta|\mu_{0},\Sigma_{0}),
\end{equation}
where the mean $\mu_{0}$ and covariance $\Sigma_{0}$ are either based on prior knowledge of the likelihood of $\theta$, if available, or set to 0 and $\sigma_{0}^{2}I$, respectively, with $\sigma_{0}^{2}$ a large number.
Based on the prior distributions, Bayesian linear regression allows us to compute the posterior distribution of the unknown parameters
given data $\mathcal{D}$ in \eqref{eq:data}. For this, we define the state matrix $X^{+} := [x_1^{+(1)};\ldots;x_1^{+(N)};\ldots;x_n^{+(1)};\ldots;x_n^{+(N)}]$
%, where
%$x^{+(i)}_j$ is the $j^{\text{th}}$ component of $x^{+(i)}$
and the \emph{design matrix} $\Phi$ of the form
\begin{equation*}
	\Phi := \begin{bmatrix}
		f_1(x^{(1)},u^{(1)}) & \hdots & f_m(x^{(1)},u^{(1)}) \\
		\vdots & \ddots & \vdots \\
		f_1(x^{(N)},u^{(N)}) & \hdots & f_m(x^{(N)},u^{(N)})
	\end{bmatrix}.
\end{equation*}
%
%\hrule
%For system \eqref{eq:sysDataGenNonlin} with zero-mean Gaussian noise $w_k\sim\mathcal{N}(\cdot|0, \Sigma_w)$ with a known covariance matrix $\Sigma_w\in\mathbb{R}^{n\times n}$ we define a prior probability distribution of the weight vector $w:=[\theta_1;\ldots;\theta_n]$ as
%\begin{equation}
%	p(w)=\mathcal{N}(w|m_{0},\Sigma_{0}).
%\end{equation}
%We define the state matrix $X^{+} := [x^{+(1)};\ldots;x^{+(N)}]$.
The posterior distribution can be computed as
\begin{align}%\label{eq:posterior}
	p(\bar\theta\given \mathcal D) & = \mathcal N(\bar\theta|\mu_{N}, \Sigma_{N}),\quad\text{with}\nonumber\\
	\mu_{N} &:= \Sigma_{N} (\Sigma_{0}^{-1} \mu_0 + (\Sigma^{-1}\otimes\Phi\T) X^+),\label{eq:mean}\\
	\Sigma_{N}^{-1} &:= \Sigma_{0}^{-1} + \Sigma^{-1} \otimes \Phi\T\Phi,
	\label{eq:covariance}
\end{align}
% \Oliver{I think this should be $\mu_{N} := \Sigma_{N} (\Sigma_{0}^{-1} \mu_0 + ({\Sigma}^{-1}\otimes\Phi\T) \xp)$?}
%where the mean and covariance are given by
and with $\otimes$ denoting the \emph{Kronecker product} \cite{Bishop2006ML}.
%\hrule
%The posterior distributions can be computed as
%\begin{equation*}%\label{eq:posterior}
%	p(\theta_j\given X^{+}_j) = \mathcal N(\theta_j|m_{N,j}, \Sigma_{N,j}),\quad j\in\{1,\ldots,n\},
%\end{equation*}
%where the mean and covariance are given by
%\begin{align}
%	m_{N,j}&:=\Sigma_{N,j}(\Sigma_{0,j}^{-1}m_{0,j}+\sigma_w^{-2}\Phi\T X^{+}_j) \text{ and}\label{eq:mean}\\
%	\Sigma_{N,j}^{-1}&:=\Sigma_{0,j}^{-1}+\sigma_w^{-2}\Phi\T\Phi,\label{eq:covariance}
%\end{align}respectively \cite{Bishop2006ML}.
Note that the parameter estimate $\hat\theta$ is set to the mean, i.e., 
%we have $\hat\theta=[\mu_{N,1},\ldots,\mu_{N,n}]\T$ for $\mu_N:=[\mu_{N,1};\ldots;\mu_{N,n}]$ with $\mu_{N,i}\in\mathbb{R}^{m}$, $i\in\{1,\ldots,n\}$. 
we construct matrix $\hat\theta=[\mu_{N,1},\ldots,\mu_{N,n}] \in \mathbb{R}^{m \times n}$ by concatenating the vectors $\mu_{N,i}\in\mathbb{R}^{m}$, $i\in\{1,\ldots,n\}$.

\subsection{Credible set}
For a desired confidence bound $(1-\alpha)\in(0,1)$ we obtain the corresponding credible set
\begin{align}
		&\Theta = \{\theta\in\mathbb{R}^{m\times n}\given
		(\bar\theta-\mu_{N})\T \Sigma_{N}^{-1}(\bar\theta-\mu_{N})\label{eq:confidenceSetNL}\\ &\hspace{160pt}
		\leq n\cdot\chi^{-1}(1-\alpha| n)\},\nonumber
\end{align}
where
$\chi^{-1}(p| \nu):=\inf\{\zeta:\,p\leq\chi(\zeta|\nu)\}$
%$\chi^{-1}(p| \nu):=\{x:\chi(x|\nu)=p\}$
denotes the inverse cumulative distribution function
% \pdfmargincomment{[Birgit:] Not a fan of this abbreviation, since it is often used to denoted infimum.}
or \emph{quantile function} of the \emph{chi-squared} distribution $\chi(\zeta|\nu):=\int_{0}^{\zeta}\frac{t^{(\nu-2)/2}e^{-t/2}}{2^{\nu/2}\Gamma(\nu/2)}dt$
%\frac{1}{\Gamma(p/2)}\gamma(\frac{p}{2},\frac{y}{2})$
with $\nu$ degrees of freedom and the \emph{Gamma function} $\Gamma(\nu):=\int_{0}^{\infty}t^{\nu-1}e^{-t}dt$ \cite{Chew1966ConfidenceMVN}.
% Here, we denote the gamma function as $\Gamma(\cdot)$ and the lower incomplete gamma function as $\gamma(\cdot)$.
%
%\new{
%From \eqref{eq:confidenceSetNL} we deduce credible sets $\Theta_j\ni\theta_j$, $j\in\{1,\ldots,n\}$, that are more conservative, namely
%\begin{align}
%		%	\begin{split}
%			&\Theta_j = \{\theta_j\in\mathbb{R}^{m}\given
%			%		\forall j=\{1,\ldots,n\}:\,
%			%
%			(\theta_j-m_{N,j})\T \Sigma_{N,j}^{-1}(\theta_j-m_{N,j})\label{eq:confidenceSetNL_indivi}\\ &\hspace{170pt}
%			\leq \chi^{-1}(1-\alpha| n)\},\nonumber
%			%	\end{split}
%\end{align}
%with covariance matrices $\Sigma_{N,j}^{-1}\in\mathbb{R}^{m\times m}$ defined via
%\begin{equation*}
%	\Sigma_{N}^{-1} = \begin{bmatrix}
%		\Sigma_{N,1}^{-1} & &\\
%		& \ddots &\\
%		&& \Sigma_{N,n}^{-1}
%	\end{bmatrix},
%\end{equation*}
%where all off-diagonal matrices are omitted. \Oliver{I don't like this formulation.}
%}
%\Sadegh{why do you need this at all? Why do you call it `confidence sets'? Eliminate if you don't need. The line of reasoning is not correct. You need to use some sort of projection.}

%\smallskip
%\color{black}
In the next section, we pave the way for answering Problem~\ref{prob:ssr} on the class of scLTL specifications by an abstraction-based control design scheme. We provide a new simulation relation that enables parameter-independent control refinement from an abstract model to the original concrete model that belongs to a parameterized class.

\section{Control Refinement via Sub-Simulation Relations}\label{sec:framework}
%\Birgit{It might be good to change the structure of this section. Instead of having A-F, make A and B headings, since these are known (not new) things.}
In order to answer Problem~\ref{prob:ssr} we employ the concept of simulation relations. In essence, simulation relations allow to compute a controller for a nominal model $\Mh$ and specification $\psi$, and transfer the guarantees of satisfaction to the original model $\M$ by quantifying their similarity.
Using the results in Sec.~\ref{sec:BLR}, we construct an abstract model $\Mh$, based on which we wish to design a controller:
\begin{equation}
	\label{eq:nom_dynamics}
	\Mh:
	\left\{ \begin{array}{ll}
		\hat x_{k+1} &= \hat f(\hat x_k,\hat u_k;\hat\theta) + \hat w_k,\quad \hat w_k\sim \hat p_{\hat w}(\cdotx),\\
		\hat y_k &= \hat h(\hat x_k).
	\end{array} \right.
\end{equation}
This abstract model could for example be $\M(\theta)$ but for a given valuation of parameters $\theta = \hat\theta$, i.e., $\Mh:=\M(\hat\theta)$ if $\hat h\equiv h$, or any other model constructed by space reduction or discretization.
Furthermore, we construct a set of models $\{\M(\theta)$ with $\theta\in\Theta\}$ using the credible set computed via the approach in Sec.~\ref{sec:BLR} or any other approach that solves Problem~\ref{prob:confset}.

The systems $\Mh$ in \eqref{eq:nom_dynamics} and $\M(\theta)$
%\new{in \eqref{eq:model}}
with a given fixed $\theta$ can equivalently be described by a general Markov decision process, studied previously for formal verification and synthesis of controllers \cite{haesaert2017verification}. Given $\hat x_k, \hat u_k$, we can model the stochastic state transitions of  $\Mh$ with a probability kernel $\hat \Tr(\cdotx|\hat x_k,\hat u_k)$ that is computed based on $\hat f$ and $\hat p_{\hat w}(\cdotx)$ (similarly for $\M(\theta)$). This leads us to the representation of the systems as general Markov decision processes, which are defined next.

%\subsection{General Markov decision processes}\label{sec:MDPs}

%\Oliver{Talk about both models here.}
%With the probability kernel $\Tr(\cdot|x_k,u_k)$, we have reformulated the recursive notion of the stochastic dynamical system $\M$ into a Markov decision process.% also denoted here by $\M$ and is defined next.}
%\red{Mention that $\theta$ is unknown for policy. Differentiate this paper from previous works by highlighting the differences.}

%\Sadegh{replace `action' with `input'.} \Oliver{done}
%
%\Sadegh{Consider replacing $\A$ with $\mathbb U$.} \Oliver{done}
%
%\Sadegh{Trace vs path vs execution vs sequence. I suggest we use input sequence, state execution, output trace.} \Oliver{done}
%
%\Sadegh{I am getting rid of $z$ and $\Z$ and replace them by $y,\Y$.} \Oliver{done}
%\Sadegh{combine the two definitions?}
%\begin{definition}[Markov decision process (MDP)]
%An MDP is a tuple $\M=(\X,x_0,\A,\Tr )$ with %\\[-1.2em]
%$\X$ the state space with states $x\in\X$; % as its elements;
%$x_0\in\X$ the initial state;
%$\A$  the input space with inputs $u\in\A$;
%and
%%	\item $\pi$, the initial probability measure $\pi:\mathcal{B}(\X)\rightarrow [0,1]$;
%$\Tr:\X\times\A\times\mathcal B(\X)\rightarrow[0,1]$, a probability kernel assigning to each state $x\in \X$ and input $u\in \A$ pair a probability measure $ \Tr(\cdotx| x,u)$ on $(\X,\mathcal B(\X))$. %\hfill$\qed$ % and
%%$\h:\X\rightarrow\Y$, a measurable output map.
%\end{definition}

\begin{definition}[General Markov decision process (gMDP)]
	A gMDP is a tuple $\M\!=\!(\X,x_0,\A, \Tr, h,  \Y)$ 
	with %\\[-1.2em]
	$\X$, the state space with states $x\in\X$; % as its elements;
	$x_0\in\X$, the initial state;
	$\U$,  the input space with inputs $u\in\A$;
	%	\item $\pi$, the initial probability measure $\pi:\mathcal{B}(\X)\rightarrow [0,1]$;
	$\Tr:\X\times\A\times\mathcal B(\X)\rightarrow[0,1]$, a probability kernel assigning to each state-input pair $(x,u)\in\X\times\U$ a probability measure $ \Tr(\cdotx| x,u)$ on $(\X,\mathcal B(\X))$; %\hfill$\qed$ % and
	%$\h:\X\rightarrow\Y$, a measurable output map.
	$\Y$, the output space; % with outputs $y\in\Y$;
	and $h:\X\rightarrow\Y$, a measurable output map.
	%We focus on
	A metric $\mathbf d_\Y$ decorates the output space $\Y$. % \hfill$\qed$  %decorated with .
\end{definition}
%%We denote the class of all MDPs with the same metric output space $\Y$ as $\mathcal{M}_\Y$.
%We indicate the input sequence of an MDP $\M$ by $\acs:= \ac{0},\ac{1},\ac{2},\ldots$
%%$u_{\bullet}:\mathbb N\rightarrow\A$ is given, where $\mathbb N:=\{0,1,2,\ldots\}$ is the set of natural numbers.
%and we define its (finite) \emph{executions} as sequences of states $\xs=\x{0},\x{1},\x{2},\ldots$ (respectively, $\xs_N=\x{0},\x{1},\x{2},\ldots, \x{N}$) initialised with the initial state  $\xin$ of $\M$ at $k=0$.
In each execution, the consecutive state $x_{k+1}\in\X$
is obtained as a realization $x_{k+1}\sim\Tr\left(\cdotx| x_k, u_k \right)$ of the controlled Borel-measurable stochastic kernel. % $\Tr\left(\cdot\mid x_t, u_t \right)$.
%Thus, an execution of $\M$ is a time sequence of these states $x_{\bullet}:\mathbb N\rightarrow\X$. % is referred to as a path or execution.
Note that for a parametrized gMDP $\M(\theta)$ its transition kernel also depends on $\theta$ denoted as $\Tr\left(\cdotx| x_k, u_k; \theta \right)$.
%
%As in Eq.~\eqref{eq:nom_dynamics}, we can assign an output mapping $h:\X\rightarrow \Y$ and a metric $\mathbf d_{\Y}$ to the MDP $\M$ to get a general MDP.
%
%\begin{definition}[General Markov decision process (gMDP)]
%A gMDP is a tuple $\M\!=\!(\X,x_0,\A, \Tr, h,  \Y)$ that combines an MDP $(\X,x_0,\A, \Tr)$ with the output space $\Y$ and a measurable output map $h:\X\rightarrow\Y$.
%%We focus on
%A metric $\mathbf d_\Y$ decorates the output space $\Y$. % \hfill$\qed$  %decorated with .
%\end{definition}
%The gMDP semantics are directly inherited from those of the MDP.
Output traces of the gMDP are obtained as mappings of (finite) state executions, namely
$\mathbf y:= y_0, y_1, y_2,\ldots$ (respectively, $\mathbf y_N:= y_0, y_1, y_2,\ldots, y_N$),
where $y_k= h(x_k)$.

Analogous to representing systems as gMDPs, satisfaction of scLTL specifications can be checked using their alternative representation as deterministic finite-state automata.
%, defined next.

%\subsection{Deterministic finite-state automata}
\begin{definition}[Deterministic finite-state automaton (DFA)]
\label{def:dfa}
A DFA is a tuple $\mathcal A = \left(Q,q_0,\Sigma,F,\trans\right)$, where $Q$ is the finite set of locations of the DFA, $q_0\in Q$ is the initial location, $\Sigma$ the finite alphabet, $F\subset Q$ is the set of accepting locations, and $\trans: Q\times\Sigma\rightarrow Q$ is the transition function.
\end{definition}

For any $n\in\mathbb{N}$, a word $\boldsymbol{\omega}_n = \omega_0\omega_1\omega_2\ldots\omega_{n-1}\in\Sigma^n$ is accepted by a DFA $\mathcal A$ if there exists a finite run $q = (q_0,\ldots,q_n)\in Q^{n+1}$ such that $q_0$ is the initial location, $q_{i+1} = \trans(q_i,\omega_i)$ for all $0\leq i \le n$, and $q_n\in F$.

%\bigskip

\subsection{Control refinement}
%\Sadegh{Could we state the next para in more general from, why only nominal, it could be obtained from discretization.}
Having defined the concepts of gMDPs and DFAs, consider a set of models $\{\M(\theta)$ with $\theta\in\Theta\}$ and suppose that we have chosen an abstract (nominal) model $\Mh$ based on which we wish to design a \emph{single} controller and quantify the satisfaction probability over all models $\M(\theta)$ in the set of models. In this section, 
%(\Birgit{Not this section. In the next section, right? Move to the next section (5.4C)})
 we first formalize the notion of
%\emph{coupled state evolution} of the abstract system
a \emph{state mapping}
and an \emph{interface function}, that together form the control refinement. Then, we investigate the conditions under which a single controller for $\Mh$ can be refined to a controller for all $\M(\theta)$ independent of the parameter $\theta$.
Note that this is the central contribution of this work, %since it allows us to not only
since instead of synthesizing a parametrized controller $\Ca(\theta)$ for $\M(\theta)$
%by using the respective parameter $\theta$ as would be possible based on \cite{haesaert2020robust}, it allows us to
we synthesize a single controller $\Ca$ that works for all models $\M(\theta)$ in the set of models $\{\M(\theta)$ with $\theta\in\Theta\}$.
% \red{(@Sofie: Can we say this like that?)}
%\red{[sofie :] changed it.}
This leads us to the novel concept of \emph{sub-probability couplings} and simulation relations.

\medskip
%that enables we can use a controller for the nominal model $\hat\M$ and refine it to the larger set of models $\{\M(\theta)$ with $\theta\in\Theta\}$.
%In the next subsection, we will detail what we mean with a control refinement.

%Consider   $ \hat\Ca\times \hat\M  \satisfies \psi$ for all $\theta\in\Theta$ with probability at least $\hat p_\psi$ .
%How can we design a robust $\Ca$ based on $\Ca$ such that
%$\Ca\times \hat\M(\theta) \satisfies \psi$ for all $\theta\in \Theta$ with probability at least $p_\psi$.
Consider the gMDP $\Mh=(\Xh,\xh{0},\Ah,\Trh,\hat h,\Yh)$
%\Birgit{with dynamics \eqref{eq:nom_dynamics}}
as the abstract model, the gMDP $\M(\theta)=(\X,x_0,\A,\Tr(\theta),h,\Y)$ referred to as the concrete gMDP, and an abstract controller $\Cah$ for $\Mh$.
To refine the controller $\Cah$ on $\Mh$ to a controller for $\M(\theta)$, we define a pair of interfacing functions consisting of a \emph{state mapping} that translates the states $ x\in\X$  to the states $\xh{} \in\Xh$  and an \emph{interface function} that refines the inputs $\uh{}\in\Ah$ to the control inputs $u\in\A$.

\noindent{\bfseries Interface function:}
%\Xh
We define an interface function \cite{girard2009hierarchical,girard2011approximate} as a Borel measurable stochastic kernel
$\InF: \Xh \times\X\times\Ah\times\borel{\A}\rightarrow[0,1]$ such that
any input $\ach{k}$ for $\Mh$ is refined to an input $\ac{k}$ for $\M(\theta)$ as
\begin{equation}\label{eq:input_refi}
	\ac{k}\sim \InF(\cdotx\,|\xh{k}, x_k, \uh{k}).
\end{equation}
% \Oliver{In the following we use a functional mapping! The output space should be either $\U$ or $[0,1]$.}

\noindent{\bfseries State mapping:} We can define the state mapping in a general form as a stochastic kernel $\Refi$ that maps the current state $\xh{k}$ and input $\ach{k}$ to the next state $\xh{k+1}$ of the abstract model.
The next state $\xh{k+1}$ has the distribution specified by $\Refi$ as
\begin{equation*}
	\xh{k+1} \sim \Refi(\cdotx |\xh{k}, \x{k}, \x{k+1},\ach{k}).
\end{equation*}
This state mapping is coupled with the concrete model via its states $\x{k}, \x{k+1}$ and depends implicitly on $\ac{k}$ through Eq.~\eqref{eq:input_refi}.

\begin{figure}[htp]
	\centering
	\scalebox{0.9}{
		\includegraphics{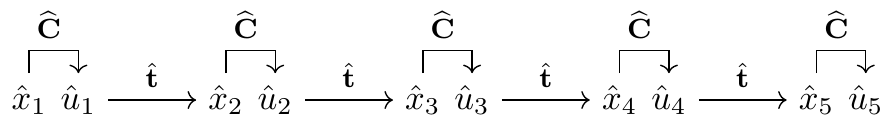}
	}
	\caption{Controller $\Cah$ on the abstract model $\Mh$.}
	\label{fig:abstract_control}
\end{figure}

\begin{figure}[htp]
	\centering
	\scalebox{0.9}{
		\includegraphics{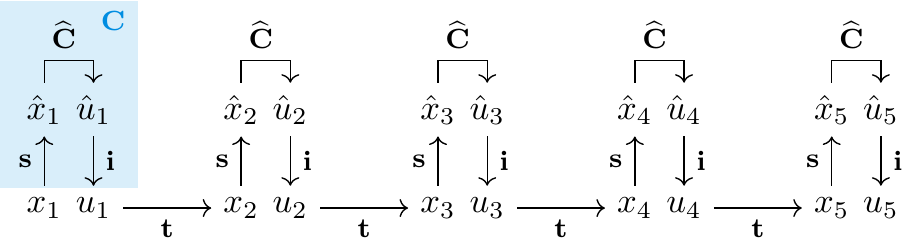}
	}
	\caption{Control refinement that uses the abstract controller $\Cah$ to control the concrete model $\M(\theta)$. }\label{fig:contrl_ref}
\end{figure}

%\red{Input/control refinement, stick with one}
Fig.~\ref{fig:abstract_control} illustrates how the abstract controller $\Cah$ defines a control input $\uh{k}$ as a function of the abstract state $\xh{k}$. Based on the state mapping $\Refi$ and the interface function $\InF$, the abstract controller $\Cah$ can be refined to a controller for $\M(\theta)$ as depicted in Fig.~\ref{fig:contrl_ref}.
In this figure, the states $\x{k}$ from the concrete model $\M(\theta)$ are mapped to the abstract states $\xh{k}$ with $\Refi$, the control inputs $\ach{k}$ are obtained using $\Cah$ and $\xh{k}$, and these inputs are then refined to control inputs $\ac{k}$ for $\M(\theta)$ using the interface function $\InF$.

%For an uncertain model $\M$ and a given abstract model $\Mh$, we are interested in the conditions under which a controller $\Cah$ designed for $\Mh$ can be refined to a controller $\Ca$ for $\M$ and which stochastic properties can be preserved during this refinement.
%%
%\input{partial_RFF}
%%
%Consider a stochastic kernel $\Refi$ that maps current states and inputs to the state update of the abstraction model.
%That is, the abstract state $\xh{k+1}$ can be updated using
%\begin{equation}
%	\xh{k+1} \sim \Refi(\cdot |\xh{k}, \x{k}, \x{k+1},\ach{k}).
%\end{equation}
%\red{Refinement not consistent with figure.}\\
%%
%This stochastic kernel leads to a control refinement of any controller $\Cah$ designed for $\widehat\M$ to a controller  for $\M$ as depicted in Figure~\ref{fig:aux_model_3}.  Furthermore, any input $\ach{k}$ can be refined to $\ac{k}$ using $\InF$.

We now question under which conditions on the interface function and the models the refinement is valid and preserves the satisfaction probability. This is addressed in the next subsection.

\subsection{Valid control refinement and sub-simulation relations}\label{sec:partialStoch}

Before diving into the definition of a valid control refinement that is also amenable to models with parametric uncertainty, we introduce a relaxed version of approximate simulation relations   \cite{haesaert2017verification} based on sub-probability couplings.% with a minor relaxation on the coupling of probability distributions (Sec.~\ref{sec:partialStoch}).

We define a sub-probability coupling with respect to a given relation $\mathcal R\subset \Xh\times\X$ as follows.
%In \cite{haesaert2017verification}, we have defined the coupling of two probability distributions with respect to a given relation $\mathcal R\subset \X_1\times\X_2$. Similarly, we trivially consider sub-probability couplings that  allocate a probability mass to $\mathcal R$.

\begin{definition}[Sub-probability coupling]
\label{def:submeasure_lifting}
	Given $\po \in\mathcal P(\X)$, $\hat\po\in \mathcal P(\Xh)$,  $\mathcal R\subset\Xh\times\X$, and a value $\delta\in[0,1]$, we say that a sub-probability measure $\W$ over $(\Xh\times\X , \mathcal B(\Xh\times\X))$ with $\W(\Xh\times\X)\geq 1-\delta$ is a \emph{sub-probability coupling} of $\hat\po$ and $\po$ over $\rel$ if
	\begin{itemize}
%		\item[a)] the support of $\W$ is a subset of $\rel$, $\operatorname{supp}(\W)\subset\rel$, that is, its probability mass is located on $\rel$,
		\item[a)] $\W(\Xh\times\X)=\W(\R)$, that is, the probability mass of $\W$ is located on $\rel$;%\Oliver{Is "mass" or "density" correct? }
		\item[b)]   for all measurable sets $A\subset \Xh$, it holds that
		$\W(A\times \X)\leq\hat\po(A)$;
		and
		\item[c)]  for all measurable sets $A\subset \X$, it holds that $\W(\Xh\times A)\leq\po(A)$.
	\end{itemize}
%\red{We say that $(\hat\po, \po)$ is in the lifted simulation relation $\bar \rel_\delta$, or equivalently $\hat\po\bar\rel_\delta\po$, if there exists a sub-probability coupling $\W$ satisfying the above conditions.}
%\pdfmargincomment[color=red]{We never use the notation introduced in this last sentence. Remove? I checked this again!!! I don't think we made any error. Instead, we are always saying that there is a "sub-probability coupling of phat and p over R", so basically the same but in prose. I'd suggest cutting it. Optionally, we can substitute it throughout the paper and the proofs.} %\hfill$\qed$
\end{definition}
Intuitively, conditions (b)-(c) state that a sub-probability coupling $\W$ of $\hat\po$ and $\po$ lives on the product space $\Xh\times\X$, and its marginals can not exceed the measures $\hat\po$ and $\po$, respectively. According to condition (a), $\W$ is fully supported on the subset $\rel$, i.e., it captures only the probability mass that $\hat\po$ and $\po$ mutually put on $\rel$.
Note that condition (a) of Definition~\ref{def:submeasure_lifting} implies that this probability mass $\W(\rel)\geq1-\delta$. 
%\Birgit{Nice clarifications!}

%\Oliver{It might be interesting to have the following in the text to show how you can go from sub-probability measures to commonly known probability measures.}
\begin{theorem}[Recovering probability coupling]\label{thm:fullcoup}
	For a sub-probability coupling $\sW$ as in Definition~\ref{def:submeasure_lifting}, we can complete $\sW$ to get a probability coupling $\fW: \mathcal{B}(\mathbb{\hat{X}} \times \mathbb{X}) \rightarrow [0,1]$ that couples the probability measures $\hat{p}$ and $p$:
	%	 its \emph{full coupling} $\fW: \mathcal{B}(\mathbb{\hat{X}} \times \mathbb{X}) \rightarrow [0,1]$
	%(c.f. $\delta$-lifting in \cite{haesaert2017verification}, Definition~8)
	%by completing it as follows:
	\begin{align}
%		\begin{split}
			&	\fW (d\xhp\!\times d\xp) = \sW(d\xhp\!\times d\xp) + \label{eq:fullcoupling}\\  &\hspace{15pt}\tfrac{1}{1-\sW(\R)}\big(p(d\xp)-\sW(\Xh \times d\xp)\big)\big(\hat{p}(d\xhp)-\sW(d\xhp\! \times \X)\big).\nonumber
%		\end{split}
	\end{align}
%	where $d\xh{}\subset \Xh$ and $d\x{}\subset \X$ are measurable sets.
	In other words, \eqref{eq:fullcoupling} satisfies the conditions in \cite[Definition~8]{haesaert2017verification}.
\end{theorem}
%	\smallskip
%\noindent\textbf{Recover probability measure
%%	Probability vs. sub-probability measures/Recover probability measure
%	:}
The first term of the coupling \eqref{eq:fullcoupling} puts only weight on $\R$, while the second term assigns the remaining probability mass in an independent fashion. Note that the factor $\tfrac{1}{1-\sW(\R)}$ normalizes the remaining probability mass such that $\fW$ satisfies the coupling properties $\fW(d\xhp\! \times \X) = \hat{p}(d\xhp)$ and $\fW(\Xh \times d\xp) = p(d\xp)$
% for all measurable sets $d\xh{} \subset \Xh$, $d\x{} \subset \X$
(cf. $\delta$-lifting in \cite[Definition~8]{haesaert2017verification}).
\begin{proof}
	First, we show that $\fW$ is a probability measure by showing that $\fW$ is fully supported on $\Xh\times\X$, i.e., $\fW (\Xh\times\X)= 1$:
	\begin{equation*}
			\fW (\Xh\times\X) = \sW(\Xh\times\X) + \tfrac{1}{1-\sW(\R)}\big(1-\sW(\Xh \times \X)\big)^2 = 1,
	\end{equation*}
	where we utilized the property of the sub-probability coupling as in Definition~\ref{def:submeasure_lifting}(a).
	The coupling properties are easily derived by calculating the marginals of $\fW$, i.e.,
	\begin{align*}
		\fW (d\xhp\!\times\X) &= \sW(d\xhp\!\times\X) +\tfrac{1}{1-\sW(\R)}
		\big(1-\sW(\Xh \times \X)\big)\\
		&\hspace{40pt}
		\times\big(\hat{p}(d\xhp)-\sW(d\xhp\! \times \X)\big), \\
		&= \hat{p}(d\xhp), \quad \\%\text{and}\\
		\fW (\Xh\times d\xp) &= \sW(\Xh\times d\xp) +\tfrac{1}{1-\sW(\R)}
		\big(p(d\xp)-\sW(\Xh \times d\xp)\big)\\
		&\hspace{40pt}
		\times\big(1-\sW(\Xh \times \X)\big), \\
		&= p(d\xp).
	\end{align*}
	Thus, $\fW$ satisfies all the conditions of \cite[Definition~8]{haesaert2017verification}.
\end{proof}

%\hrule
%\begin{align*}
%	&p(A)\equiv\W_{full}(\Xh \times A)\\
%	&=\W(\Xh \times A) + \tfrac{1}{1-\W(\R)}\left((p(A)-\W(\Xh \times A))(\hat{p}(\Xh)-\W(\Xh \times \X))\right)\\
%	&\text{With $\hat{p}(\Xh)=1$ and $\W(\R)=\W(\Xh\times\X)$}\\
%	&= \W(\Xh \times A) + (p(A)-\W(\Xh \times A)) \text{ qed.}
%\end{align*}
%\hrule
\begin{remark}
	For the parametrized case, i.e., $\hat\po(\cdotx)$ and $\po(\cdotx|\theta)$, the sub-probability coupling $\W(\cdotx|\theta)$ may likewise depend on $\theta$.
	Furthermore, although we define a sub-probability coupling $\W$ as a probability measure over the probability spaces $\X$ and $\Xh$, we use it in its kernel form $\Wt$ in the remainder of this paper. We obtain a particular probability measure $\W$ from $\Wt$ for a fixed choice of $(\hat x,x,\ach{})$ (cf. Table~\ref{tbl:notation}).
\end{remark}
\begin{table}
	\caption{Notation used for the transition kernels and measures associated to the different models. 
		%By fixing the conditioning variables of the transition kernel (in bold), we obtain a concrete probability measure describing the distribution of the consecutive state. For the composed system $\Mh\times\M(\theta)$ we additionally indicate the lower bound via the sub-probability coupling. \Birgit{I am not a fan of putting this explanation in the caption here. Perhaps add it as a remark in the text, or remove completely.}
	}
	\label{tbl:notation}
	\centering
	\begin{tabular}{|c|c|c|}
		\hline 
		&&\\[-1em]
		%System 
		Model & Transition kernel & Associated measure \\[.1em]
		\hline
		&&\\[-1em]
		$\M(\theta)$ & $\Tr(d\xp|x,\InF(\cdotx|\xh{},\x{},\ach{});\theta)$ & $t(d\xp|\theta)$ \\[.1em]
		\hline
		&&\\[-1em]
		$\Mh$ & $\Trh(d\xhp|\hat x,\ach{})$ & $\hat t(d\xhp)$ \\[.1em]
		\hline
		&&\\[-1em]
		$\Mh\times\M(\theta)$ & $\Tr_\times(d\xhp\!\times d\xp|\hat x,x,\ach{};\theta)$ & $t_\times(d\xhp\!\times d\xp|\theta)$ \\[.1em]
%		&&\\[-1em]
%		& $\geq\Wt(d\xhp\!\times d\xp|\hat x,x,\ach{};\theta)$ & $\geq\W(d\xhp\!\times d\xp|\theta)$ \\[.1em]
		\hline
	\end{tabular}
%	\begin{tabular}{|c|c|c|c|}
%		\hline 
%		&&&\\[-1em]
%		& $\M(\theta)$ & $\Mh$  & $\Mh\preceq^{\delta}_\eps\M(\theta)$ \\[.1em]
%		\hline
%		&&&\\[-1em]
%		Kernel & $\Tr(d\xp|x,\InF(\xh{},\x{},\ach{});\theta)$ & $\Trh(d\xhp|\hat x,\ach{})$  & $\Wt(d\xhp\!\times d\xp|\hat x,x,\ach{})$  \\[.1em]
%		\hline
%		&&&\\[-1em]
%		Measure & $t(d\xp|\theta)$ & $\hat t(d\xhp)$ & $\W(d\xhp\!\times d\xp)$ \\[.1em]
%		\hline
%	\end{tabular}
\end{table}

%We say that a kernel $\Wt:\Y\times\borel{\Y}\rightarrow \mathbb{R}_{>0}$  is a sub-probability kernel if \todo{[complete this]}.  \color{black}
Let us now define $(\varepsilon, \delta)$-sub-simulation relations for stochastic systems, where $\varepsilon$ indicates the error in the output mapping and $\delta$ indicates the closeness in the probabilistic evolution of the two systems. %, which is a definition similar to $(\varepsilon, \delta)$-stochastic simulation relatio\Xhns. In contrast to the latter, our similarity definition will now use sub-probability couplings (c.f., Definition~\ref{def:submeasure_lifting}).
%\color{purple!60!blue}
%\Sadegh{Oliver, check the hat notation on all symbols to unify the usage of hat and widehat for each symbol. For example, we should only use one of the two cases of $\hat{\M}$ and $\Mh$, and one of the two cases of $\widehat{\X}$ and $\Xh$.}
\begin{definition}[($\varepsilon,\delta$)-sub-simulation relation (SSR)]
\label{def:subsim}
	Consider two gMDPs
	$\M\!=\!(\X,x_0,\A, \Tr, h, \Y)$ and $\Mh\!=\!(\Xh,\xh{0},\Ah, \Trh, \hat h,  \Y)$, a measurable relation $\rel\subset \Xh\times\X$, and an interface function $\InF: \Xh \times\X\times\Ah\times\borel{\A}\rightarrow[0,1]$. If there exists a sub-probability kernel  $\Wt(\cdotx|\xh{},x, \uh{})$
	%\begin{itemize}\itemsep=0pt
	%	\item[I.] an interface function \(\InF :\Xh\times\X\times\Ah \rightarrow \A\)
	%	\item[II.] a sub-probability kernel  $\Wt(\cdot|\xh{},x, \uh{})$
	% \end{itemize}
	such that
	\begin{itemize}\itemsep=0pt
		\item[(a)] $(\hat x_{0}, x_{0})\in \rel$;
		\item[(b)] for all $(\hat x,x)\in\rel$ and $\ach{}\in \Ah$, $\Wt(\cdotx|\hat x,x,\ach{})$ is a sub-probability coupling  of $\hat \Tr(\cdotx|\hat x,\ach{})$ and $\Tr(\cdot|x,\InF(\cdotx|\xh{},\x{},\ach{}))$ over $\rel$  with respect to $\delta $ (see Definition~\ref{def:submeasure_lifting});
		\item[(c)] $\forall (\hat x,x)\in\rel: \mathbf d_\Y(\hat h(\hat x),h(x)) \leq \varepsilon$;
	\end{itemize}
then $\Mh$ is in an $(\varepsilon,\delta)$-SSR with $\M$ denoted by \mbox{$\Mh\preceq^{\delta}_\eps\M$}.
\end{definition}

%\Oliver{Keep either the example or the remark.}
\begin{remark}[Intuition]
%	To illustrate the SSR concept, consider 
%%	two gMDPs $\M$ and $\Mh$ as 
%	the setup in Definition~\ref{def:subsim}, with identical initial states $\hat x_0=x_0$ and the interface function $\hat u=u$ for simplicity.
	To illustrate the SSR concept, consider two gMDPs $\M$ and $\Mh$ in an SSR $\Mh\preceq^{\delta}_\eps\M$.
	Then, Definition~\ref{def:subsim} puts three conditions on the composed system $\Mh\times\M$ evolving on the product space $\Xh\times\X$. Note that $\rel$ defines a subset of $\Xh\times\X$.
	According to condition (a), both systems start in $\rel$ upon initialization. 
	In any subsequent time step, once in $\rel$ and applying the control inputs $\hat u$ and $u\sim\InF(\cdotx)$, respectively, condition (b) certifies that the systems stay in $\rel$ for the next time step with a probability at least $(1-\delta)$.
	Finally, given the two systems are in $\rel$, the corresponding outputs $\hat y=\hat h(\hat x)$ and $ y= h( x)$ are $\varepsilon$-close.
\end{remark}
%\new{Equivalently to Definition~\ref{def:submeasure_lifting}}, the SSR
%\red{[Sofie]:  Some comments
%	Looking back at the definitions, Definition 6 is an alternative for the known stochastic simulation relation. That is if a stochastic simulation relation exists than there also exist this subsimulation relation, and the existence of a subsimulation relation implies the existence of a stochastic simulation relation.
%	However, Definition7 which is based on Definition6 is a relaxation of the control refinement based on the original stochastic simulation relation.
%Note that this also shows why Theorem 1 is quite trivial.
%}\Oliver{Last bit to be incorporated or note once more here could be nice. @Sadegh: Leaving this with you.}
\begin{remark}
%[Previous definitions]
Both Definitions~\ref{def:submeasure_lifting}-\ref{def:subsim} are technical relaxations of the definitions of $\delta$-lifting and of $(\varepsilon,\delta)$-stochastic simulation relations provided in \cite{haesaert2017verification}. These relations quantify the similarity between two models by bounding their transition probabilities with $\delta$ and their output distances with $\varepsilon$.
%	\pdfmargincomment{Mention that this new coupling is a partial coupling?}

%These new extended definitions are essential in the next section to achieve control refinement that holds for uncertain stochastic systems. Note that these definitions are fully equivalent to those in \cite{haesaert2017verification} for stochastic systems that are not subject to parametric uncertainty. \Sofie{Dont mention this yet bc we're not addressing this in these defs yet}
%\Sadegh{It is not clear whether we allow r and v to depend on the parameter or not. What about the state mapping? We should be clear about this and make sure we emphasize this aspect. It is not clear now.}
\end{remark}
For a model class $\M(\theta)$, where $\hat \Tr(\cdotx|\hat x,\ach{})$ and $\Tr(\cdotx|x,\InF(\cdotx|\xh{},\x{},\ach{});\theta)$, the above definition allows us to have an interface function $\InF$ that is independent of $\theta$ but a sub-probability kernel $\Wt(\cdotx|\xh{},x, \uh{};\theta)$ which may depend on $\theta$.
In order to solve Problem~\ref{prob:ssr}, we require the state mapping $\Refi(\cdotx|\xh{}, \x{}, \xp, \ach{})$
%\pdfmargincomment{What is the consensus about the two-sided notation with and without indices?}
to also be independent of $\theta$. This leads us to a condition under which the state mapping $\Refi$ gives a valid control refinement as formalized next.
%\todo{Instead of calling this a valid control refinment can we call this an ($\varepsilon, \delta$)-accurate control refinement?}

%\red{(@Sofie: I omitted the explicit dependence on $\theta$ in most of the definitions, since they also hold when there is no dependence on $\theta$. Nevertheless, it might be more useful for understanding to add the dependence in the following definition. What do you think?)[Sofie:] I would leave the definition as is. However, after the def. you could also give the equation explicitly for the parameterized case. I think that would be more clear.}
\begin{definition}[\bfseries Valid control refinement]
\label{def:validrefinement}
	Consider an interface function $\InF: \Xh \times\X\times\Ah\times\borel{\A}\rightarrow[0,1]$ and a sub-probability kernel  $\Wt$ according to Definition~\ref{def:subsim}.
%	\pdfmargincomment[color=red]{Definition 5?}
	We say that a state mapping $\Refi: \Xh\times\X\times\X\times\Ah\rightarrow\mathcal P(\Xh)$ % which gives realizations of the abstract model's state  $x_{1, k+1}\sim \Refi(\cdot |x_{1,k}, x_{2,k}, x_{2,k+1},a_{1,k})$.
	defines a \emph{valid control refinement} if the composed transition probability measure
	%\footnote{We have dropped the dependence on $\xh{k}, \x{k}, \ach{k}$ for brevity.} \Sadegh{Is this $\Ca$ or $\Cah$. The composition doesn't come across very well specially that we have eliminated the conditions and the controller. Perhaps we should refer to the figure above.}
	\begin{align}
%		\begin{split}
			\Tr_\times&(d\xhp \times d\xp|\xh{}, \x{}, \ach{}):=\label{eq:compprobmeas}\\
			&\Refi(d\xhp| \xh{}, \x{}, \xp, \ach{}) \Tr(d\xp|\x{},\InF(\cdotx|\xh{}, \x{}, \ach{}))\nonumber
			%&\ac{k}\sim\InF(\xh{k}, \x{k}, \ach{k})
%		\end{split}
	\end{align}
	 upper-bounds the sub-probability coupling $\Wt$, namely
	 \begin{equation}
	 \label{eq:sub_prob}
	\Tr_\times(A|\xh{}, \x{}, \ach{})\geq \Wt(A|\xh{}, \x{}, \ach{}),
	\end{equation}
	for all measurable sets $A\subset \Xh\times \X$, $(\xh{}, \x{})\in\rel$, and $\ach{}\in\Ah$.
\end{definition}
Note that for a model class $\M(\theta)$ that is in relation with a model $\Mh$ the right-hand side of \eqref{eq:sub_prob} can depend on $\theta$ and the left-hand side can depend on $\theta$ only via the dynamics of $\M(\theta)$ represented by the kernel $\Tr$, i.e.,
\begin{equation*}
	\int\!\!\!\!\int_{(\xhp{},\xp{})\in A}\!\!\!\!\!\!\!\!\!\!\!\!\!\!\!\!\!\!\!\!\!\!\Refi(d\xhp{}| \xh{}, \x{}, \xp\!, \hat u) \Tr(d\xp{}|\x{},\InF(\cdot|\xh{}, \x{}, \ach{});\theta)\!\geq\!\Wt(A|\xh{}, \x{}, \ach{};\theta).
\end{equation*}
The next theorem states that similar to the $(\varepsilon, \delta)$-approximate simulation relation defined in \cite{haesaert2017verification}, there always exists at least one valid control refinement for our newly defined SSR.
\begin{theorem}[Existence]
\label{thm:exist_refine}
For two gMDPs $\Mh$ and $\M(\theta)$ with $\Mh\preceq^{\delta}_\eps\M(\theta)$, there always exists a valid control refinement.
\end{theorem}
%\Oliver{Have the proof here or in the appendix?}
%\color{blue!70!black}
\begin{proof}
	If  $\Mh\preceq^{\delta}_\eps\M(\theta)$, then this implies that there also exists a probability kernel $\fWt(d\xhp\times d\xp|\cdotx;\theta)$ (cf. Theorem~\ref{thm:fullcoup}) satisfying the conditions for an $(\varepsilon, \delta)$-approximate probabilistic simulation relation
%	\pdfmargincomment[color=red]{Shouldn't this be a $(\varepsilon, \delta)$-approximate probabilistic relation?}
	(cf. \cite[Definition~9]{haesaert2017verification}). 
 Let $\Tr_\times=\fWt$.
We can then find a state mapping $\Refi$ as the conditional kernel (which defines a $t$-almost surely unique measurable map for every $(\hat x,x,\hat u)$ \cite[Corollary 3.1.2]{borkar2012probability}) obtained from $\Tr_\times$ such that
% We can then find a state mapping $\Refi$ as a conditional kernel $\fWt'(d\xhp|\xp,\cdotx)$ \new{(\cite[Corollary 3.1.2]{borkar2012probability})} obtained from $\fWt(d\xhp\times d\xp|\cdotx;\theta)$, that is, 
% \[\new{\Refi(d \xhp
% 	|\xh{}, \x{}, \xp,\ach{}):= \fWt'(d \xhp |\xh{}, \x{},\xp,\ach{})},\]
% 	where the right term is conditioned on $\xp$ such that
	\begin{equation*}
	\Tr_\times(d\xhp\!\!\times\! d\xp |\xh{}, \x{},\ach{};\theta)\!=\!	\Refi(d\xhp|\xh{}, \x{},\xp\!,\ach{})\Tr(d\xp|\x{},\InF(\cdot);\theta).%&&\qedhere
	\end{equation*}
Note that the existence of $\fWt$ implies the existence of a sub-probability coupling minorizing $\Tr_\times$, hence satisfying \eqref{eq:sub_prob}.
%
% \Birgit{Here, you should prove that a valid control refinement exists but you do not use the conditions defined in Def. 8. Why not? You could add the following: \newline 
% Following \eqref{eq:compprobmeas}, this implies that the composed probability measure equals 
% 		\begin{align}
% 		\begin{split}
% 			\Tr_\times&(d\xh{k+1} \times d\x{k+1}|\xh{k}, \x{k}, \ach{k};\theta):=\\
% 			&\Refi(d\xh{k+1}| \xh{k}, \x{k}, x_{k+1}, \ach{k};\theta) \Tr(d\x{k+1}|\x{k},\InF(\cdotx|\xh{k}, \x{k}, \ach{k});\theta) = \\
% 		& 	\fWt(d\xh{k+1}\times d\x{k+1} |\xh{k}, \x{k},\x{k+1},\ach{k};\theta).
% 		\end{split} \label{eq:vbarTemp}
% 	\end{align}
% It can readily be concluded from \eqref{eq:fullcoupling} that a completed kernel (as in \eqref{eq:vbarTemp}) always upper-bounds a sub-probability kernel, hence
%  the composed transition probability measure upper-bounds the sub-probability coupling. Therefore, condition \eqref{eq:sub_prob} in Def. \ref{def:validrefinement} is satisfied and $\Refi$ is a valid control refinement. 
% }
\end{proof}
%\color{black}

%\pdfmargincomment{I moved the following two sentences here:}
%The new extended Definitions~\ref{def:submeasure_lifting}-\ref{def:subsim} are essential to achieve control refinement that holds for uncertain stochastic systems. Note that the above theorem states that our new framework fully recovers the results of \cite{haesaert2017verification} for non-parametric models.

Note that in general there is more than one valid control refinement
%as formalized Definition~\ref{def:validrefinement}
for given $\Mh$ and $\M(\theta)$ with $\Mh\preceq^{\delta}_\eps\M(\theta)$. This allows us to choose an interface function $\InF$ not dependent on $\theta$.
%\Sofie{[Here it would be nice to point out that there could be more than one control refinement.]} \Oliver{Is this not the case in  \cite{haesaert2017verification}?}\Oliver{Bc this allows us to choose the on e not dependen on theta}
%\Sadegh{DFA vs scLTL, which one to give?} \Oliver{done}
%
The above theorem states that our new framework fully recovers the results of \cite{haesaert2017verification} for non-parametric models.

Although Definition~\ref{def:validrefinement} gives a sufficient condition for a valid refinement, it is not a necessary condition. For specific control specifications, one can also use different refinement strategies such as the one presented in \cite{haesaert2018temporal}, where information on the value function and relation is used to refine the control policy. %\Sadegh{@Sofie: add a couple of sentences to hint on what kind of refinement is proposed there, otherwise not informative.}
%\pdfmargincomment{(@Sofie: Is this a valid refinement also for the case of a parametric stochastic system? If yes, we can keep it and somehow say that value functions are explained later (one of the reviewers noted that), otherwise, I think it doesn't add much value.)}

In the next theorem, we establish that our new similarity relation is transitive. This property is very useful when the abstract model is constructed in multiple stages of approximating the concrete model. We exploit this property in the case study section.

\begin{theorem}[Transitivity]
\label{thm:transitive}
Suppose $\Mt\preceq^{\delta_1}_{\eps_1}\Mh$ and $\Mh\preceq^{\delta_2}_{\eps_2}\M$. Then, we have $\Mt\preceq^{\delta}_{\eps}\M$ with $\delta = \delta_1+\delta_2$ and $\eps = \eps_1+\eps_2$.
\end{theorem}
%\color{blue!70!black}
\begin{proof}
	This proof is similar to the one given in \cite{haesaert2017verification} applied to sub-probability couplings.
	Consider three gMDPs $\M\!=\!(\X,x_0,\A, \Tr, h,  \Y)$, $\Mh\!=\!(\Xh,\hat x_0,\hat \A, \hat \Tr,\hat  h,   \Y)$, and $\Mt\!=\!(\tilde\X,\tilde x_0,\tilde\A, \tilde\Tr, \tilde h,  \Y)$. Given $\Mt\preceq^{\delta_1}_{\eps_1}\Mh$ and $\Mh\preceq^{\delta_2}_{\eps_2}\M$ with $\R_1,\sWt_1$ and $\R_2,\sWt_2$, respectively, as formalized in Definition~\ref{def:subsim},
	define the relation $\R\subset\tilde{\X}\times\X$ as $\R:=\{ (\tilde{x},x)\in\tilde{\X}\times\X \given \exists\hat{x}\in\Xh: (\tilde{x},\hat{x})\in\R_1, (\hat{x},{x})\in\R_2\}$.
	%	$\M$ and $\Mh$ are in an SSR with $\delta$ and $\varepsilon$, i.e.,  , as defined in Definition~\ref{def:subsim}.

	\medskip
	\noindent\emph{$\varepsilon$-deviation:}
	From the definition of $\rel$, we have that  $\forall(\xt{},x)\in\R$ $\exists \hat x\in\hat\X:(\tilde{x},\hat{x})\in\R_1, (\xh{},{x})\in\R_2$. Given $(\tilde x, x, \hat x)$ and the mutual output metric $\out{\cdotx}$, we can bound the output error from above as
	\begin{align*}
		\out{ \tilde{h}( \tilde{x}),h(x)} &\leq \out{ \tilde{h}( \tilde{x}), \hat{h}(\hat{x})}+ \out{ \hat{h}( \hat{x}),h(x)},\\
		&\leq \varepsilon_{1} + \varepsilon_{2}.
	\end{align*}

	\medskip
	\noindent\emph{$\delta$-deviation:}
	We start by completing the sub-probability couplings $\sWt_1$ and $\sWt_2$ to probability kernels $\fWt_1$ and $\fWt_2$ that couple $(\tilde{\Tr},\Trh)$ and $(\Trh,\Tr)$, respectively, (cf. $\delta$-lifting in \cite{haesaert2017verification}, Definition~8) using Eq.~\eqref{eq:fullcoupling}. Utilizing the proof in \cite{haesaert2017verification}, App.~D.2., we get that there exists a probability kernel $\fWt$ over $(\tilde{\X}\times\X , \mathcal B(\tilde{\X}\times\X))$
	%\pdfmargincomment{@Sadegh: Further terms necessary bc kernel?}
	that couples $\tilde{\Tr}$ and $\Tr$ over $\R$. Furthermore, we get that for all $z\in\Z$, where $\Z:=\{ (\xt{},\x{},\ut{}) \given (\xt{},\x{})\in\R \text{ and } \ut{}\in \tilde{\U}\}$, we have $\fWt(\R| z)\leq\delta_1+\delta_2$.
	From Eq.~\eqref{eq:fullcoupling} we get that for all measurable sets $d\xtp\subset\tilde{\X}$, $d\xp\subset\X$ we have $\fWt(d\xtp\times d\xp)\geq\sWt(d\xtp\times d\xp)$. Hence, we conclude that there exists an SSR $\Mt\preceq^{\delta}_{\eps}\M$ with $\delta\leq\delta_1+\delta_2$.
\end{proof}
%\color{black}

\subsection{Temporal logic control with sub-simulation relations}
For designing controllers to satisfy temporal logic properties expressed as scLTL specifications, we employ the representation of the specification as a DFA $\mathcal A = \left(Q,q_0,\Sigma,F,\trans\right)$ (see Definition~\ref{def:dfa}).
%, where $Q$ is the finite set of locations of the DFA, $q_0\in Q$ is the initial location, $\Sigma$ the finite alphabet, $F\subset Q$ is the set of accepting locations, and $\trans: Q\times\Sigma\rightarrow Q$ is the transition function. \Oliver{Repetitive, since now defined in Definition~\ref{def:dfa}: ..., we employ the representation of the specification as a DFA $\mathcal A = \left(Q,q_0,\Sigma,F,\trans\right)$ (c.f. Definition~\ref{def:dfa}).}
We then use a robust version of the dynamic programming characterization of the satisfaction probability \cite{haesaert2020robust} defined on the product of $\Mh$ and $\mathcal A$
and obtain a data-driven bound on the satisfaction probability in Theorem~\ref{thm:satProb}. We now summarize the details of the characterization presented in \cite{haesaert2020robust}, %provide the details of this characterization 
that encode the effects of both $\varepsilon$ and $\delta$.

Denote the $\varepsilon$-neighborhood of an element $y\in\Y$ as
\[B_\varepsilon(\hat y) :=\{y \in \Y|\,  \mathbf{d}_\Y\left(y,\hat y\right)\leq \varepsilon\}, \]
%
% stationary % you don't need this.
and a Markov policy $\pi:\Xh\times Q\rightarrow \mathcal P(\Ah)$.
Similar to dynamic programming with perfect model knowledge \cite{Sutton2018RL}, we utilize value functions $\Vb_l^\pi:\Xh\times Q\rightarrow [0,1]$, $l\in\{0,1,2,\ldots\}$ that represent the probability of starting in $(\hat x_0, q_0)$ and reaching the accepting set $F$ in $l$ steps. These value functions are connected recursively via operators associated with the dynamics of the systems.
%However, instead of computing $V_N^\mu$ explicitly, we calculate a value function $\Vb_N^\mu:\Xh\times Q\rightarrow [0,1]$ and retrieve the true probability in a consecutive step.
%\Oliver{(@Sofie: I think this is correct but it isn't nice.)}\red{[Sofie]: This paragraph is not clear to me. What does "resembling" mean here? I wouldnt know what the difference would be between the different value functions.  }\Oliver{Do you have an idea of how to phrase in a better way?}\Oliver{Switched bar}
%\pdfmargincomment{Is this understandable now?}

Let the initial value function $\Vb_0\equiv 0$. Define the \emph{$(\eps,\delta)$-robust operator} \( \mathbf T_{\eps,\delta}^{\pi}\), acting on value functions as
\begin{align}
	\notag \textstyle \mathbf T_{\eps,\delta}^{\pi} (\Vb)(\hat x,q)\!:=\!\Lim&\Big(\!\!\int_{\hat\X} \min_{q^{+\!}\in \trans_\varepsilon(q,\xhp)}\!\!\max\left\{\1_{F}(q^{+\!}), \Vb(\xhp,q^{+\!})\right\}\\
	&\hspace{1cm}\times \Trh(d\xhp|\hat x,\pi)-\delta\Big),\label{eq:robustoperator}
\end{align}
%\Oliver{This is inconsistent w.r.t. $\theta$ and leads to artificially conservative (impossible) trajectories to be considered in contrast to e.g. \cite{lew2021sampling}. We would get a $\delta(\theta)$ and then need to take the worst case w.r.t. $\theta\in\Theta$ in the Bellman operator.}
with $\trans_\varepsilon(q,\xhp):=\{\trans(q,\alpha)\mbox{ with }\alpha \in\mathcal L(B_\varepsilon(\hat h (\xhp)))\}$. The function $\Lim:\mathbb R\rightarrow [0,1]$ is the truncation function with $\Lim(\cdotx) := \min(1,\max(0,\cdotx))$, and $\ind_F$ is the indicator function of the set $F$.
%Note that $\delta$ accounts for the parametric uncertainty.\Sofie{[Prev. sentence is unclear in this context. Please remove.]}
The value functions are connected via $\mathbf T_{\eps,\delta}^{\pi}$ as
\begin{equation*}
	\Vb_{l+1}^\pi = \mathbf T_{\eps,\delta}^{\pi}(\Vb_l^\pi),\quad l\in\{0,1,2,\ldots\}.
\end{equation*}
Furthermore, we define the \emph{optimal $(\eps,\delta)$-robust operator}
% $\mathbf T_{\eps,\delta}^{\ast}$ as
\begin{equation*}
%	\label{eq:optimalrobustoperator}
	\mathbf T_{\eps,\delta}^{\ast} (\Vb)(\hat x,q):=\sup_{\pi} \mathbf T_{\eps,\delta}^{\pi} (\Vb)(\hat x,q).
\end{equation*}
The outer supremum is taken over Markov policies $\pi$ on the product space $\Xh\times Q$.
% \Oliver{(@Sofie: Isn't the Markov policy defined over $\Xh$ only?)}\red{[Sofie]: This is a Markov policy for the composed stochastic system with state $\Xh\times Q$.}. \Oliver{Then the local definition $\mu:\Xh\rightarrow \mathcal P(\Ah,\mathcal B(\Ah))$ is wrong, right?}
%\pdfmargincomment{Do we need to note anything bc the Markov policy was originally introduced to map from the state space and not the product space (see Sec.~\ref{sec:MDPs})?}
Via \cite[Cor.~4]{haesaert2020robust}, we can now define the lower bound on the satisfaction probability by looking at the limit case $l\rightarrow\infty$, as summarized in the following proposition.%

\begin{proposition}[($\varepsilon,\delta$)-robust satisfaction probability]
\label{prop:delepsreach_infty}
Suppose $\widehat\M\preceq^{\delta}_\varepsilon\M$ with $\delta>0$, and the specification being expressed as a DFA  $\mathcal{A}$.
Then, for any $\M$ we can construct $\Ca$ such that the specification is satisfied by $\Ca\times\M$ with probability at least $\mathcal S_{\varepsilon,\delta}^\ast$. This quantity is the \emph{($\varepsilon,\delta$)-robust satisfaction probability} defined as
\begin{equation*}
%\label{eq:opt_sat_rob_2v}
\mathcal S_{\varepsilon,\delta}^\ast
:=\min_{ q^{+\!}}\in \trans_\varepsilon(q_0,\hat x_0)\!\!\max\left\lbrace\ind_F(q^{+\!}), \Vb_\infty^{\ast}(\hat x_0,q^{+\!})\right\rbrace,
\end{equation*}
where $\Vb_\infty^{\ast}$ is the unique solution of the fixed-point equation
\begin{equation}
\label{eq:fixed_point}
\Vb_{\infty}^\ast = \mathbf T^\ast_{\eps,\delta} (\Vb_{\infty}^\ast),
\end{equation}
obtained from $\Vb_\infty^\ast :=\lim_{l\rightarrow\infty} (\mathbf T^\ast_{\varepsilon,\delta})^{l}(\Vb_0)$ with $\Vb_0 =0$.
The abstract controller $\Cah$ is the stationary Markov policy $\pi^\ast$
% \new{$\boldsymbol{\mu}^\ast=(\mu^\ast,\mu^\ast,\mu^\ast,\ldots)$}
% \red{$\mu^\ast$} \Oliver{Shouldn't this be bold?}
 that maximizes the right-hand side of \eqref{eq:fixed_point}, i.e., $\pi^\ast=\arg \sup_{\pi} \mathbf T_{\eps,\delta}^{\pi} (\Vb_\infty^\ast)$.
 %\pdfmargincomment[color=red]{Is the formula for getting mustar correct?}
% 	\red{[Sofie: this is not very precise]} \Oliver{Refer to the optimal operator equation instead?}
The controller $\Ca$ is the refined controller obtained from the abstract controller $\Cah$, the interface function $\InF$, and the state mapping $\Refi$.
%\red{(@Sofie: As discussed, this proposition only holds like this for $\delta\neq 0$ and infinite-horizons. Otherwise, we don't necessarily retrieve a unique solution nor a stationary policy. Add sth to address this.)}\Sofie{[the main issue is indeed the $\delta>0$. Stationarity is not an issue, even if the specification has a bounded time horizon, by computing it over an infinite horizon the non stationary  part of the policy will be encoded via the finite states of the DFA.  ]}\Oliver{Sadegh and I will take a look at this.}\Oliver{For journal version}
\end{proposition}
%\Sofie{[I need to take another look at this Proposition later.]}
%The proof follows similar steps as in \cite{haesaert2020robust}.

\begin{remark}[Finite-horizon specifications]
	%Proposition~\ref{prop:delepsreach_infty} holds only for
	%the discounted case, i.e.,
	%$\delta>0$ and infinite horizon specifications.
	For a finite horizon specification $\psi$ and $\delta\ge 0$, a finite number of value functions $\Vb_{l+1}^\ast =\mathbf T^\ast_{\varepsilon,\delta}(\Vb_l)$, $l\in\{0,1,2,\ldots,n_\psi\}$ are computed with $\Vb_0 =0$ and $n_\psi$ being the horizon of the specification. This will also give a non-stationary control policy.
%	Convergence of Eq.~\eqref{eq:fixed_point} cannot be guaranteed for $\delta=0$ or finite horizons.
%	For $\delta=0$ or finite horizons, no claims about the convergence of $\Vb_k^\ast :=(\mathbf T^\ast_{\varepsilon,\delta})^{k}(\Vb_0)$ can be made.
%	don't necessarily retrieve a unique solution nor a stationary policy.
%	Whilst the non-stationarity would be captured by the DFA, non-uniqueness will lead to \red{[Is this even an issue?]}.
\end{remark}

\begin{remark}[Bounded state space]
	To compute solutions for systems with unbounded support over a bounded state space $\X$,
	we add a sink state to the composed system $\Mh\times\mathcal A$ to capture the transitions leaving the bounded set of states.
%	the specification $\psi$ is augmented with an additive safety specification such that all trajectories leaving $\X$ violate the specification. \Birgit{I am not a fan of this remark. Alternative: To compute solutions for systems with unbounded support
%		over a bounded state space $\X$, one generally adds a sink state to capture transitions that leave the bounded set of states.}
\end{remark}

\begin{theorem}[Lower bound on satisfaction probability]\label{thm:satProb}
Given the dataset $\mathcal D$ in \eqref{eq:data}, let $\Theta$ be a credible set of $\theta^*$ with confidence level $(1-\alpha)$ (Definition~\ref{def:credible_set}).
Let $\Ca$ be the refined controller with $(\varepsilon,\delta)$-robust satisfaction probability $\mathcal S_{\varepsilon,\delta}^\ast$ for all $\theta\in\Theta$ (Proposition~\ref{prop:delepsreach_infty}). Then, we can conclude that
\begin{equation*}
	\mathbb P(\Ca\times\M(\theta^\ast)\satisfies \psi\, \vert\, \mathcal D) \ge \mathcal S_{\varepsilon,\delta}^\ast (1-\alpha).
\end{equation*}
\end{theorem}
\begin{proof}
	This result follows from the $(\varepsilon,\delta)$-robust satisfaction probability $\mathcal S_{\varepsilon,\delta}^\ast$ from Proposition~\ref{prop:delepsreach_infty} and the results of
	Sec.~\ref{sec:BLR}.
%	, we get that for the dataset $\mathcal D$, the system under the designed controller satisfies the specification with probability at least $\mathcal S_{\varepsilon,\delta}^\ast (1-\alpha)$.
	%It is possible to combine $\mathcal S_{\varepsilon,\delta}^\ast$ formulated in Proposition~\ref{prop:delepsreach_infty} with the confidence $(1-\alpha)$ used in the construction of the credible set $\Theta$ to get a lower bound on the satisfaction probability the specification $\psi$. In other words, we can use the law of total probability to get
%	This can be shown 
	Using the law of total probability we have
	\begin{align*}
		\mathbb P&(\Ca\times\M(\theta^\ast)\satisfies \psi\, \vert\, \mathcal D)\\
		& =
		\mathbb P(\Ca\times\M(\theta^\ast)\satisfies \psi\,\vert\, \theta^\ast\in\Theta,\mathcal D) \mathbb P(\theta^\ast\in\Theta)\\
		& \quad + \mathbb P(\Ca\times\M(\theta^\ast)\satisfies \psi\,\vert\, \theta^\ast\not\in\Theta,\mathcal D) \mathbb P(\theta^\ast\notin\Theta),\\
		& \ge \mathcal S_{\varepsilon,\delta}^\ast (1-\alpha).\qedhere
	\end{align*}
\end{proof}
\begin{remark}[Tightening the robust satisfaction probability]
	Once a controller $\Ca$ has been obtained, we can compute a less conservative robust satisfaction probability than $\mathcal S_{\varepsilon,\delta}^\ast$ from Proposition~\ref{prop:delepsreach_infty} via
	\begin{equation*}
		\mathbb P(\Ca\times\M(\theta^\ast)\satisfies \psi\, \vert\, \mathcal D) \!
		= \!\! \int_{\theta\in\mathbb{R}^{m\times n}} \hspace{-25pt}
		\mathbb P(\Ca\times\M(\theta)\satisfies \psi\,\vert\, \theta,\mathcal D) \mathbb P(\theta\vert\mathcal D).
		%	& \quad + \mathbb P(\Ca\times\M(\theta^\ast)\satisfies \psi\,\vert\, \theta^\ast\not\in\Theta,\mathcal D) \mathbb P(\theta^\ast\notin\Theta),\\
		%	& \ge \mathcal S_{\varepsilon,\delta}^\ast (1-\alpha).
	\end{equation*}
	Note that instead of considering the worst-case parameters $\theta\in\Theta$,
	 %by considering all parameters $\theta\in\Theta$ equally with a probability of 1, 
	 we may obtain a likelihood function $ \mathbb P(\theta\vert\mathcal D)$ spanning a probability distribution over individual values of $\theta$ given $\mathcal D$.
\end{remark}

\section{Simulation Relations for Nonlinear Systems}
\label{sec:establish_relation}
%\color{blue!70!black}
In this section, we apply the previously defined concepts
to construct simulation relations and show how to answer Problems~\ref{prob:confset} and \ref{prob:ssr}.
%, first on a simple linear system, and then on general nonlinear systems.
%Whilst we focus on nonlinear systems that are linear in the unknown parameters, the outlined approach is applicable to general nonlinear systems by choosing an appropriate parameter estimation method.
%We start by inferring a credible set of the true parametrization from state-input data and derive a valid control refinement.
%For the first part, we employ Bayesian linear regression, for which we reproduce the central results.
%We showcase our approach for two different robust parameter estimation methods, acknowledging that they can be replaced by any other suitable methods providing robust parameter estimates with the corresponding credible sets.
%Furthermore, whilst we restrict ourselves to systems with Gaussian noise here, the approach is generally applicable to other distributions. \new{Refer to \cite{Schon2023GMM}, providing further an extended approach based on approximating arbitrary continuous noise distributions using finite Gaussian mixture models.} 
Furthermore, whilst we restrict ourselves to systems with known Gaussian noise here, the approach is generally applicable to other (uncertain) distributions, as detailed in \cite{Schon2023GMM}, where it is shown how to extend the approach based on approximating arbitrary continuous noise distributions using finite Gaussian mixture models.

% !TEX root =  main.tex
%\subsection{Data-driven parameter estimation}\label{sec:BLR}
\smallskip
Consider a nonlinear system $\M(\theta^\ast)$ as in Eq.~\eqref{eq:sysDataGenNonlin}.
% $\theta\T=[\theta_1,\ldots,\theta_n]\T\in\Theta\subset\mathbb{R}^{n\times m}$, i.e.,
%%\pdfmargincomment{[Birgit:] You switched notation here. In the preliminaries you use mathbb R. I believe this R is used for simulation relation only. Besides that it is a bit surprising that you start using spaces here, instead of when you first introduce the model. But I guess that is fine, if you only need them at this point.}
%\begin{equation}
%%	\label{eq:sysDataGenNonlin}
%	\M(\theta): \left\{ \begin{array}{ll}
%		x_{k+1} &= \theta\T f(x_k, u_k) + w_k\\
%		y_k &= h(x_k),%+D_vv_k,
%	\end{array} \right.
%\end{equation}
%where $x\in\mathbb{R}^n$, and the functions $h$, $\hat h$, and $f:\X\times\U\rightarrow\mathbb{R}^{m}$ are known. \Oliver{Does it make more sense to first do the identification and then construct the nominal model?}
Let data $\mathcal{D}$ as in \eqref{eq:data} be obtained from this unknown true system $\M(\theta^\ast)$.  
Based on $\mathcal{D}$, we approximate $\theta^\ast$ with an estimate $\hat\theta$ and construct a credible set $\Theta$
% that contains the true parameters $\theta^\ast$ with a given confidence $(1-\alpha)\in(0,1)$, i.e., $\P(\theta^\ast\in\Theta)\geq 1-\alpha$ (cf. Fig.~\ref{fig:confSetSketch}). Let $\hat\theta\in\Theta$. 
as described in Sec~\ref{sec:BLR}.
For the fixed parameter estimate $\hat\theta$ we construct the nominal model 
\begin{equation}
	\label{eq:sysInfNonlin}
	\Mh: \left\{ \begin{array}{ll}
		\hat x_{k+1} &= \hat\theta\T f(\hat x_k, \hat u_k) + \hat w_k,\\
		\hat y_k &= \hat h(\hat x_k),
	\end{array} \right.
\end{equation}
where we abbreviate $\Mh:=\M(\hat{\theta})$.
%\pdfmargincomment{[Birgit:] I am not a big fan of this abbreviation, since in $\M(theta)$ you clearly see that the model depends on $\theta$, however, in $M hat$ you do not see that this model also depends on the parameter $theta hat$. Perhaps we can use $M hat(theta hat)$ instead? Or do you want to make clear that M(theta) can be a set of models and M hat is just one model?}.
%\Oliver{Update the following based on the new covariance results.}
We assume that the noise $w_k,\hat w_k\sim \mathcal N(\cdotx|0,\Sigma)$ has Gaussian distribution with a full rank covariance matrix $\Sigma\in\mathbb{R}^{n\times n}$ and $\hat h\equiv h$.
%We denote by $\sigma_w^2$ and $I$ the noise variance and identity matrix, respectively. 
%Note that $f:=[f_1,\ldots,f_m]\T$ can be a comprehensive library of functions.

%\subsection{Robust parameter estimation}\label{sec:simRelNonLinEst}

%%\color{black}
%\begin{figure}[h]
%	\centering
%	%	\includegraphics[width=.7\columnwidth]{confSetSketch.png}
%	\begin{tikzpicture}[scale=.5]
%		\draw plot[smooth, tension=.8] coordinates {(-3.5,0.5) (-1,3) (4.5,3.5) (2,1) (1,-2) (-3,-1) (-3.5,0.5)};
%		
%		\node at (-2.5,-1) [black,right] {$\Theta$};
%		
%		\node at (0,1+.5) [black,right] {$\hat\theta$};
%		\fill (0,1) circle [radius=.1];
%		
%		\node at (1.8,2.5+.5) [cherryred,right] {$\theta^\ast$};
%		\fill[cherryred] (1.8,2.5) circle [radius=.1];
%	\end{tikzpicture}
%	\caption{The unknown true parameters $\theta^\ast$ are approximated with an estimate $\hat\theta$ and are robustly contained within a credible set $\Theta$ with a confidence of $(1-\alpha)$.}
%	\label{fig:confSetSketch}
%\end{figure}
%\color{blue!70!black}

%\subsection{Valid control refinement}
\smallskip
To find a valid control refinement for the systems $\M$ and $\Mh$ in \eqref{eq:sysDataGenNonlin} and \eqref{eq:sysInfNonlin}, respectively, we first write their transition kernels.
The probability of transitioning from a state $x$ with input $u$ to a state $x^+\in S\subset\X$ is given by
\begin{equation*}
	\P(x^+\in S|x,u) = \int_{S} \Tr(dx^+|x,u),
\end{equation*}
where the kernel $\Tr$ is given by
\begin{equation}
	\begin{split}
		\Tr(\dxp|x,u;\theta) &= \mathcal N(\dxp|\theta\T f(x, u), \Sigma),\\
		&	= \int_w \delta_{\theta\T f(x, u)+w}(\dxp) \mathcal N(dw|0, \Sigma).
	\end{split}\label{eq:tr_nonlin}
\end{equation}
%\pdfmargincomment{Birgit: Also explain what Sigma $\Sigma$ itself represents. This has not been introduced in the paper yet. Same holds for $m$. Also make sure that it is not an issue that in  the beginning of this section (IVA) you denote  the variance as $\sigma^2_wI$.}
Similarly, the stochastic transition kernel of $\Mh$ is
\begin{align}
	\begin{split}
		\hat\Tr(\dxhp|\hat x,\hat u)& = \mathcal N(\dxhp|\hat \theta\T f(\hat x, \hat u), \Sigma),\\
		&	= \int_{\hat w} \delta_{\hat \theta\T f(\hat x,\hat u)+\hat w}(\dxhp) \mathcal N(d\hat w|0,\Sigma).
	\end{split}
	\label{eq:tr_hat}
%	\fhdel{\dxhp|\hat w}&:=\delta_{\hat \theta\T f(x, u)+\hat w}(\dxhp).
%	\label{eq:fhdel_nonlin}
\end{align}
%Note that $\fdel{\cdot}$ and $\fhdel{\cdot}$ are dependent on $x,u$, and $\hat x,\hat u$, respectively, omitted here for conciseness.

We can rewrite the model in Eq.~\eqref{eq:sysDataGenNonlin} as
\begin{equation*}
	x_{k+1} =\underbrace{\hat\theta\T f(x_k, u_k)}_{\text{nominal dynamics}} + \underbrace{(\theta-\hat\theta)\T f(x_k, u_k)+w_k}_{\text{disturbance}}.
\end{equation*}
Note that the disturbance part consists of a deviation caused by the unknown parameters $\theta$ and a deviation caused by the noise $w_k$.
Let us assume that we can bound the former as 
\begin{equation*}
	%\label{eq:bound_offset}
	\|(\theta-\hat\theta)\T f(x, u)\|\leq d(x, u),\quad  \forall x, u,  \theta.%\pdfmargincomment{Comma before "for all" symbol?}
	%	\label{eq:disturbance_bound}
\end{equation*}
%\color{blue!70!black}
% \Oliver{Why do we use the subscript $k$ for the interface function but not state mapping, for example?}
Using the interface function $\ac{}\sim\InF(\cdotx|\hat x, x, \hat u):=\ach{}$ and the noise coupling
\begin{align}
	& \hat w \equiv  \gamma(x, u,\theta;\hat\theta)+w, \,\,\text{ with }\nonumber\\
	& \gamma(x, u,\theta;\hat\theta) :=
	(\theta-\hat\theta)\T f(x,u),\label{eq:gamma_paraId_nonlin}
\end{align}
we get the state mapping
\begin{equation}
%	\label{eq:statemap_paraId_nonlin}
\label{eq:statemap_nonlin}
	\xhp = \xp+\hat\theta\T (f(\hat x,\hat u)-f(x,u)),
\end{equation}
that is both deterministic and not dependent on $\theta$, i.e., 
\begin{equation*}
%	\label{eq:statemap_nonlin}
	\xhp \sim \Refi(\cdotx |\xh{}, \x{}, \xp,\ach{}) := \delta_{\xp+\hat\theta\T\! (f(\hat x,\hat u)-f(x,u))}(d\xhp),
\end{equation*}
To establish an SSR $\widehat{\M}\preceq^{\delta}_\eps\M(\theta)$, we select the identity relation
\begin{equation}
	\label{eq:relation_nonlin}
	\R:=\{ (\hat x,x)\in\Xh\times\X \given x=\hat x\}.
\end{equation}
Condition (a) of Definition~\ref{def:subsim} holds by setting the initial states $\hat x_{0} = x_{0}$. Condition (c) is satisfied with $\eps=0$ since both systems use the same output mapping. For condition (b), we define the sub-probability coupling $\Wt(\cdotx|\theta)$ over $\R$:
% \Oliver{I think I would add the dependence of $\gamma(\theta)$ and say that dropped 'some' arguments of X and X to lighten...}
{\allowdisplaybreaks
\begin{align}
	%	\begin{split}
		&\Wt(d\xhp\!\times d\xp|\theta)\!=\!\!  \int_{\hat w}\int_w \!
		\delta_{\hat \theta\T\! f(\xh{}, \uh{})+\hat w}(\dxhp)
		\delta_{\theta\T\! f(x, u)+w}(\dxp)\nonumber \\
		&\hspace{0pt}\times \delta_{\offset(\theta)+w}(d\hat w) \min\{\mathcal N(dw|0,\Sigma),\mathcal N(dw|-\offset(\theta),\Sigma)\},
		\label{eq:subprobcoup_nonlin}
		%	\end{split}
\end{align}
}
that takes the minimum of two probability measures. Note that we dropped some arguments of $\Wt$ and $\offset$ to lighten the notation.

%\Oliver{Update the theorem and proof based on new covariance.}
\begin{theorem}[Establishing an SSR]
	\label{thm:paraId_nonlin}
	Let data $\mathcal{D}$ be given.
	Then, the nonlinear models \eqref{eq:sysDataGenNonlin}-\eqref{eq:sysInfNonlin} are in an SSR
	$\widehat{\M}\preceq^{\delta}_\eps\M(\theta)$
%	, where $\hat\theta=[m_{N,1},\ldots,m_{N,n}]\T$  \pdfmargincomment{[Birgit:] You can move this (theta hat is the mean you obtain from the data) to somewhere before (just before this Theorem is fine, just to clarify that you have now obtained an abstract model, that is the nominal parameter obtained from data). This should be clear before you introduce the Theorem. }% with \eqref{eq:mean}
	with the interface function $\ac{k} = \ach{k}$, relation \eqref{eq:relation_nonlin}, and sub-probability kernel \eqref{eq:subprobcoup_nonlin}.
	For a given confidence level $(1-\alpha)$ 
%	 \pdfmargincomment{[Birgit:] It is not clear to me why this confidence level is given and where you need this. I did not see this in the equations before. Or I missed it and you should emphasize it more.} 
	 the state mapping \eqref{eq:statemap_nonlin} defines a valid control refinement with $\eps=0$ and
	\begin{align}
		\delta(\hat x, \hat u)= 1-2\cdot\mathrm{cdf}\left\lbrace -\frac{\sqrt{r}}{2}\norm{f(\hat x,\hat u)}
			\right\rbrace,\label{eq:delta_ddnonlinear}\\
			r:=\norm{\Sigma^{-1}}\norm{\Sigma_{N}} \cdot n\cdot\chi^{-1}(1-\alpha| n)\nonumber,
	\end{align}
	where $\cdf{\cdotx}$ is the cumulative distribution function of a Gaussian distribution, i.e., $\cdf{\zeta}:= \int_{-\infty}^\zeta \frac{1}{\sqrt{2\pi}}\exp(-\beta^2/2)d\beta$, covariance matrix $\Sigma_{N}$ via \eqref{eq:covariance}, and $\Sigma$ is the covariance of the noise in \eqref{eq:sysDataGenNonlin}.
	%radius $r^2:=\new{n\cdot}\chi^{-1}(1-\alpha| n)$, and $\lambda_{max}(\Sigma)$ is the maximum eigenvalue of $\Sigma$.
%	, and $\Theta$
%	, $\offset$ 
%	as defined in \eqref{eq:confidenceSetNL}.
%	 and \eqref{eq:gamma_paraId_nonlin}, respectively.
%\Birgit{Not a fan of the formulation of this Theorem. How do you associate values of $\epsilon$ and $\delta$ to a control refinement? Why not prove that If $\epsilon =0$ and $\delta = ...$, then we get an SSR and a valid control refinement? I reformulated this Theorem and proof in my thesis.}
	\end{theorem}
%	\begin{proof}
		The proof of Theorem~\ref{thm:paraId_nonlin} has been deferred to App.~\ref{app:proofSSRnonlinear}.
%		Analogously, we get $\gamma(x, u,\theta;\hat\theta)=0$ for all $\theta\in\Theta$, $(x,u)\in\X\times\U$. Using $\Theta=\{\theta\in\R^\mathfrak{p}\given p(\theta_i\given Y_i) \geq \alpha, \forall i=1,\ldots,n\}$ and recognizing that $\cdf{}$ is a monotone function
%		we obtain \eqref{eq:delta_ddnonlinear}.
%	\end{proof}
	 Note that this theorem gives a $\delta$ that depends on $(\xh{k},\ach{k})$,
	 thus providing a less conservative result compared to a constant global $\delta$.
	 %\Oliver{My original intent why I split the dimensions in \eqref{eq:delta_ddnonlinear} was that that is what I used before where the noise was without covariance. Technically there exists a less conservative $\delta$ where you look at every dimension individually if the noise has no covariance. That is the one I used in the case studies. Should I add a remark about this?}
%	  as compared to a constant global $\delta$, which would be more conservative.
%	 \pdfmargincomment{Check which one is correct and also make sure that correct in figures.}. 
%	 The upper bound of this quantity over $\X\times\U$ could be easily obtained to get a constant global $\delta$, which is more conservative.  \Oliver{..}
%	 \pdfmargincomment{[Birgit:] Why would you want to do this if this is more conservative? Is it easier/faster to use in the next computations?}
	
\begin{figure}
		\centering
		\includegraphics[width = 0.85\columnwidth]{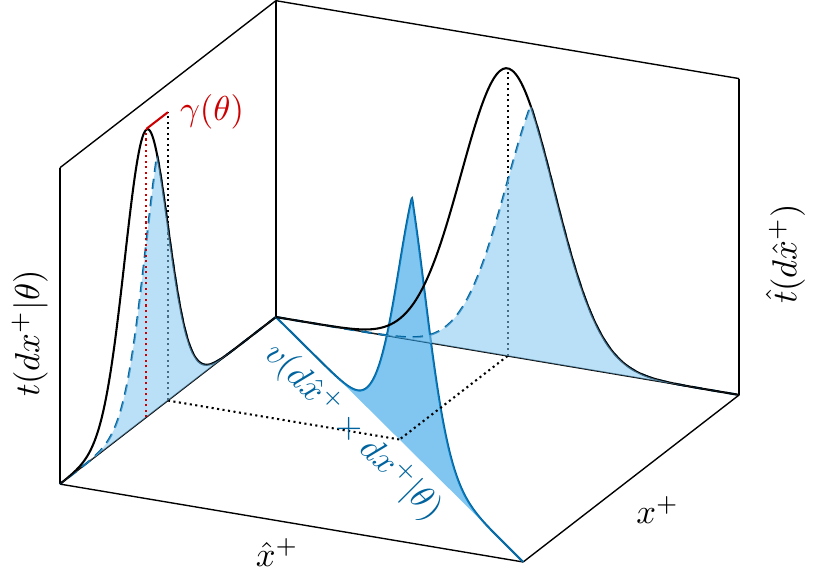}
		\caption{2D representation of the sub-probability coupling $\W$ for the stochastic transition kernels $\hat t$ and $t$ given some $(\xh{},\x{},\ach{})$. Displayed is the coupling over the product space $\Xh\times\X$ as well as its marginals (cf. Definition~\ref{def:submeasure_lifting}(b)-(c)). 
			%		\Oliver{Could also add residual coupling probability if desired.}
}
		\label{fig:subprobcoup}
  
	\end{figure}

%	\color{blue!70!black}
	\medskip
	\noindent\textbf{Visualization of the sub-probability coupling:}
	We can reduce Eq.~\eqref{eq:subprobcoup_nonlin} using the definition of $\offset$ in \eqref{eq:gamma_paraId_nonlin} to
	\begin{align*}
		&\Wt(d\xhp\times d\xp|\theta) = \int_{\hat w} \delta_{\xhp}(d\xp)
		\delta_{\hat \theta\T\! f(x, u)+\hat w}(\dxhp)\\
		&\hspace{90pt}\times \min\{\mathcal N(d\hat w|\offset(\theta),\Sigma),\mathcal N(d\hat w|0,\Sigma)\},
	\end{align*}
	which is
	 illustrated in Fig.~\ref{fig:subprobcoup} in the 2D plane.
	This coupling is constructed based on the sub-probability coupling of the two noise distributions
	\begin{align*}
		&\bar{{w}}( d\hat w\times d w|\theta) := \delta_{\offset(\theta)+w}(d\hat w) \\
		&\hspace{80pt}\times \min\{\mathcal N(dw|0,\Sigma),\mathcal N(dw|-\offset(\theta),\Sigma)\},
	\end{align*}
	%\Oliver{Deal with bold notation.}
	where $\offset$ quantifies the offset between the two distributions. 
	%The noise coupling is illustrated in Fig.~\ref{fig:gaussplot} in two dimensional space.
%	The resulting sub-probability coupling \eqref{eq:subprobcoup_nonlin} is
	%\input{Gaussplot}
	\color{black}
	\medskip
%	\color{blue!70!black}
%	\begin{remark}
	
			With the approach presented in this paper, parametric uncertainty is compensated by shifting the noise distributions relative to each other  by the offset $\offset$ in the sub-probability coupling. Hence, parametric uncertainty can only be compensated on state variables that are perturbed with noise.
%			 \pdfmargincomment{[Birgit:] I am not sure if I would make this a remark, or just as 'normal' text.}
%			\pdfmargincomment{Birgit: This is due to the fact that we need Bw to be invertible. So the pseudo-inverse does not work, right? This is confusing, since earlier you mention that if Bw is not invertible, then we take the pseudo-inverse.}
%	\end{remark}
%	\begin{remark}[Bounded support]
%			In some cases it is of interest to consider stochastic systems with bounded support, e.g., by truncating the Gaussian noise distributions or uniform support.
%			Whilst for unbounded support we require $(\Xh,\X)$ to be unbounded (cf. App.~\ref{app:proofSSRnonlinear}), this restriction can be relaxed for bounded support, allowing them to be bounded sets as well.
%	\end{remark} % Version with regression for input state data

%%%%%%%%%%%%%%%%%%%%%%%%%%%%%%%%%%%%%%%%%%%%%%%%%%%%%%%%%%%%%%%%%%%%%%%%%%%%%%%%
\section{Case studies}
\label{sec:case_study}
%Now that we established the theoretical foundations of designing a valid control refinement for a parametric stochastic system, thus able to design a controller such that the controlled system satisfies a given temporal logic specification with at least probability $p$ (c.f. Problem~\ref{prob:prob1}),
We demonstrate the effectiveness of our approach on a linear system and the nonlinear discrete-time version of the Van der Pol Oscillator.
\subsection{Linear system with complex specification}
%\red{Need to update all results to without measurement noise!}

\begin{figure}
	\centering
	\begin{tikzpicture}
		\node[state, initial] (q0) {$q_0$};
		\node[state, right of=q0, xshift=1.5cm] (q1) {$q_1$};
		\node[state, accepting, right of=q1, xshift=1.5cm] (q2) {$q_F$};
		\path[-stealth]
		(q0) edge[loop above] node{$\neg P_1$} (q0)
		(q0) edge[below] node{$ P_1$} (q1)
		(q1) edge[loop above] node{$\neg P_2 \wedge\neg P_3$} (q1)
		(q1) edge[bend right, above] node{$P_2$} (q0)
		(q1) edge[above] node{$P_3 \wedge\neg P_2$} (q2)
		(q2) edge[loop above] node{$1$} (q2)
		(q0) edge[bend right, below] node{$P_1\wedge P_3$} (q2)
		;
	\end{tikzpicture}
	\caption{DFA corresponding to the specification of the package delivery problem, where $q_F$ is the accepting location.}
	\label{fig:dfa_delivery}
\end{figure}
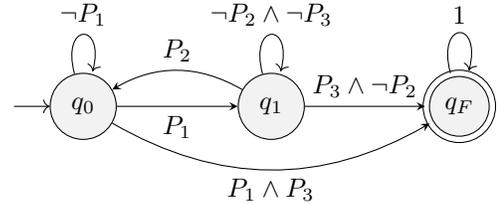

Consider the uncertain linear system in \eqref{eq:sysDataGenNonlin} with
$f(x_k, u_k) = \left[ x_k,u_k \right]\T$, $h(x_k)=x_k$,
%, where $A:=[\theta_1,0.3;0.2,0.7]$, $B:=[\theta_2,0;0,1.4]$, $\sigma_w^2:=1$, and $C:=[1,0;0,1]$, describing an agent translating in a 2D space.
and $\Sigma=\sqrt{0.1}I_2$, describing an agent moving in a 2D space.
% and $v_k\sim \mathcal N(0,\sqrt{0.1}I)$
Define the state space $\X = [-6,6]^2\subset\mathbb{R}^2$, input space $\U = [-1,1]^2$, and output space $\Y = \X$.
The goal is to compute a controller for a package delivery problem.
For this, the controller should navigate the agent to pick up a parcel at region $P_1$ and deliver it to target region $P_3$. If the agent passes the region $P_2$ on its path, it loses the package and must pick up a new one at $P_1$. This behavior is captured by the specification $\psi = \Event(P_1 \andltl (\notltl P_2 \Until P_3))$. Note that the corresponding DFA presented in Fig.~\ref{fig:dfa_delivery} contains a backward loop from location $q_1$ and is hence not expressible as a reach-avoid specification over the state space only.
Here, we consider
$P_1 =[3,6]\times[-2.5,1]$, $P_2 =[-1,1]\times[-4,3]$,
and $P_3 =[-6,-3]\times[-6,-3]$.

\begin{figure}
	\centering
	\includegraphics[width=0.85\columnwidth]{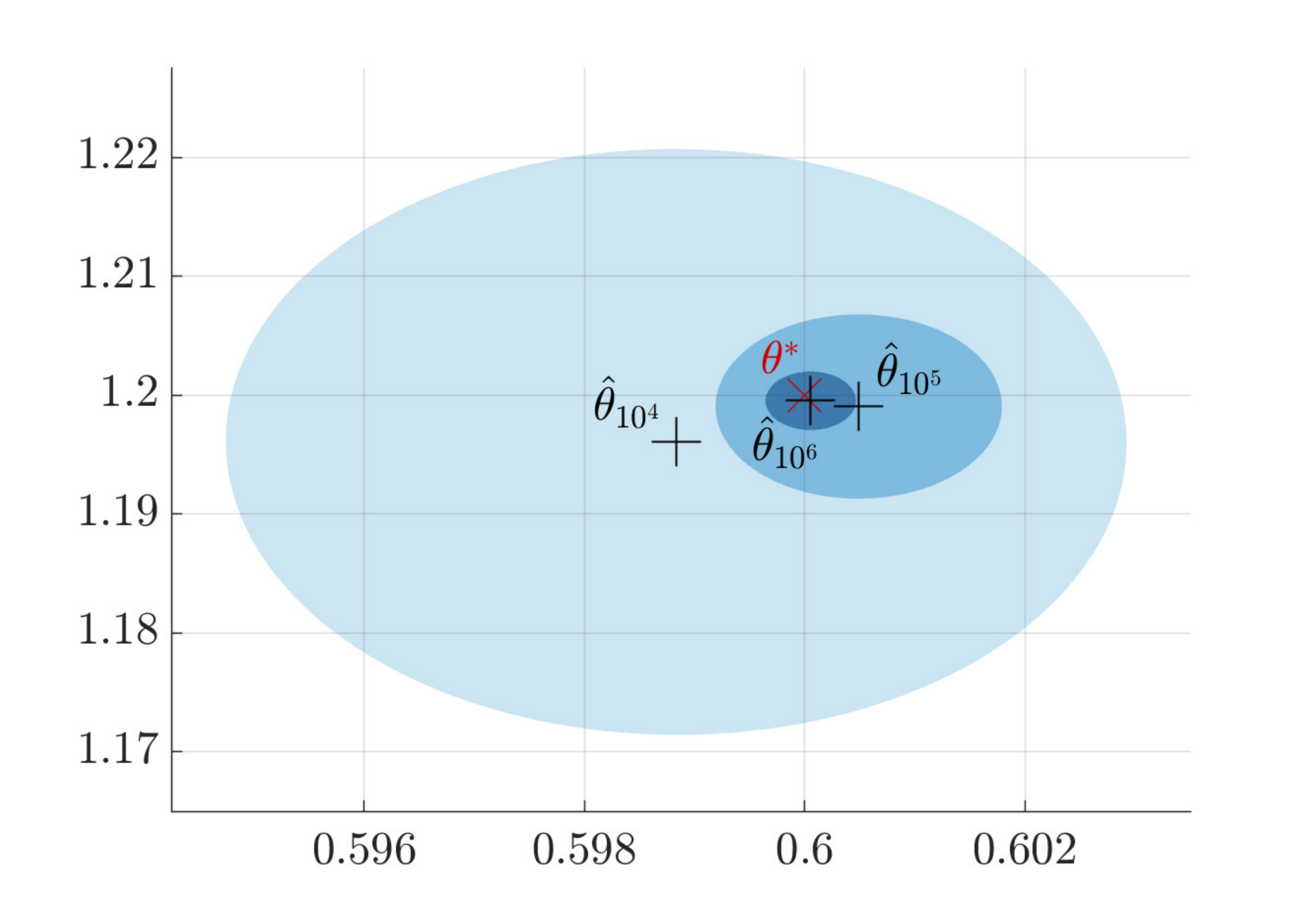}
	\caption{Contracting credible set $\Theta$ and parameter estimate $\hat\theta$ for data sizes $N=10^4,10^5,10^6$ (light to dark blue) with an associated confidence bound of $(1-\alpha)=0.9$.}
	\label{fig:confSetDelivery}
\end{figure}

\begin{figure}
	\centering
	\includegraphics[width=0.85\columnwidth]{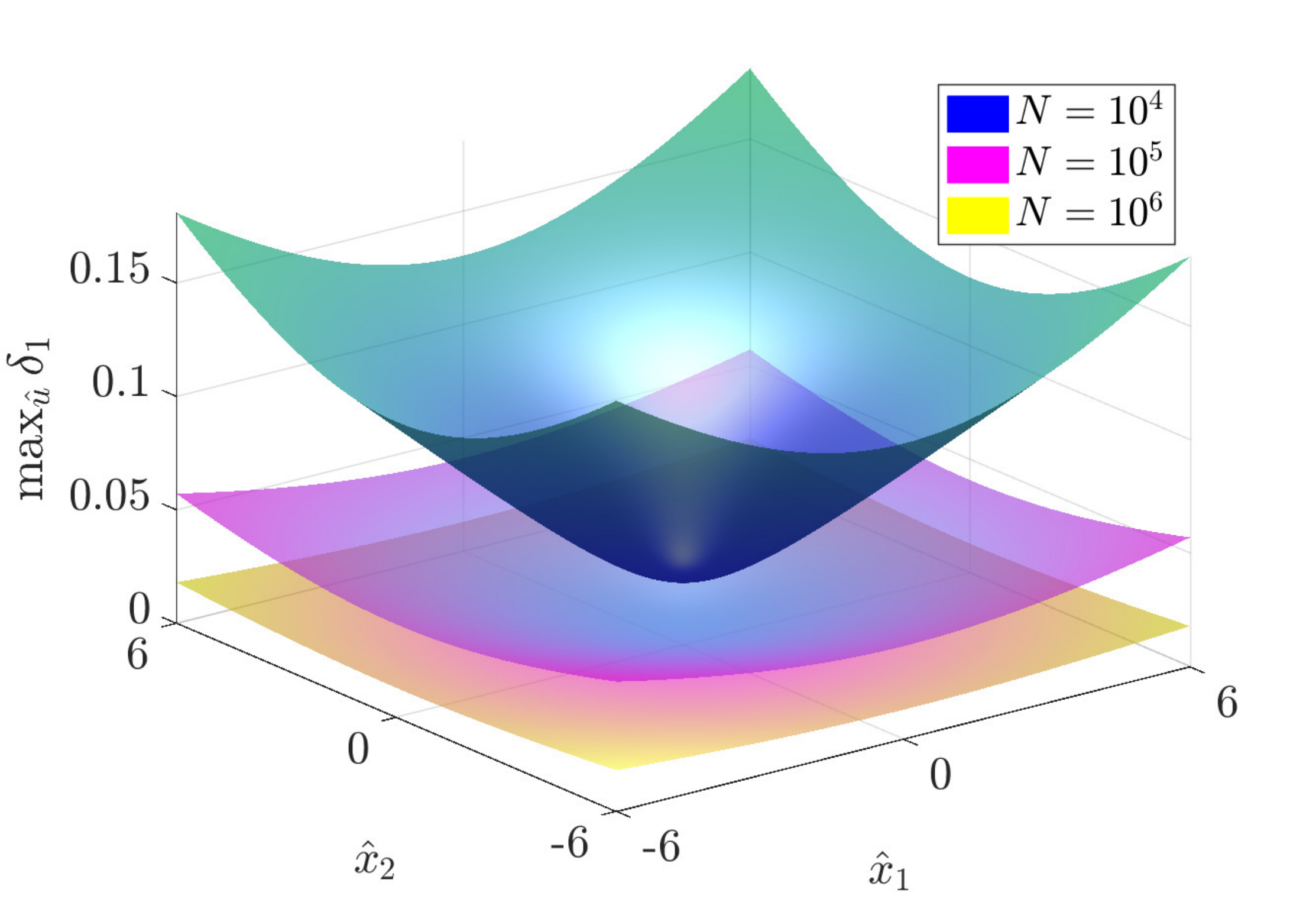}
	%	\vspace{0.1cm}
	%	\includegraphics[width=\columnwidth]{resultsDelivery_1e5.eps}\\
	%	\vspace{0.1cm}
	%	\includegraphics[width=\columnwidth]{resultsDelivery_1e6.eps}
	\caption{Maximum value of $\delta_1$ with respect to $\hat u$ as a function of the initial state $(\hat x_1,\hat x_2)$ for $N=10^4,10^5,10^6$ (top to bottom) for the package delivery case study with a confidence bound of $(1-\alpha)=0.9$.}
	\label{fig:delta1_Delivery}
 \vspace{-0.3cm}
\end{figure}

\begin{figure}
		\centering
		\includegraphics[width=.8\columnwidth]{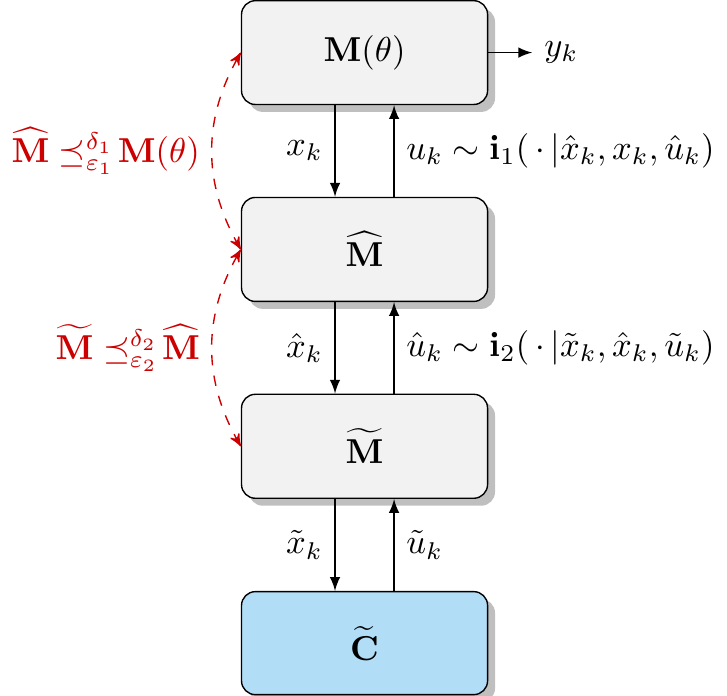}
		\caption{Two layers of abstraction with two simulation relations. The first relation is an SSR between the set of feasible models $\{\M(\theta)|\theta\in\Theta\}$ and the continuous-space nominal model $\Mh$, compensating the parametric uncertainty. The second relation is an approximate simulation relation between the nominal model and its discrete abstraction as in \cite{haesaert2020robust}, compensating the discretization error.
%		 \Sadegh{change $t$ to $k$, always pay attention to the time index. Also add index 1 and 2 for the two interface functions.}
	 }
		\label{fig:SimRelsSetup}
\end{figure}

\begin{figure}
	\centering
	\includegraphics[width=0.9\columnwidth]{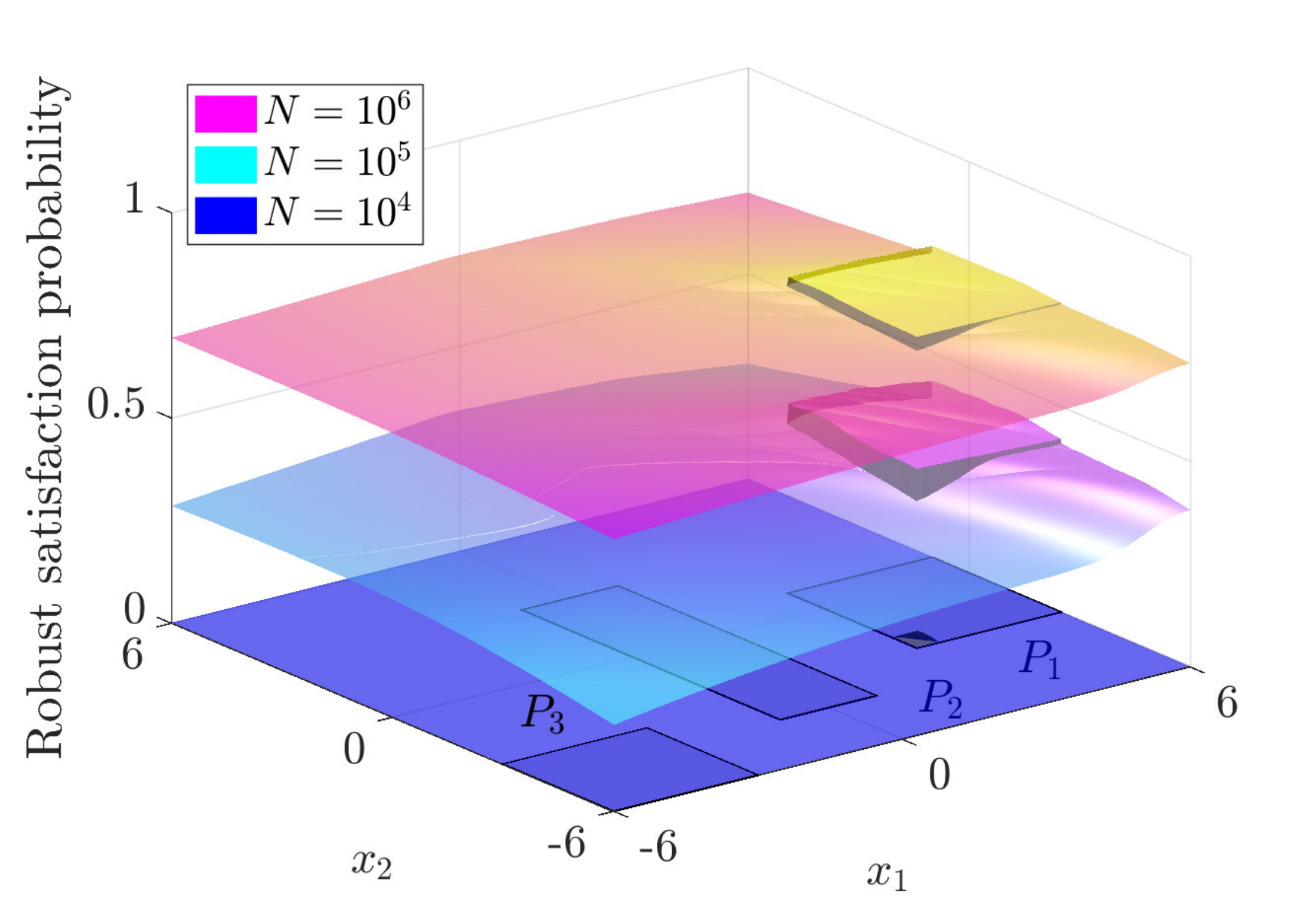}
	%	\vspace{0.1cm}
	%	\includegraphics[width=\columnwidth]{resultsDelivery_1e5.eps}\\
	%	\vspace{0.1cm}
	%	\includegraphics[width=\columnwidth]{resultsDelivery_1e6.eps}
	\caption{Robust satisfaction probability as a function of the initial state for $N=10^4,10^5,10^6$ (bottom to top) for the package delivery case study with a confidence bound of $(1-\alpha)=0.9$.}
	\label{fig:satProb_Delivery}
 \vspace{-0.3cm}
\end{figure}

\begin{figure}
	\centering
 	\includegraphics[width=0.85\columnwidth]{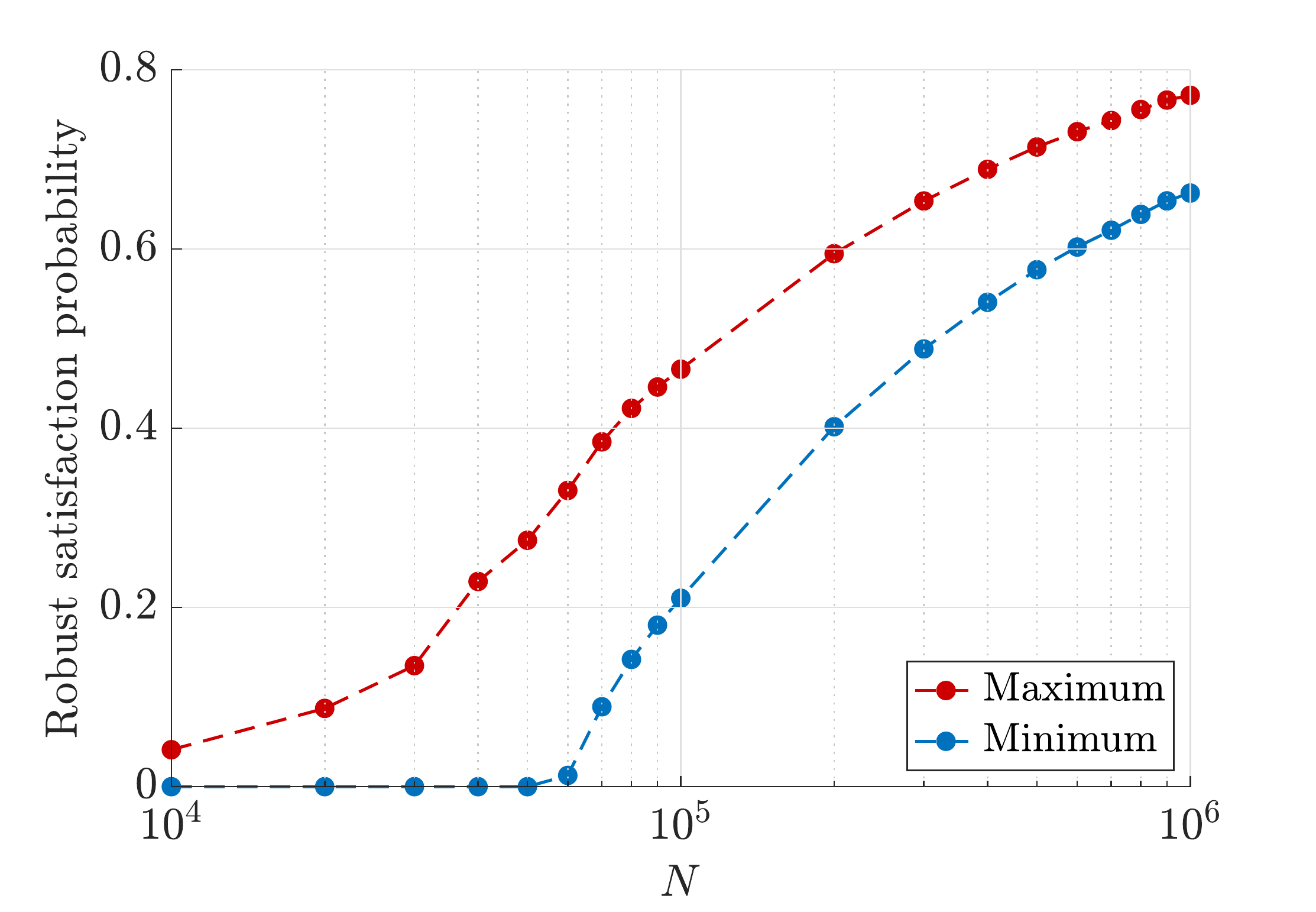}
	\caption{Maximum and minimum of the robust satisfaction probability with respect to the initial state as a function of the size of the dataset $N$.
 % \Oliver{Re-render. Font in legend.}
%		\Oliver{Could be nice to add the true satisfaction probability via MCMC as well! However, this would be for a single initial point.}
%	\Oliver{Add more data points.}
%	 \Sadegh{Add legend for the figure. Put legend max and min for the two graphs. The same thing with figure 8 and 10, put N=... for each plot.}
	}
	\label{fig:satisfProbPkgDelBounds}
\end{figure}

For estimating the parameters, we sample uniform excitations u from $\U$, set the prior distribution to $\mathcal N(\theta_j|0,10I_4)$, $j\in\{1, 2\}$, and draw $N$ data points from the true system with parametrization $\theta^\ast = \left[0.6,0.3,1.2,0;0.2,0.7,0,1.4\right]\T$.
Using Bayesian linear regression we obtain an estimate $\hat\theta$ and a credible set $\Theta$
(displayed for $(\theta_{11}, \theta_{33})$ in Fig.~\ref{fig:confSetDelivery})
for a lower confidence bound of 0.9 as outlined in Sec.~\ref{sec:BLR}.
Note that the credible set is contracting for increasing amounts of data.
We select $\Mh=\M(\hat\theta)$ in \eqref{eq:sysInfNonlin} with $\hat h\equiv h$ to be the nominal model and get $\eps_1=0$.
$\delta_1$ is computed using Eq.~\eqref{eq:delta_ddnonlinear} as a function of the state and input. The value of $\max_{\hat u}\delta_1$ is displayed in Fig.~\ref{fig:delta1_Delivery}.
Note that $\delta_1$ grows rapidly for states away from the central region, thus a global upper bound on $\delta_1$ would result in poor lower bounds on the satisfaction probability.
The following steps are performed using the toolbox SySCoRe \cite{vanhuijgevoort2023syscore}.
We compute a second abstract model $\Mt$ by discretizing the space of $\Mh$,  i.e., $\Xh$ into $2500^2$ and $\Uh$ into $5$ partitions, and
%. Then, we use the results of \cite{haesaert2020robust} to 
get ${\Mt}\preceq^{\delta_2}_{\eps_2}\Mh$ with $\eps_2=0.034$ and $\delta_2=0.004$. Thus, using the transitivity property in Theorem~\ref{thm:transitive}, we have $\Mt\preceq^{\delta}_{\eps}\M(\theta)$ with $\delta = \delta_1+\delta_2$ and $\eps = \eps_1 + \eps_2$ (cf. Fig.~\ref{fig:SimRelsSetup}).
The robust probability of satisfying the specification with this $\eps$ and $\delta$ is computed based on Theorem~\ref{thm:satProb} and is depicted in Fig.~\ref{fig:satProb_Delivery} as a function of the initial state of $\M(\theta)$.
Fig.~\ref{fig:satisfProbPkgDelBounds} shows the maximum and minimum of the robust satisfaction probability with respect to the initial condition for increasing data sizes.
%Note that the offset between the maximum and 1 for $N\rightarrow\infty$ converges to a value dependent on $(\varepsilon_1,\delta_2,\varepsilon_{2})$.
This shows that the robust satisfaction probability is improved by increasing the amount of data.

\begin{figure}
	\centering
	\includegraphics[width=0.85\columnwidth]{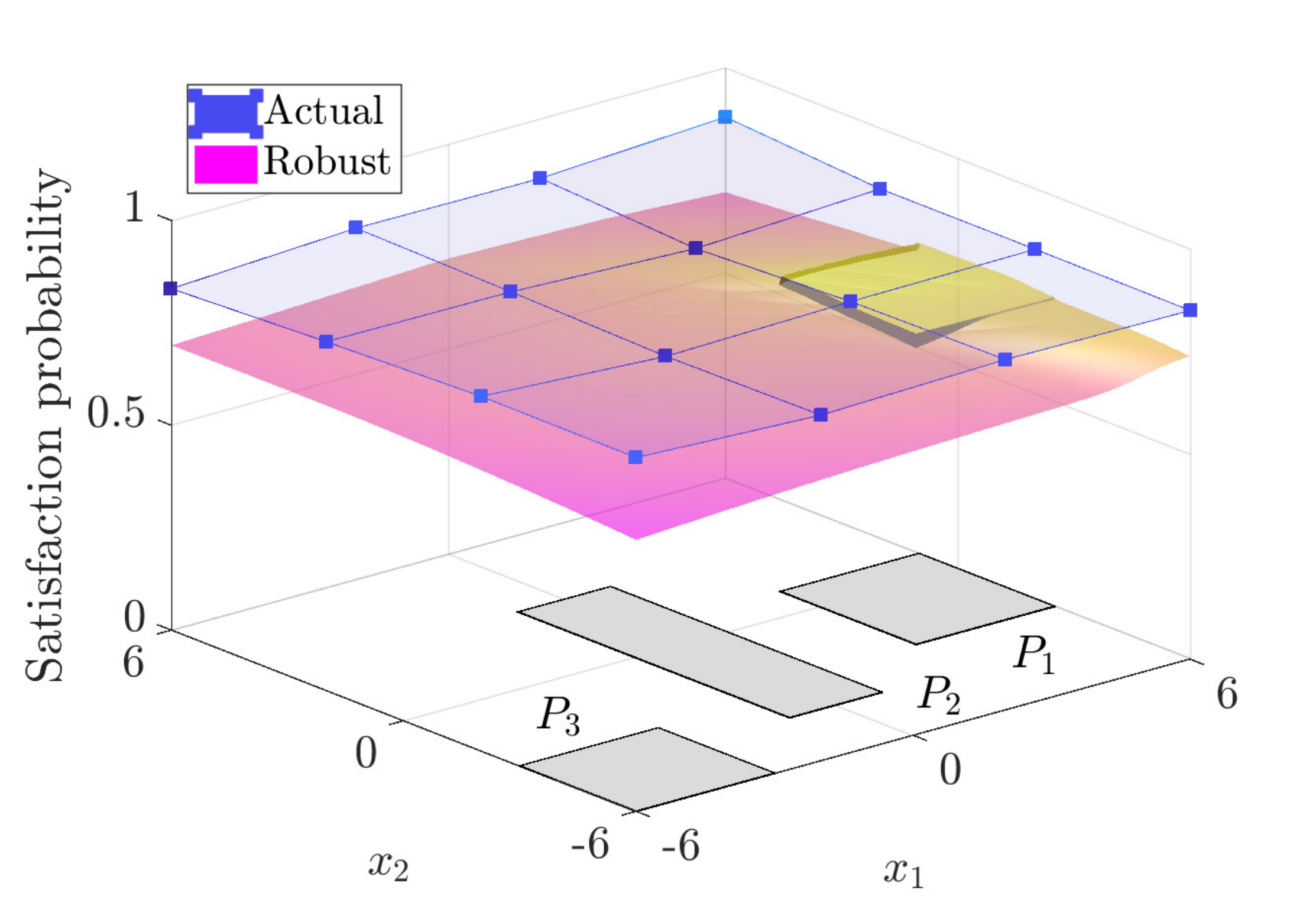}
	%	\vspace{0.1cm}
	%	\includegraphics[width=\columnwidth]{resultsDelivery_1e5.eps}\\
	%	\vspace{0.1cm}
	%	\includegraphics[width=\columnwidth]{resultsDelivery_1e6.eps}
	\caption{Comparison of the robust satisfaction probability (in red) and the actual satisfaction probability estimated via Monte Carlo simulation as a function of the initial state for $N=10^6$ for the package delivery case study.
%		\Sadegh{Update the figure and put the actual satisfaction probability. Was it always one? Run larger number of simulations.}
	%We have used the same confidence bound $(1-\alpha)=0.9$ for both the Monte Carlo estimation and our approach.
	}
	\label{fig:satProb_Delivery_MC}
 \vspace{-0.3cm}
\end{figure}

To visualize the conservatism in the computation of the robust satisfaction probability, we take the controller synthesized by our approach with data size $N=10^6$ and apply it to the system. We compare the true satisfaction probability estimated using Monte Carlo simulation with the robust satisfaction probability for several representative initial states. We run 500 simulations per initial state with a maximum length of 60 time steps.
%\Oliver{Is this correct numbers?! YES, checked 27.12.23}
	%We compare the robust satisfaction probability obtained with $N=10^6$ against the satisfaction probability of the system under the controller obtained  approximated using Monte Carlo simulation
	%	\footnote{$10^\red{5}$ samples per representative point with max. length of 10 time steps.}
	Fig.~\ref{fig:satProb_Delivery_MC} shows the robust satisfaction probability (in red) alongside the actual satisfaction probability (in blue) estimated via Monte Carlo simulation using Chebychev's inequality with a confidence level of 90\%.
Note that even for the case of $N\rightarrow\infty$ the actual satisfaction probability can not generally attain the value of $1$ due to the intrinsic stochastic effects of the system. Only the uncertainty associated with the model parameters 
%	\Birgit{I prefer \emph{model parameters} instead of \emph{system knowledge}} 
	can be reduced by sampling more data.
 
\begin{figure}
	\centering
	\includegraphics[width=.65\columnwidth]{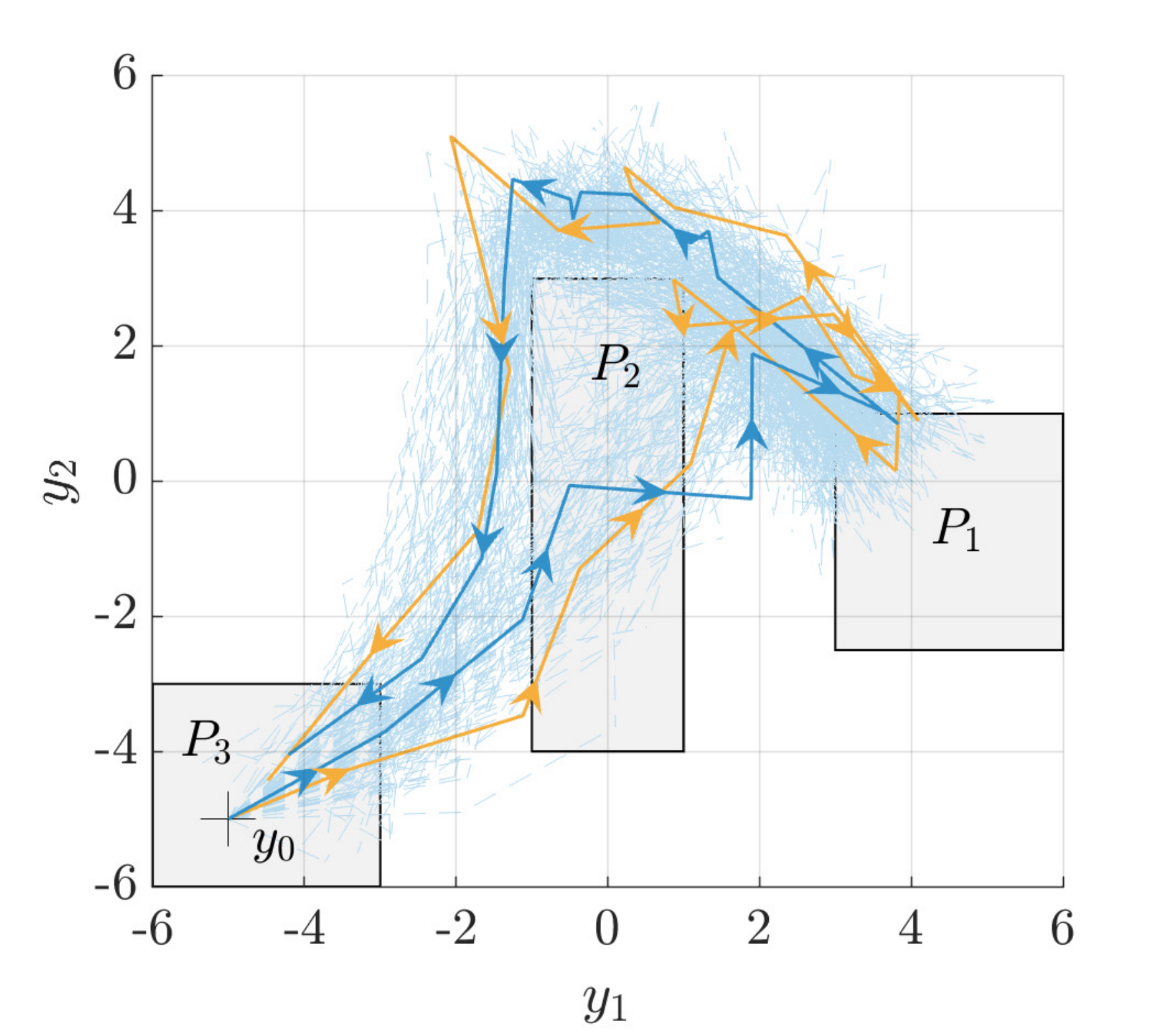}
	\caption{100 example trajectories of the package delivery case study with two trajectories satisfying the specification highlighted. Note that the yellow agent loses a parcel in $P_2$ and has to take an additional loop.}
	\label{fig:delivery_Sim}
 \vspace{-0.3cm}
\end{figure}

%\new{Results for varying amounts of data are given in Table~\ref{tbl:resultsDelivery}.}
In Fig.~\ref{fig:delivery_Sim}, 100 example trajectories of the controlled system initialized at $x_0=[-5, -5]\T$
%for $N=60\red{?!}$\pdfmargincomment{Two uses for $N$!} time steps
are shown, 99 of which satisfy the specification. This emphasizes the capability of the proposed approach of being able to synthesize controllers guaranteeing satisfaction of complex temporal logic specifications over infinite-horizon runs.

%\Oliver{Compare to Samual Goyan code based on interval MDP. Ours should be better}
\begin{comment}
\new{
Only few approaches with similar capabilities have been proposed.
The authors of \cite{Dutreix2022IMDP} design controllers for uncertain systems using \emph{interval Markov decision processes} (IMDP).
Whilst the IMDP is assumed to be given based on reachable set computations, the subsequent dynamic programming step involves finding so-called \emph{greatest permanent winning components} of the IMDP and computing bounds on the associated reachability probabilities.
The proposed iterative algorithm involves substantial optimizations and is highly computationally expensive, as demonstrated on a case study with monotone dynamics.
}
\end{comment}

%\begin{figure*}
%	\centering
%	\includegraphics[width=.32\textwidth]{resultsPkgDel_5e4.eps}
%	\hfill
%	\includegraphics[width=.32\textwidth]{resultsPkgDel_1e5.eps}
%	\hfill
%	\includegraphics[width=.32\textwidth]{resultsPkgDel_5e5.eps}
%	\caption{\red{Draft. Will be updated.} Lower bound on the satisfaction probability and values of $\delta_1$ as a functions of the initial state for $N=5\cdot10^4,10^5,5\cdot10^5$ (left to right) for the package delivery case study with a confidence of $1-\alpha=0.9$.}
%	\label{fig:delta1ANDsatProb_Delivery}
%\end{figure*}
%\new{\begin{remark}
%		If $\delta$ is dependent on both $x$ and $u$,
%%		, for example when $\gamma$ is input dependent,
%		a state-input-dependent upper bound $d(x,u)$ has to be calculated and $\delta$ is optimized w.r.t. $\mu$ when applying the $(\eps,\delta)$-robust operator in \eqref{eq:robustoperator}. \Oliver{Remove? I know that a reviewer was asking for CDC how this is computed and that for us $\delta(x,u)$.}
%\end{remark}}

\subsection{7-dimensional building automation system}\label{sec:bas}
We demonstrate the scalability of the presented approach on a 7D affine system of a building automation system, showcasing its compatibility with model-order reduction. The system dynamics can be rewritten similar to \eqref{eq:sysDataGenNonlin} as
\begin{equation*}
	\begin{array}{ll}
		x_{k+1} &= \begin{bmatrix}
		    \theta & A
		\end{bmatrix}\T f(x_k, u_k) + w_k,\\
		y_k &= \begin{bmatrix}
		    1&0&0&0&0&0&0
		\end{bmatrix} x_k,
	\end{array}
\end{equation*}
where $\theta\in\mathbb{R}^{8\times 2}$ is unknown and $f(x_k, u_k) = [x_{k};u_k]\T$, a matrix $A\in\mathbb{R}^{8\times 5}$,
and the noise distribution $w_k\sim \mathcal N(\cdotx|0,\Sigma)$ are known. It is worth mentioning that $\Sigma$ is non-diagonal.
The system performs on a state space $\X \subset \mathbb R^7$, input space $\U \subset\mathbb R$, and output space $\Y\subset \mathbb R$.
We refer to \cite{cauchi2018benchmarks} and \cite[Sec.~4.3]{vanhuijgevoort2023syscore} for more details. 
The objective is to synthesize a controller maintaining the temperature in the first zone of the building (represented by $y_k$) at $20\pm0.5^{\circ} C$ for $6$ consecutive time steps. 

\begin{figure}
	\centering
	\includegraphics[width=0.85\columnwidth]{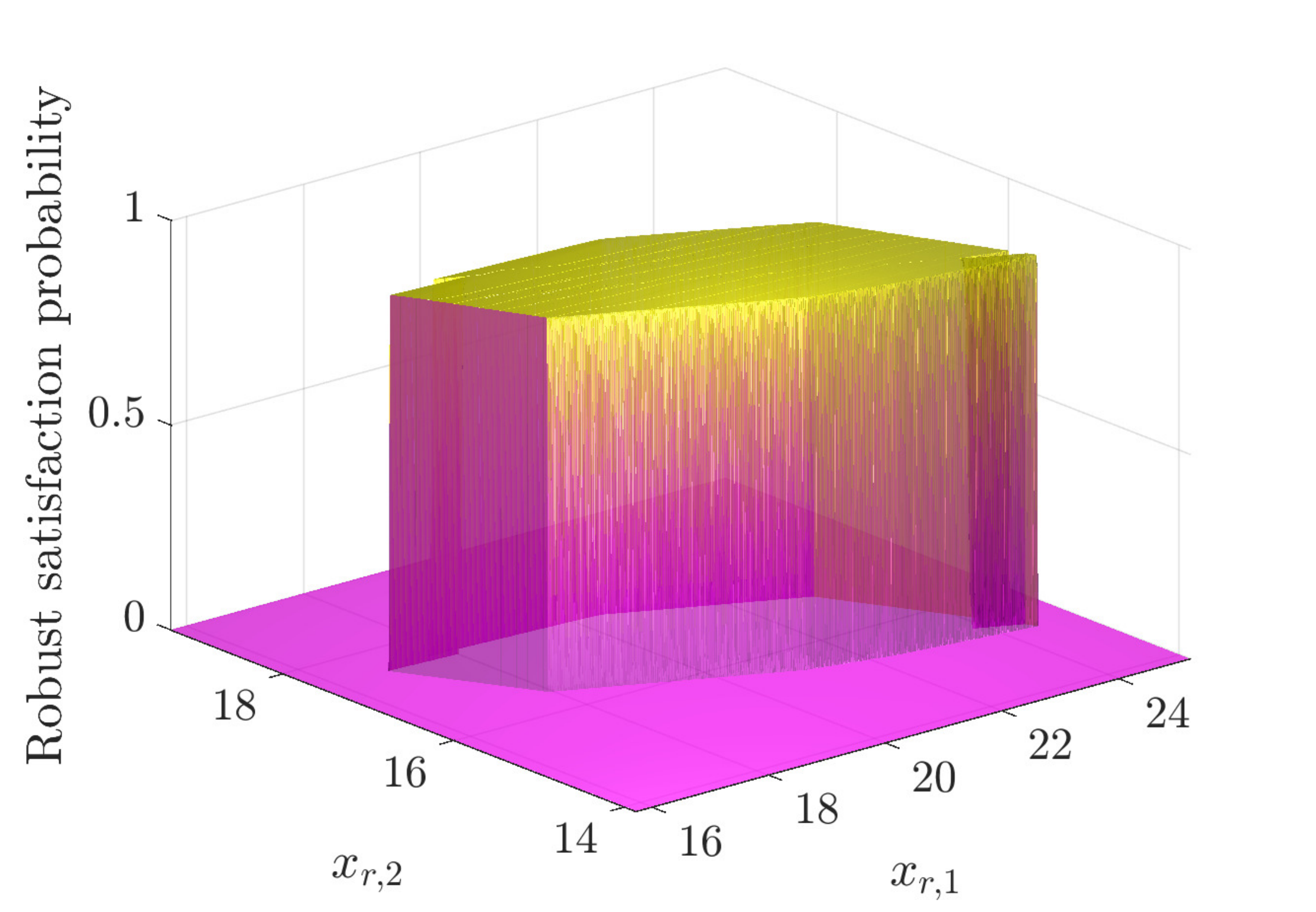}\\
	\caption{Robust satisfaction probability as a function of the reduced initial state for $N=10^6$ for the building automation system with a confidence bound of $(1-\alpha)=0.9$.}
	\label{fig:satProb_BAS}
 \vspace{-0.3cm}
\end{figure}

Following similar steps to the previous case study, we estimate the uncertain parameters $\theta$ by drawing $N=10^{6}$ data points from the true system.
Using the theoretical results from \cite{haesaert2017verification, VanHuijgevoort2020SimQuant}, we reduce the identified 7D system to a 2D system with the corresponding simulation relation.
We generate an abstraction of the reduced-order system by discretizing $\Xh$ into $3000^2$ and $\Uh$ into $3$ partitions and compute the corresponding error parameters to establish the simulation relation.
The robust probability of satisfying the specification is computed based on Theorem~\ref{thm:satProb} and is depicted in Fig.~\ref{fig:satProb_BAS} as a function of the initial state of the reduced system. Note that the results can be projected back onto the state space of the original 7D system.

\noindent\textbf{Comparison:} All steps required in obtaining the controller and robust satisfaction probability from the dataset $\mathcal{D}$ amount to less than $100$ seconds on an AMD Ryzen 5950X (single core used).
To put this into perspective with other available approaches, we consider an IMDP generated via GP regression as the state of the art \cite{Jackson2020safety,jiang2022safe}.
This approach is limited to systems with dimension-wise independent dynamics and establishes formal guarantees by leveraging concentration inequalities to obtain probabilistic error bounds for the predictions of the GP models \cite{Fiedler2021GP}.
However, not only is this approach not scalable as it is incompatible with model-order reduction and severely limited by the disadvantages of GPs themselves, but the resulting error bounds are extremely conservative.
To highlight this, consider a setup similar to ours, with $N=10^6$ data samples, confidence $(1-\alpha)=0.9$, and a state and input discretization of size $9^7$ and $3$, respectively, to obtain an abstraction of similar size to ours.
As exact inference via GPs has a computational complexity of $\mathcal{O}(N^3)$ \cite{rasmussen2006gaussian}, this renders obtaining accurate large-scale models computationally excruciating.
For sparse options, there are no results available that grant formal guarantees.
Provided we could obtain GP models with sample size $N$, the dimension-wise error bound $$\P\big(|\mu_{N,i}(x, u)-f_i(x,u; \theta^\ast)|\leq \mathcal{E}_i(x,u)\big)\geq1-\alpha,$$ via \cite[Proposition~2]{Fiedler2021GP} for dimensions $i\in\{1,\ldots, n\}$ can be extrapolated\footnote{We compute the average minimum error over $10$ runs for $N=1000,1500,\ldots,5000$ and extrapolate linearly.} to be of magnitude $[\min_{x,u}\mathcal{E}_i(x,u)]_{i=1}^n:=[3.1,3.0,3.4,3.3,3.2,3.5,3.4]$, where $\mu_{N,i}$ denotes the posterior mean function of the $i$th GP and $f:=[f_1;\ldots;f_n]$ is the true dynamics function as defined in \eqref{eq:data}.
Subsequently, constructing the IMDP from the GPs involves nonlinear optimizations w.r.t. $\X$ and the error bounds for a total of more than $68$ trillion possible transitions for each matrix. Storing one such dense matrix with double precision would take $500$TB of memory. 
In contrast, SySCoRe can leverage sparse matrices and tensor multiplications to circumvent this.
Using GPs with an active set of $2000$ data points, the same machine as before can compute $3.6$ transition optimizations per second, rendering the IMDP construction infeasible.
%computing the IMDP times out after \red{$24$ hours} with an estimated computation time of $\red{?}$ hours per transition probability matrix.
As the error bounds $\mathcal{E}_i$ are large in comparison to the size of the partitions and the noise variance, the approach yields the trivial lower bound of zero for the satisfaction probability.
This emphasizes the powerful capabilities of the proposed synthesis approach via probabilistic coupling relations w.r.t. both scalability and accuracy. 
% \red{discuss this comparison with Oliver, not happy with how things are set up. We should apply the GP approach to the whole case study, fix the size of the abstraction, and compare the computational time and the obtained results. We could also fix a lower bound on the probability for both approaches and compare the computational time and size of the abstraction.}
% \Oliver{Working on this.}

\subsection{Van der Pol Oscillator}\label{sec:vdp}
%Consider as a second example a forced, stochastically perturbed Van der Pol Oscillator in discrete time.
%\red{Revise!}
As a renown example of a nonlinear system, consider
the state evolution of the Van der Pol Oscillator in discrete time: 
%is given as
\begin{align*}
	%	\begin{split}
		x_{1,k+1} &= x_{1,k} + x_{2,k} \tau_s+w_{1,k},
%		\label{eq:vdp_dyn}
		\\
		x_{2,k+1} & = x_{2,k} + (-x_{1,k}+a(1-x_{1,k}^2)x_{2,k}) \tau+ u_k+w_{2,k},
		\nonumber
		%	\end{split}
\end{align*}
with parameter $a$ and sampling time $\tau_s = 0.1$.
%\color{blue!70!black}
The dynamics can be rewritten in the form of \eqref{eq:sysDataGenNonlin}, where
$f(x, u) = [x_{1}, x_{2}, x_{1}^2x_{2}, u]\T$,
$h(x)=x$,
and $\Sigma=0.2I_2$.
We consider the state space $\X = \mathbb R^2$, input space $\U = [-1,1]$, and output space $\Y=\X$. We want to design a controller such that the system remains inside region $P_S$ while reaching target region $P_T$, written as $\psi = P_S \Until P_T$. The regions are $P_S=[-3,3]^2$ and $P_T = [2,3]\times[-1,1]$.

\begin{figure}
	\centering
	\includegraphics[width=0.85\columnwidth]{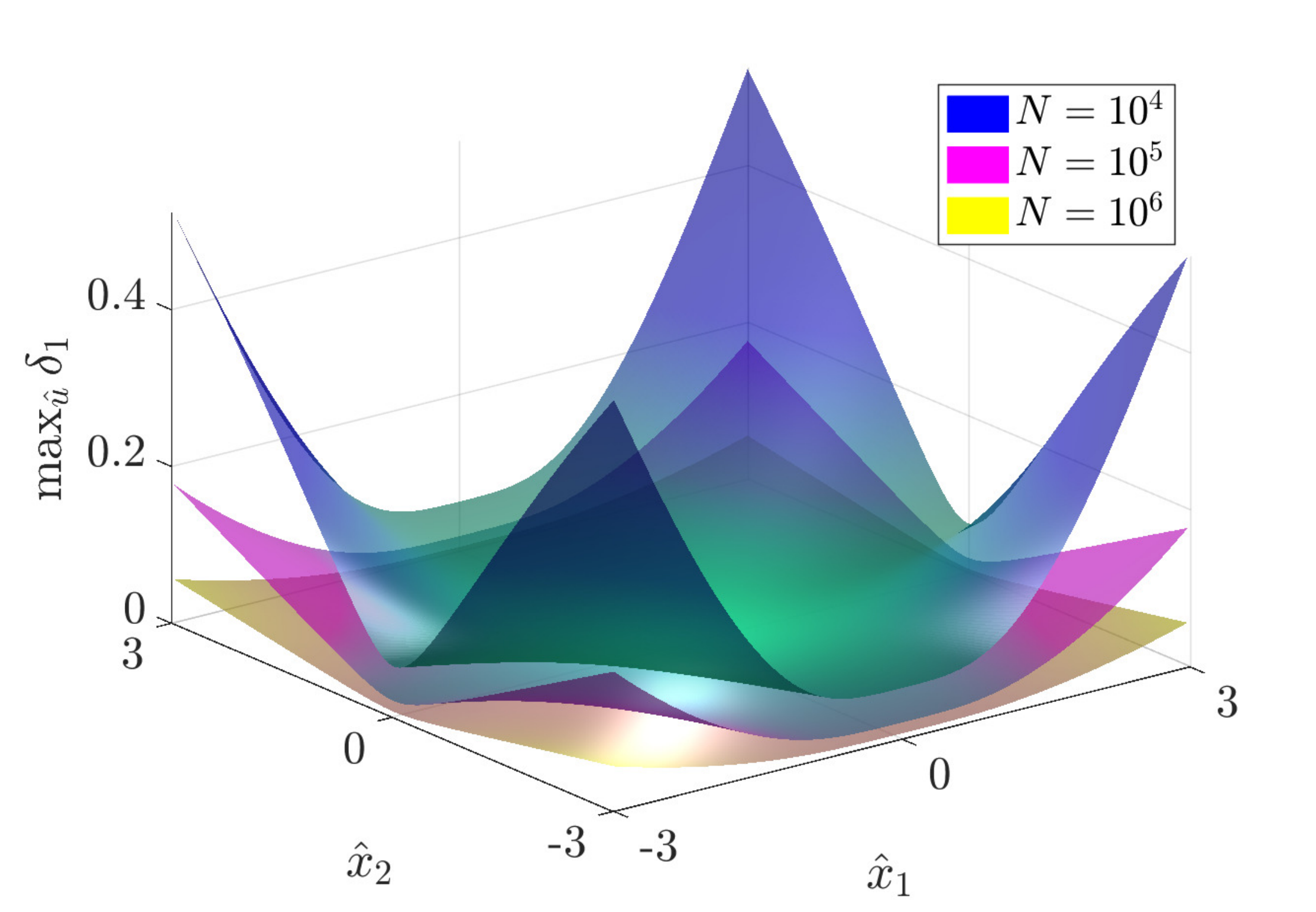}\\
	%\vspace{0.1cm}
	\includegraphics[width=0.85\columnwidth]{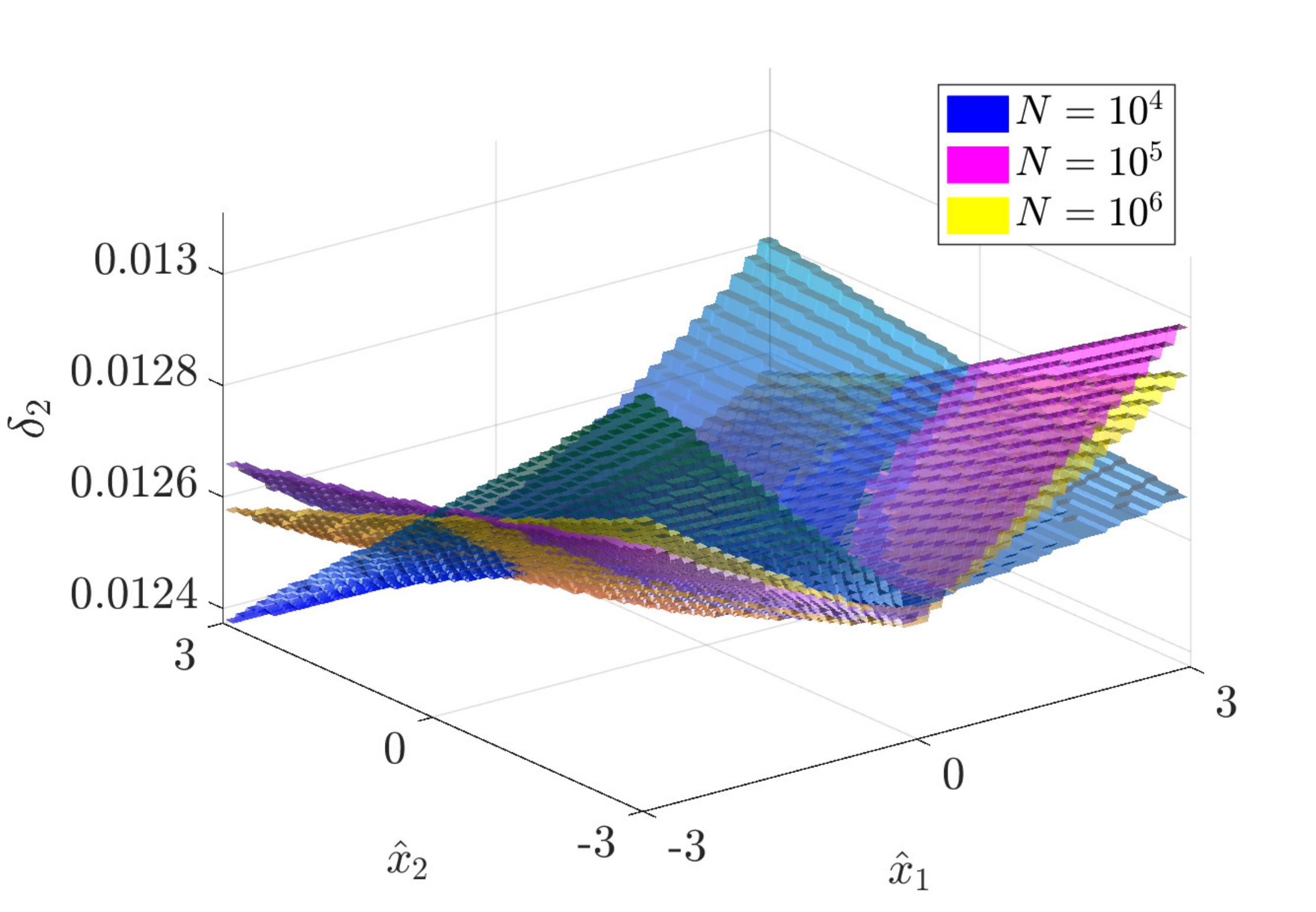}
	%\label{fig:vdPol_delta2}
	\caption{Maximum value of $\delta_1$ (top) with respect to the input as a function of the initial state, and $\delta_2$ as a function of the initial state (bottom) for $N=10^4,10^5,10^6$ for the package delivery case study with a confidence bound of $(1-\alpha)=0.9$.
%		\Sadegh{add legend}
}
	\label{fig:deltas}
 \vspace{-0.3cm}
\end{figure}

\begin{figure}
	\centering
	\includegraphics[width=0.85\columnwidth]{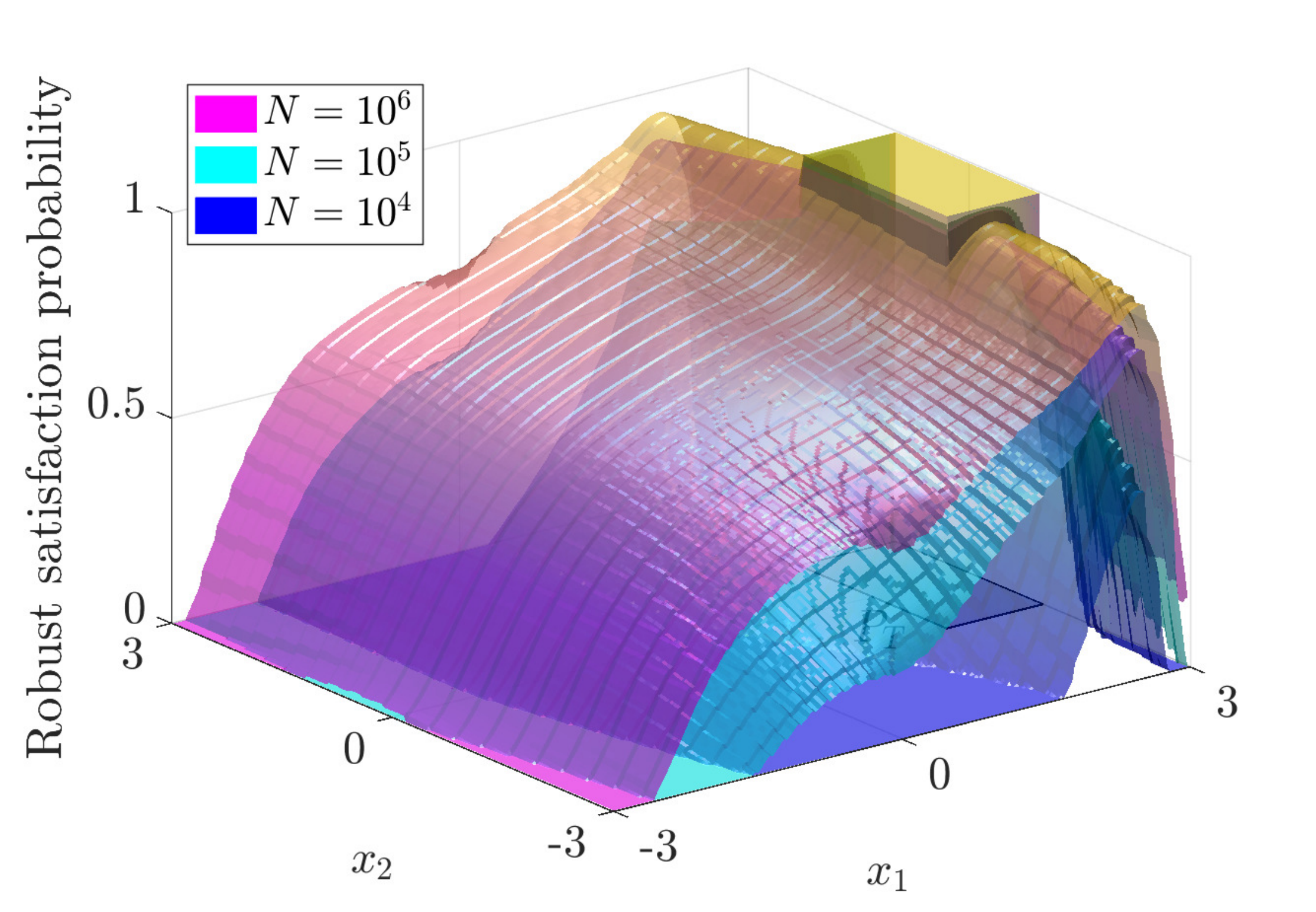}
	\caption{Robust satisfaction probability as a function of the initial state for $N=10^4,10^5,10^6$ (bottom to top) for the Van der Pol Oscillator with a confidence bound of $(1-\alpha)=0.9$.}
	\label{fig:satProb_VDP}
 \vspace{-0.3cm}
\end{figure}

For estimating the parameters, we sample uniform inputs $u$ from $\U$, set the prior distributions to $\mathcal N(\theta_j|0,10I_4)$, $j\in\{1, 2\}$, and draw $N$ data points from the true system with parametrization $\theta^\ast = [1, \tau_s, 0, 0; -\tau_s, 1+\tau_s a, -\tau_s a, 1]\T$, where $a = 0.9$ and $\tau_s=0.1$.
Following similar steps as in previous case studies, we obtain the models $\Mh, \Mt$ (discretizing $\Xh$ into $200^2$ and $\Uh$ into $5$ partitions) and corresponding simulation relations with $\eps_1=0$, $\max_{\hat u} \delta_1$ in Fig.~\ref{fig:deltas} (top), $\eps_2=0.1$, and $\delta_2$ given in Fig.~\ref{fig:deltas} (bottom).
The probability of satisfying the specification with $\eps=\eps_1+\eps_2$ and $\delta=\delta_1+\delta_2$ is computed based on Theorem~\ref{thm:satProb} and is given in Fig.~\ref{fig:satProb_VDP} as a function of the initial state of $\M(\theta)$.

\begin{comment}
\begin{figure*}[!tbp]
	\centering
	\subfloat[$\delta_{\M\rightarrow\Mh}$]{\includegraphics[width=0.8\columnwidth]{VDPDeltaOverStateSpace.pdf}\label{fig:delta}}
\quad
	\subfloat[$\delta_{\Mh\rightarrow\Mt}$]{\includegraphics[width=\columnwidth]{vdPol_delta2.pdf}\label{fig:vdPol_delta2}}
	\caption{Values of $\delta_{\M\rightarrow\Mh}$ (a) and $\delta_\absstwo$ (b) for different points in the state space. Since $\delta_{\M\rightarrow\Mh}$ is rapidly growing towards 1 for states away from the central region of the state space, it makes sense to calculate $\delta_{\M\rightarrow\Mh}$ state dependently.%\Oliver{Use either these two plots or \ref{fig:delta} and \ref{fig:vdPol_delta2}.}
	}
	\label{fig:deltas}
\end{figure*}
\end{comment}

%\begin{figure}[h]
%	\centering
	%		\def\svgwidth{\columnwidth}
	%		\input{value_function_LTI_eps0001.pdf_tex}
%	\includegraphics[width=\columnwidth]{VDPDeltaOverStateSpace.pdf}
%	\caption{Values of $\delta_{\M\rightarrow\Mh}$ for different points in the state space. Since $\delta_{\M\rightarrow\Mh}$ is rapidly growing towards 1 for states away from the central region of the state space, it makes sense to calculate $\delta_{\M\rightarrow\Mh}$ state dependently.}
%	\label{fig:delta}
%\end{figure}

%\begin{figure}[h]
%	\centering
%	\includegraphics[width=\columnwidth]{vdPol_delta2.pdf}
%	\caption{Values of $\delta_\absstwo$ for different points in the state space.}
%	\label{fig:vdPol_delta2}
%\end{figure}

\section{Discussion and Conclusions}
\label{sec:discussAndExtend}
%\pdfmargincomment{Half to one column. What are the challenges.}
% The approach presented in this paper is the first to integrate parameter identification techniques with robust abstraction-based methods for stochastic systems.
We presented a new simulation relation for stochastic systems that can establish a quantitative relation between a parameterized class of models and a simple abstract model.
With this, it extends the applicability of previous work and allows us to synthesize robust controllers from data and provide quantified guarantees for unknown nonlinear systems with unbounded noise and complex infinite-horizon specifications.
Moreover, the approach is compatible with \emph{model-order reduction} techniques and can hence be applied to higher-dimensional systems.
Properties expressed as scLTL specifications can be represented as DFAs (Definition~\ref{def:dfa}). 
	This allows us to reformulate the satisfaction of an scLTL specification as a reachability problem on the product system composed of the DFA and the abstracted system.
Note that the class of properties studied in this work subsumes signal temporal logic \cite{maler2004STL, lindemann2023risk}.
 % \red{Add a sentence somewhere that the class of properties studied in our approach subsumes signal temporal logic, and then refer to 1-2 references (a data-driven paper by Lars/Dimos and the original paper of oded Maler who introduced STL .)}\Oliver{Done. Lars doesn't have many data-driven papers and I didn't find any with him and Dimos that fit.}
The method described in this work is naturally extendable to \emph{linear temporal logic} specifications over \emph{finite-traces} (LTL\textsubscript{f}) using the native approach outlined in \cite{Wells2020ftLTL}.
LTL\textsubscript{f} semantics are useful for formalizing finite-horizon planning problems.
With the approach in \cite{Wells2020ftLTL}, our results for SSRs can be extended to LTL\textsubscript{f} by first translating the LTL\textsubscript{f} formula to an equivalent first-order logic (FOL) formula \cite{DeGiacomo2013ftLTL2FOL} and subsequently generating a minimal DFA using, e.g., the tool MONA \cite{Henriksen1995MONA}.
Thereafter, the procedure remains the same, by constructing a product MDP and solving the reachability problem (cf. \cite{haesaert2020robust}).

%\smallskip

%Extensions
% will address the following limitations of the presented approach.
Although we focus on systems with Gaussian support here, the approach is generally applicable to other (uncertain) disturbance distributions as demonstrated in \cite{Schon2023GMM}. 
%We are currently working on deriving the mathematical details that make the approach applicable to any distribution of the disturbance. \Birgit{You already referred to this in the paper. So it is no longer future work. Either mention here that we have this extension (with ref to paper) or remove here (since it is already mentioned in paper twice).}
%
In order to achieve formal guarantees when only noisy measurement data is available (i.e., when the observations are extended to have probability distributions conditioned on the state), relation \eqref{eq:relation_nonlin} has to be relaxed since the state is not observed directly. The coupling in this paper has to be adapted to accommodate this.
\bibliographystyle{abbrv}

\bibliography{references.bib}
\appendix
%\vspace{-0.5cm}
%\Oliver{Depending on how we structure the paper, I will update the proofs.}
%\color{blue!70!black}
\subsection{Proof of Theorem~\ref{thm:paraId_nonlin}}\label{app:proofSSRnonlinear}
	We first show that \eqref{eq:subprobcoup_nonlin} is a sub-probability coupling of $\hat \Tr(\cdotx|\hat x,\ach{})$ in \eqref{eq:tr_nonlin} and $\Tr(\cdotx|x,\ach{};\theta)$ in \eqref{eq:tr_hat} over $\R$ in \eqref{eq:relation_nonlin}. We show that they satisfy the conditions of Definition~\ref{def:submeasure_lifting} and derive $\delta$. %(App~\ref{app:def_sub_a_nonlin}-\ref{app:def_sub_c_nonlin}).
	%		(App.~\ref{app:def_sub_a_nonlin}-\ref{app:def_sub_c_nonlin})
	Then, we show that \eqref{eq:statemap_nonlin} is a valid control refinement.
	%(App~\ref{app:def_valCoRef_nonlin}).
	%
	To make the notation more compact, we define the tuple $z:=(\xh{},\x{},\ach{})$, where $z\in\Z$ with $\Z:=\{ (\xh{},\x{},\ach{}) \given (\xh{},\x{})\in\R \text{ and } \ach{}\in \Ah \}$.
	%we obtain concrete probability measures, namely $\Wt(\dxhp\times\dxp|z;\theta)$, $\Trh(\dxhp|z)$, and $\Tr(\dxp|z;\theta)$.
	%\Oliver{Should I write $\Trh(\dxhp|\hat x, u)$, and $\Tr(\dxp|\hat x, \hat u;\theta)$ here? I'm using $z$ because it does not matter how I denote it.}
	
	\subsubsection{Definition~\ref{def:submeasure_lifting}(a)}\label{app:def_sub_a_nonlin}
	For all $z\in\Z$ and $\theta\in\Theta$ we first show that $\Wt$ is entirely located on $\R$, i.e., $\Wt(\Xh\times\X|z;\theta) = \Wt(\R|z;\theta)$. We start by integrating $\Wt$ over $\Xh\times\X$.
	From Eq.~\eqref{eq:subprobcoup_nonlin} we get that for all $z\in\Z$ and $\theta\in\Theta$ we have 
%	\Birgit{Little bit strange that you have $x \in \X$, but not a set associated to $w$ here. }
{\allowdisplaybreaks
	\begin{align}
		&\Wt(\Xh\times\X|z;\theta)\nonumber\\&
		=  \int_{\xhp\in\Xh} \int_{\xp\in\X} \int_{\hat w\in\hat{\mathbb{W}}} \int_{w\in{\mathbb{W}}}
%			\int_{\hat w\in\Xh}\int_{w\in\X} \!
		\delta_{\hat \theta\T\! f(\xh{}, \uh{})+\hat w}(\dxhp)\delta_{\offset(\theta)+w}(d\hat w)\nonumber\\
		&\hspace{5pt}\times
		\delta_{\theta\T\! f(x, u)+w}(\dxp)  \min\{\mathcal N(dw|0,\Sigma),\mathcal N(dw|-\offset(\theta),\Sigma)\}\nonumber,\\
%	& =   \int_{\hat w}\int_w \int_{\xhp\in\Xh} \int_{\xp\in\X} \!
%		\delta_{\hat \theta\T\! f(\xh{}, \uh{})+\hat w}(\dxhp)
%		\delta_{\theta\T\! f(x, u)+w}(\dxp)\nonumber \nonumber\\
%		&\hspace{20pt}\times \delta_{\offset+w}(d\hat w) \min\{\mathcal N(dw|0,\Sigma),\mathcal N(dw|-\offset,\Sigma)\}\nonumber,\\
%		& \new{=   \int_{\hat w}\int_w \int_{\xhp\in\Xh} \int_{\xp\in\X} \!
%		\delta_{\hat \theta\T\! f(\xh{}, \uh{})+\offset+ w}(\dxhp)
%		\delta_{\theta\T\! f(x, u)+w}(\dxp)\nonumber \nonumber}\\
%		&\new{\hspace{20pt}\times  \min\{\mathcal N(dw|0,\Sigma),\mathcal N(dw|-\offset,\Sigma)\}\nonumber,}\\
		\begin{split}
			&=  \int_{\xhp\in\Xh} \int_{\xp\in\X}  \int_{w\in{\mathbb{W}}}
				\delta_{ \xp}(\dxhp)
				\delta_{\theta\T\! f(x, u)+w}(\dxp)\\
			&\hspace{20pt}\times  \min\{\mathcal N(dw|0,\Sigma),\mathcal N(dw|-\offset(\theta),\Sigma)\}.
		\end{split}
		\label{eq:intermediatecouplingequality}
%	& \geq  \int_w  \min\{\mathcal N(dw|0,\Sigma),\mathcal N(dw|-\offset,\Sigma)\},\label{eq:minNorms}
\end{align}
}
where we used \eqref{eq:relation_nonlin}, \eqref{eq:gamma_paraId_nonlin}, and the interface function $u\equiv\hat u$ in the last step.
%\red{with $\Gamma:=\int_{\hat w}\int_w \int_{\xhp\notin\Xh} \int_{\xp\notin\X} \!
%\delta_{\hat \theta\T\! f(\xh{}, \uh{})+\hat w}(\dxhp)
%\delta_{\theta\T\! f(x, u)+w}(\dxp)\nonumber \nonumber\\
%\times \delta_{\offset+w}(d\hat w) \min\{\mathcal N(dw|0,\Sigma),\mathcal N(dw|-\offset,\Sigma)\}$ (zero, e.g., for $\X=\Xh=\mathbb{R}$) and}
%where we have changed the order of integration.
%
\begin{comment}
	\begin{align*}
		&\Wt(\Xh\times\X|z;\theta) =
%		 \int_{\xhp\in\Xh} \int_{\xp\in\X} \Wt(\dxhp\times \dxp|z;\theta)\\
		 \int_{w} \int_{\xhp\in\Xh} \int_{\xp\in\X} \delta_{\hat\theta\T\! f(x,u)+w+\offset(z,\theta)}(\dxhp) \\
		 &\hspace{5pt} \times \delta_{\theta\T\! f(z)+w}(\dxp)
		\min\{\mathcal N(dw|0,\Sigma),\mathcal N(dw|-\offset(z,\theta), \Sigma)\}.
	\end{align*}
	Since by appropriate choice of $(\Xh,\X)$ we have that for all $w\sim p_w(\cdot)$ we can always find corresponding consecutive states $(\xhp, \xp)$ in $\Xh\times\X$, i.e., $\forall z\in\Z\,\forall\theta\in\Theta\,\forall w\sim p_w(\cdot)\,\exists(\xhp, \xp)\in\Xh\times\X:\, \xhp = \hat\theta\T\!f(z)+w+\offset(z,\theta) \andltl \xp=\theta\T\!f(z)+w$, 		
	\begin{equation}
		\Wt(\Xh\times\X|z;\theta) = \int_{w}\min\{\mathcal N(dw|0,\Sigma),\mathcal N(dw|-\offset(z,\theta), \Sigma)\}.\label{eq:minNorms}
	\end{equation}
	\end{comment}
		%
	Similarly, we integrate \eqref{eq:subprobcoup_nonlin} over $\R$ to get that for all $z\in\Z$ and $\theta\in\Theta$ we have
	\begin{align*}
		&\Wt(\R|z;\theta) 
%		= \int_{w} \int_{(\xhp, \xp)\in\R} \delta_{\hat\theta\T\! f(z)+w+\offset(z,\theta)}(\dxhp) \delta_{\theta\T\! f(z)+w}(\dxp)\\
%		& \min\{\mathcal N(dw|0,I),\mathcal N(dw|-\offset(z,\theta), I)\}\\
%		&= \int_{w} \int\!\!\!\!\int_{(\xhp\!,\xp{})\in\R}\!\!\!\!
%		\delta_{\hat \theta\T\! f(\xh{}, \uh{})+w+\gamma}(\dxhp)
%		\delta_{\theta\T\! f(x, u)+w}(\dxp)\nonumber\\
%		&\hspace{80pt}\times \min\{\mathcal N(dw|0,\Sigma),\mathcal N(dw|-\offset,\Sigma)\}\nonumber,\\
		 %&\!\!\delta_{\hat\theta\T\! f(\h)+w+\offset(z,\theta)}(\dxp) \delta_{\theta\T\! f(z)+w}(\dxp)\\ &\hspace{70pt}\times\min\{\mathcal N(dw|0,\Sigma),\mathcal N(dw|-\offset(z,\theta), \Sigma)\}.
		 = \int_{w\in{\mathbb{W}}} \int\!\!\!\!\int_{(\xhp\!,\xp{})\in\R}\!\!\!\!
		 \delta_{\xp}(\dxhp)
		 \delta_{\theta\T\! f(x, u)+w}(\dxp)\nonumber\\
		 &\hspace{80pt}\times \min\{\mathcal N(dw|0,\Sigma),\mathcal N(dw|-\offset(\theta),\Sigma)\}\nonumber,\\
		 &=  \int_{\xhp\in\Xh} \int_{\xp\in\X}\int_{w\in{\mathbb{W}}}
		 	\delta_{\xp}(\dxhp)
		 	\delta_{\theta\T\! f(x, u)+w}(\dxp)\nonumber\\
		 &\hspace{80pt}\times \min\{\mathcal N(dw|0,\Sigma),\mathcal N(dw|-\offset(\theta),\Sigma)\}\nonumber,
	\end{align*}
	%By similar reasoning, we have that for an offset of $\Tr(\dxhp|z;\theta)$ and $\Trh(\dxhp|z)$ 
	where the choice of $\R:=\{ (\hat x,x)\in\Xh\times\X \given x=\hat x\}$, interface $ u \equiv \hat u$, and $\offset(x,u,\theta) = (\theta-\hat\theta)\T f(x,u)$ ensures that $(\xhp,\xp{})\in\R$ for $(\xh,\x{})\in\R$. 
	With \eqref{eq:intermediatecouplingequality}
%	Therefore, the integration of two Dirac delta measures over $\R$ becomes one \red{(mind the $\Gamma$ as before)}, and
	we get $\Wt(\R|z;\theta) = \Wt(\Xh\times\X|z;\theta)$, thus Definition~\ref{def:submeasure_lifting}(a) is satisfied.
	\subsubsection{Definition~\ref{def:submeasure_lifting}(b)}\label{app:def_sub_b_nonlin}
	For any measurable set $S \subset \hat \X$, we integrate $\Wt$ in \eqref{eq:subprobcoup_nonlin} over $S\times\X$ to get that for all $z\in\Z$ and $\theta\in\Theta$ the coupling $\Wt$ satisfies condition (b) of Definition~\ref{def:submeasure_lifting}:
	\begin{align*}
		&\Wt(S \times \X|z;\theta) 
%		&=  \int_{w}  \int_{\xp\in \X} \delta_{\theta\T f(z)+w}(\dxp) \int_{\xhp\in S} \delta_{\hat\theta\T f(z)+w+\offset(z,\theta)}(\dxhp)  \min\{\mathcal N(dw|0,I),\mathcal N(dw|-\offset(z,\theta), I)\}\\
%		=  \int_{w}  \int_{\xhp\in S} \delta_{\hat\theta\T\! f(z)+w+\offset(z,\theta)}(\dxhp)\\ 
%		&\hspace{25pt}\times  \min\{\mathcal N(dw|0,I),\mathcal N(dw|-\offset(z,\theta), I)\}\\
		\leq \int_{w\in{\mathbb{W}}}  \int_{\xhp\in S} \delta_{\hat\theta\T\! f(\xh{},\uh{})+w}(\dxhp)\\ 
		&\hspace{80pt}\times  \min\{\mathcal N(dw|\offset(\theta),\Sigma),\mathcal N(dw|0, \Sigma)\},\\
		&\hspace{0pt}\leq \int_{w\in{\mathbb{W}}}  \int_{\xhp\in S} \delta_{\hat\theta\T\! f(\xh{},\uh{})+w}(\dxhp) \mathcal N(dw|0, \Sigma) = \Trh(S|z),
	\end{align*}
	where the first inequality becomes equality for unbounded $\X$.
%\pdfmargincomment{the smaller equal to is because when $\X$ is bounded and the marginal does not fully resolve}\footnotetext[1]{Due to choice of $\X,\Xh$.}
	%Note that, as before, we assumed appropriate choice of $\X$.
	 % we have $\forall z\in\Z,\forall\theta\in\Theta,\forall w\sim p_w(\cdot),\exists \xp\in \X:\, \xp = \theta\T f(z)+w$.
	%\Oliver{We need we have $\forall z\in\Z,\forall\theta\in\Theta,\forall w\sim p_w(\cdot),\exists \xp\in \X:\, \xp = \theta\T f(z)+w$.}
	\subsubsection{Definition~\ref{def:submeasure_lifting}(c)}\label{app:def_sub_c_nonlin}
	%Analogous to App.~\ref{app:def_sub_b_nonlin},
	 For any measurable set $S \subset \X$, we integrate $\Wt$ in \eqref{eq:subprobcoup_nonlin} over $\Xh\times S$ and get that for all $z\in\Z$ and $\theta\in\Theta$, $\Wt$ satisfies condition (c) of Definition~\ref{def:submeasure_lifting}:
	\begin{align*}
		\Wt(\Xh \times S|z;\theta) 
%		&=  \int_{w}  \int_{\xp\in S} \delta_{\theta\T\! f(z)+w}(\dxp)  \min\{\mathcal N(dw|0,I),\mathcal N(dw|-\offset(z,\theta), I)\}\\
		&\leq \int_{w\in{\mathbb{W}}}  \int_{\xp\in S}\!\!\!\!\!\! \delta_{\theta\T\! f(\xh{},\uh{})+w}(\dxp) \mathcal N(dw|0, \Sigma), \\
		&= \Tr(S|z).
	\end{align*}
	Hence, having proven that \eqref{eq:subprobcoup_nonlin} satisfies all conditions of Definition~\ref{def:submeasure_lifting}, \eqref{eq:subprobcoup_nonlin} is a valid sub-probability coupling.
	
	\subsubsection{Proving $\delta$ in \eqref{eq:delta_ddnonlinear}}
	\label{app:delta}
	We show that $\delta(\hat x,\hat u)$ in \eqref{eq:delta_ddnonlinear} satisfies $\Wt(\Xh\times\X|z;\theta)\geq 1-\delta(\hat x,\hat u)$. 
%	\Birgit{The way the Theorem is currenlty formulated, it seems like it is given that the two models are in an SSR. This would imply that the statement $\Wt(\Xh\times\X|z;\theta)\geq 1-\delta(\hat x,\hat u)$ is satisfied by definition. So why prove this? I suggest reformulating the theorem, such that you want to indeed prove a SSR.} 
	For this, we want to find the relation between $\delta(\xh{},\uh{})$ and $\offset(\xh{},\uh{},\theta)$. We recognize that $\Wt(\Xh\times\X|z;\theta)$ in \eqref{eq:intermediatecouplingequality} reduces to the minimum of two normal distributions
	\begin{equation*}
		\Wt(\Xh\times\X|z;\theta)
%		=   \int_{\xhp\in\Xh} \int_{\xp\in\X}\int_{\hat w}\int_w \!
%		\delta_{\hat \theta\T\! f(\xh{}, \uh{})+\hat w}(\dxhp)
%		\delta_{\theta\T\! f(x, u)+w}(\dxp)\nonumber\\
%		&\hspace{20pt}\times \delta_{\offset+w}(d\hat w) \min\{\mathcal N(dw|0,\Sigma),\mathcal N(dw|-\offset,\Sigma)\}\nonumber,\\
%		%	& =   \int_{\hat w}\int_w \int_{\xhp\in\Xh} \int_{\xp\in\X} \!
%		%		\delta_{\hat \theta\T\! f(\xh{}, \uh{})+\hat w}(\dxhp)
%		%		\delta_{\theta\T\! f(x, u)+w}(\dxp)\nonumber \nonumber\\
%		%		&\hspace{20pt}\times \delta_{\offset+w}(d\hat w) \min\{\mathcal N(dw|0,\Sigma),\mathcal N(dw|-\offset,\Sigma)\}\nonumber,\\
%		& \new{=   \int_{\hat w}\int_w \int_{\xhp\in\Xh} \int_{\xp\in\X} \!
%			\delta_{\hat \theta\T\! f(\xh{}, \uh{})+\offset+ w}(\dxhp)
%			\delta_{\theta\T\! f(x, u)+w}(\dxp)\nonumber \nonumber}\\
%		&\new{\hspace{20pt}\times  \min\{\mathcal N(dw|0,\Sigma),\mathcal N(dw|-\offset,\Sigma)\},}
		 \geq  \int_{w\in{\mathbb{W}}}  \min\{\mathcal N(dw|0,\Sigma),\mathcal N(dw|-\offset(\theta),\Sigma)\},%\label{eq:minNorms}
	\end{equation*}
	where we used the fact that the integral of the Dirac delta measure over the whole domain is at most equal to one.
	In fact, the inequality becomes an equality if the spaces $(\Xh,\X)$ are unbounded.
	Due to the symmetric property of the normal distribution, we get
	\begin{equation}
	\label{eq:half-space}
	\Wt(\Xh\times\X|z;\theta) =   2\int_{A} \mathcal N(dw|0,\Sigma),
	\end{equation}
	 with $A = \{w\in\mathbb{W}\,|\, \gamma\T\Sigma^{-1}(2w+\gamma)\le 0\}$. Note that the set $A$ is a half-space obtained by simplifying the inequality $\mathcal N(dw|0,\Sigma) \le \mathcal N(dw|-\gamma,\Sigma)$. The computation in \eqref{eq:half-space} can be interpreted as $\Wt(\Xh\times\X|z;\theta) = 2\mathbb P(\gamma\T\Sigma^{-1}(2w+\gamma)\le 0)$ with $w$ having normal distribution with mean zero and covariance $\Sigma$. This amounts to integrating a multivariate normal distribution over a half space. Note that in this case, $2\gamma\T\Sigma^{-1}w$ is a one-dimensional normal random variable with zero mean and variance $4\gamma\T\Sigma^{-1}\gamma$. Thus,
	    \begin{align*}
	\Wt(\Xh\!\times\!\X|z;\!\theta) \!=  2\!\cdot\!\cdf{\!\!\frac{-\gamma\T\Sigma^{-1}\gamma} {\sqrt{4\gamma\T\Sigma^{-1}\gamma}}\!\!}\!=\! 2\!\cdot\!\cdf{\!\!-\frac{1}{2}\!\sqrt{\gamma\T\Sigma^{-1}\gamma}\!}\!.
	\end{align*}
	 Therefore, we have $\Wt(\Xh\times\X|z;\theta)\geq 1-\delta(\hat x,\hat u)$, with
	 \begin{align}
	 \label{eq:delta_global}
		\delta(\hat x,\hat u) & := 1 - 2\cdot\cdf{-\frac{1}{2}\sqrt{\zeta(\hat x,\hat u)}},\\
		\zeta(\hat x,\hat u) & := \sup_{\theta\in\Theta}\left(\offset(\xh{},\uh{},\theta)\T\Sigma^{-1}\offset(\xh{},\uh{},\theta)\right).\nonumber
	\end{align}
	 %
	%refer to Lem.~6 in \cite{VanHuijgevoort2020SimQuant} to get a lower bound $\bar{\delta}(\hat x,\hat u)$, i.e.,
	Note that we calculate a $\delta$ dependent on $(\hat x,\hat u)$ to reduce the conservatism of our approach.
	Given the definition of $\gamma$ and the characterization of $\Theta$ in \eqref{eq:confidenceSetNL}, the computation of $\zeta(\hat x,\hat u)$ is through the optimization
	 \begin{align}
	 \label{eq:zeta}
		\zeta(\hat x,\hat u) & := \sup_\theta\,\,f(\hat x,\hat u)\T(\theta-\hat\theta)\Sigma^{-1}(\theta-\hat\theta)\T f(\hat x,\hat u)\\
		& \text{s.t.}\quad (\Delta\bar\theta)\T \Sigma_{N}^{-1}(\Delta\bar\theta)
		\leq n\cdot\chi^{-1}(1-\alpha| n),\nonumber
	\end{align}
	where $\Delta\bar\theta:=[\theta_1-\hat\theta_1;\ldots;\theta_n-\hat\theta_n]$. This is a quadratically constrained quadratic program, which is convex and can be solved efficiently using interior point methods. Here, we compute an upper bound for the optimal solution, which is sufficient for the purpose of getting a correct bound for $\delta$:
	\begin{align*}
	f(\hat x,&\hat u)\T(\theta-\hat\theta)\Sigma^{-1}(\theta-\hat\theta)\T f(\hat x,\hat u)\\
		&  = \Delta\bar\theta\T(I_n \otimes f(\hat x,\hat u)\T)\T \Sigma^{-1} (I_n \otimes f(\hat x,\hat u)\T) \Delta\bar\theta,\\
		& \le \norm{I_n \otimes f(\hat x,\hat u)\T}^2\norm{\Sigma^{-1}} \norm{\Delta\bar\theta}^2,\\
		&= \norm{ f(\hat x,\hat u)}^2 \norm{\Sigma^{-1}} \norm{\Delta\bar\theta}^2,\\
		& \le \norm{ f(\hat x,\hat u)}^2 \norm{\Sigma^{-1}}\frac{n\cdot\chi^{-1}(1-\alpha| n)}{\lambda_{min}(\Sigma_{N}^{-1})},\\
		& = \norm{ f(\hat x,\hat u)}^2 \norm{\Sigma^{-1}}\norm{\Sigma_{N}} \cdot n\cdot\chi^{-1}(1-\alpha| n).
	\end{align*}
	In the above, we have used the following (in)equalities:
	$x\T Ax\le \norm{A} \norm{x}^2$ and $x\T Ax\ge \lambda_{min}(A) \norm{x}^2$ for any vector $x$ and symmetric positive semi-definite matrix $A$; $\norm{AB}\le \norm{A}\cdot\norm{B}$ and $\norm{A\otimes B} = \norm{A}\cdot\norm{B}$ for any two matrices $A$ and $B$; and $\norm{I_n} = 1$. Therefore, we have
	\begin{align*}
		\sqrt{\zeta(\hat x,\hat u)}\le\norm{f(\hat x,\hat u)}\sqrt{r},
	\end{align*}
with $r:=\norm{\Sigma^{-1}}\norm{\Sigma_{N}} \cdot n\cdot\chi^{-1}(1-\alpha| n)$. Substituting this into \eqref{eq:delta_global} completes the proof of the expression for $\delta$ and shows that condition (b) of Definition~\ref{def:subsim} is satisfied.
Recall that condition (a) of Definition~\ref{def:subsim} holds by setting the initial states $\hat x_{0} = x_{0}$ and that condition (c) is satisfied with $\eps=0$ since both systems use the same output mapping.
It follows that $\widehat{\M}\preceq^{\delta}_\eps\M(\theta)$ with $\delta(\hat x,\hat u)$ in \eqref{eq:delta_ddnonlinear} and $\varepsilon=0$.

	\subsubsection{Definition~\ref{def:validrefinement}}\label{app:def_valCoRef_nonlin}
	It follows directly from  $\widehat{\M}\preceq^{\delta}_\eps\M(\theta)$ and Theorem~\ref{thm:exist_refine} that there exists a valid refinement.
	We find a composed probability measure $\Tr_\times$ of the state mapping in  \eqref{eq:statemap_nonlin} using \eqref{eq:compprobmeas} that bounds the sub-probability coupling $\Wt$ in \eqref{eq:subprobcoup_nonlin} from above for all $z\in\Z$ and $\theta\in\Theta$:
	\begin{align*}
		\Tr_\times(d\xhp\times d\xp|z,\theta) &=
%		\int_w \int_{\hat w} \delta_{\xp+\hat\theta\T\! (f(\hat z)-f(z))}(\dxhp)  \delta_{\theta\T\! f(z)+w}(\dxp)  \delta_{\offset(z,\theta)+w}(d\hat w) \mathcal N(dw|0, I)\\
%		&=
		\int_{w\in{\mathbb{W}}}\int_{\hat w\in\hat{\mathbb{W}}} \delta_{\hat\theta\T\! f(\xh{},\uh{})+w+\offset(\theta)}(\dxhp)\\ 
		&\hspace{25pt}\times  \delta_{\theta\T\! f(\xh{},\uh{})+w}(\dxp)  
%		\delta_{\offset(z,\theta)+w}(d\hat w)
		 \mathcal N(dw|0, \Sigma),\\
		&\geq \Wt(d\xhp\times d\xp|z;\theta).
	\end{align*}
	This concludes the proof of Theorem~\ref{thm:paraId_nonlin}.

\begin{IEEEbiography}[{\includegraphics[width=1in,height=1.25in,clip,keepaspectratio]{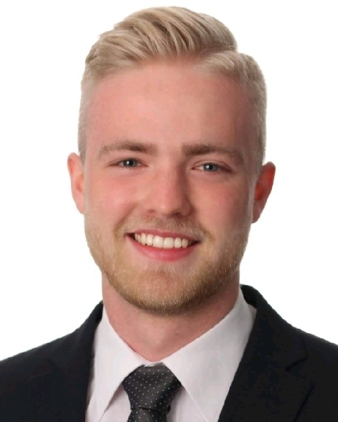}}]{Oliver Sch\"on} received the B.Sc. and M.Sc. degrees in mechanical engineering from Paderborn University, Paderborn, Germany, in 2018 and 2021, respectively. He is currently pursuing the Ph.D. degree in computer science in the School of Computing, Newcastle University, Newcastle upon Tyne, United Kingdom.
Prior to joining the School of Computing, he was a Research Assistant with the Department of Control Engineering and Mechatronics, Heinz Nixdorf Institute, Paderborn, Germany.
His research interests include the advance of model-based and data-driven correct-by-design synthesis for complex uncertain systems.
%development of efficient model-based and data-driven methods for controller synthesis of cyber-physical systems.
%
Mr. Sch\"on is a recipient of the Deutschlandstipendium Scholarship in 2015 and 2016, Erasmus+ European Scholarship in 2020, and Ph.D. Studentship of the School of Computing, Newcastle University in 2022.\\[-1cm]\mbox{ }
\end{IEEEbiography}

\begin{IEEEbiography}[{\includegraphics[width=1in,height=1.25in,clip,keepaspectratio]{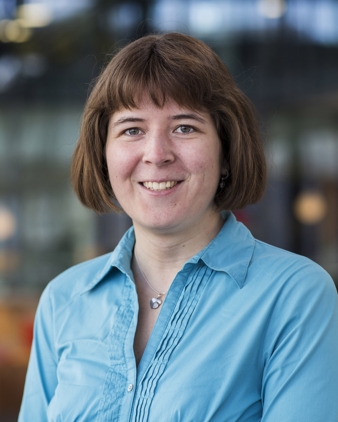}}]{Birgit van Huijgevoort} is a Postdoctoral Researcher at the Max Planck Institute for Software Systems, Germany. She received her B.Sc. degree cum laude in Electrical Engineering (automotive) in 2016 and her M.Sc. degree cum laude in Systems \& Control in 2018 from Eindhoven University of Technology (TU/e). In 2023, she received the Ph.D. degree from TU/e. 
Her research interests are in the field of correct-by-design control synthesis for uncertain cyber-physical systems with respect to temporal logic specifications.
\\[-1cm]\mbox{ }
\end{IEEEbiography}

\begin{IEEEbiography}[{\includegraphics[width=1in,height=1.25in,clip,keepaspectratio]{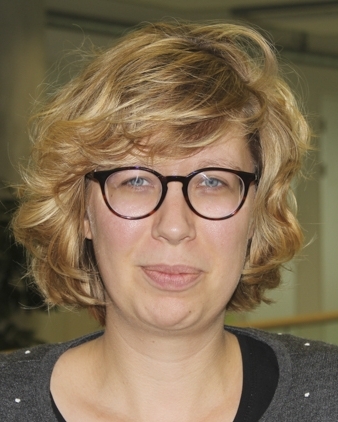}}]{Sofie Haesaert} is an Assistant Professor at the Control Systems group, Department of Electrical Engineering, Eindhoven University of Technology (TU/e), The Netherlands. From 2017 to 2018, she was a Postdoctoral Scholar at Caltech. She received her Ph.D. from TU/e in 2017. She received her B.Sc. degree cum laude in mechanical engineering in 2010 at the Delft University of Technology. In 2012, she received her M.Sc. degree cum laude in systems \& control at the Delft University of Technology, The Netherlands.
Her research interests are in the identification, verification, and control of cyber-physical systems for temporal logic specifications and performance objectives.
%
%Dr. Haesaert is a recipient of the €4m Award from the European Research Council for the project “SymAware” in 2022.
\\[-1cm]\mbox{ }
\end{IEEEbiography}

\begin{IEEEbiography}[{\includegraphics[width=1in,height=1.25in,clip,keepaspectratio]{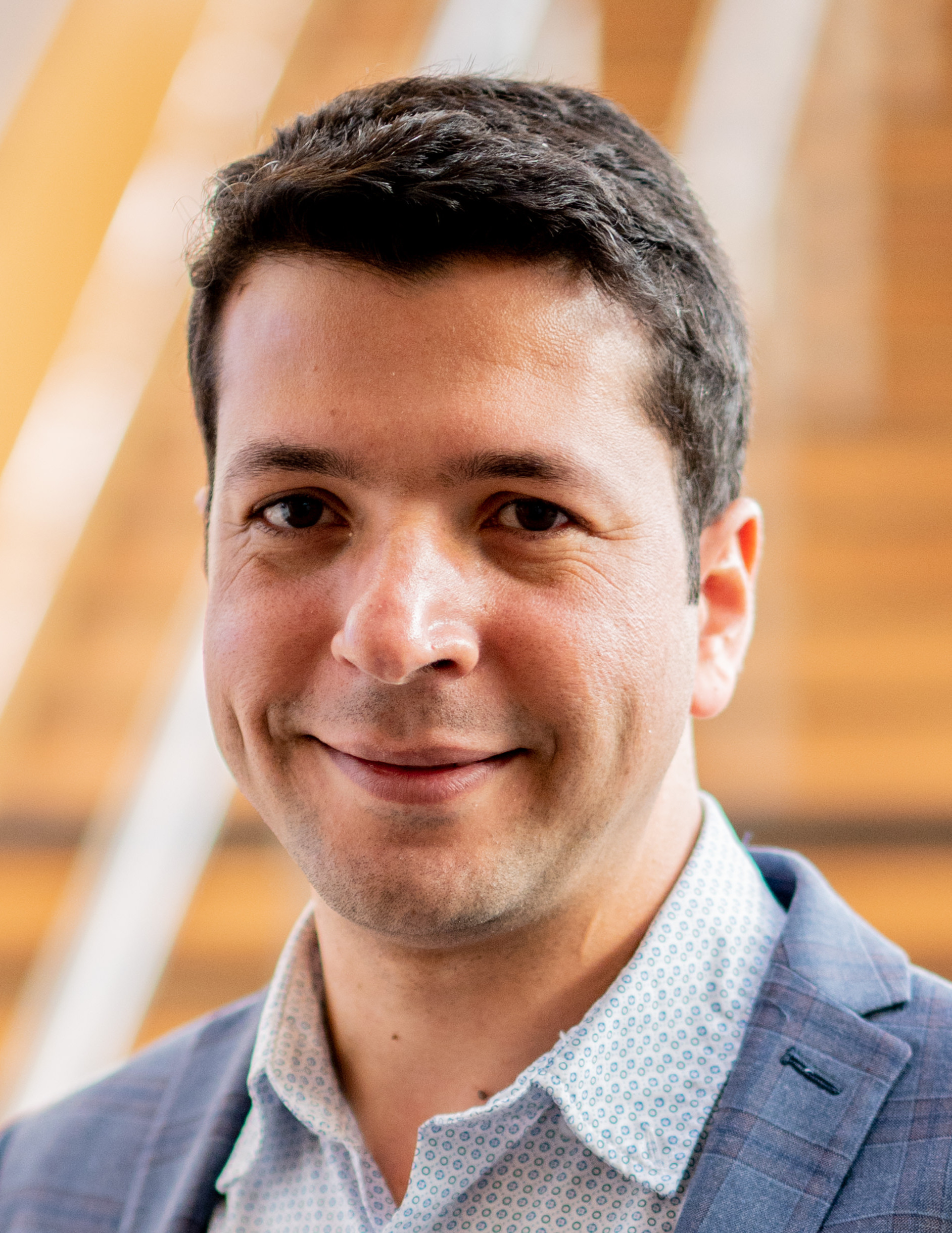}}]{Sadegh Soudjani} is a Research Group Leader at the Max Planck Institute for Software Systems, Germany. Previously, he was the Director of the AMBER Group at Newcastle University, United Kingdom, and Professor in Cyber-Physical Systems at Newcastle University.
He received the B.Sc. degrees in mathematics and electrical engineering, and the M.Sc. degree in control engineering from the University of Tehran, Tehran, Iran, in 2007 and 2009, respectively.
He received the Ph.D. degree in systems and control in November 2014 from the Delft Center for Systems and Control at the Delft University of Technology, Delft, the Netherlands.
Before joining Newcastle University, he was a Postdoctoral Researcher at the Department of Computer Science, University of Oxford, United Kingdom, and at the Max Planck Institute for Software Systems, Germany.

% Dr. Soudjani is the recipient of
% the ERC Consolidator Grant in 2023,
% the New Investigator Award from the UK EPSRC Research Council in 2021,
% Newcastle Teaching Award in 2020,
% QEST Best Paper Award in 2018,
% and
% DISC Best PhD Thesis Award in 2015.
% %
% His research interests are formal model-based and data-driven synthesis, abstraction, and verification of complex dynamical systems with application in cyber-physical systems, particularly, involving smart grids and energy networks.
\\[-1cm]\mbox{ }
\end{IEEEbiography}

\end{document}